\documentclass[12pt,letterpaper]{article}
\usepackage{epsfig,rotating,setspace,latexsym,amsmath,epsf,amssymb,amsfonts,bm,theorem,cite,caption,subcaption,enumerate,longtable,accents}
\usepackage{algorithm,algorithmic,graphicx,epsf,authblk,epstopdf,url,color,multirow}
\usepackage{mathtools,comment,tabularx}
\usepackage{comment}

\newtheorem{theorem}{Theorem}

\newtheorem{remark}{Remark}

\newtheorem{example}{Example}
\newenvironment{Proof}[1]{\medskip\par\noindent{\bf Proof:\,}\,#1}{{\mbox{\,$\blacksquare$}\par}}

\newcolumntype{Y}{>{\centering\arraybackslash}X}

\allowdisplaybreaks

\title{Fully Robust Federated Submodel Learning in a Distributed Storage System\thanks{This work was supported by ARO Grant W911NF2010142.}}

\author{Zhusheng Wang \qquad Sennur Ulukus\\
	\normalsize Department of Electrical and Computer Engineering\\
	\normalsize University of Maryland, College Park, MD 20742\\
	\normalsize  \emph{zhusheng@umd.edu} \qquad \emph{ulukus@umd.edu}}

\begin{document}
\date{}
\maketitle

\vspace*{-1.0cm}

\begin{abstract}
We consider the federated submodel learning (FSL) problem in a distributed storage system. In the FSL framework, the full learning model at the server side is divided into multiple submodels such that each selected client needs to download only the required submodel(s) and upload the corresponding update(s) in accordance with its local training data. The server comprises multiple independent databases and the full model is stored across these databases. An eavesdropper passively observes all the storage and listens to all the communicated data, of its controlled databases, to gain knowledge about the remote client data and the submodel information. In addition, a subset of databases may fail, negatively affecting the FSL process, as FSL process may take a non-negligible amount of time for large models. To resolve these two issues together (i.e., security and database repair), we propose a novel coding mechanism coined ramp secure regenerating coding (RSRC), to store the full model in a distributed manner. Using our new RSRC method, the eavesdropper is permitted to learn a controllable amount of submodel information for the sake of reducing the communication and storage costs. Further, during the database repair process, in the construction of the replacement database, the submodels to be updated are stored in the form of their latest version from updating clients, while the remaining submodels are obtained from the previous version in other databases through routing clients. Our new RSRC-based distributed FSL approach is constructed on top of our earlier two-database FSL scheme which uses private set union (PSU). A complete one-round FSL process consists of: 1) an FSL-PSU phase where the union of the submodel indices to be updated by the selected clients in the current round is determined, 2) an FSL-write phase where the updated submodels are written back to the databases, and 3) additional auxiliary phases where sufficient amounts of necessary common randomness are generated at both server and client sides. Our proposed FSL scheme is also robust against database drop-outs, client drop-outs, client late-arrivals, as well as any manipulations of database information by an active adversary who corrupts the communicated data. 
\end{abstract}

\section{Introduction}
In traditional federated learning (FL), a large number of clients collaboratively train a machine learning model in the server such that the local training data of each participating client is kept private from others \cite{FL, FL_Yangconcept, SecAgg}. In traditional FL, each client downloads and then updates the full learning model by using its own limited data, which can be inefficient in terms of the communication and computation overhead at both server and client sides, especially when the clients are resource-constrained mobile devices. To overcome this shortcoming, a new learning framework named federated submodel learning (FSL) is put forward in \cite{FSL}. In the FSL framework, the full learning model at the server is divided into multiple submodels according to the characteristics of the learning model data. By matching its local training data with the model characteristics, each participating client only needs to download the required submodel(s) from the server and then upload the corresponding submodel updates. There are two fundamental problems that can be abstracted out of the FSL framework: The first problem is, how can each client download its desired submodels without disclosing these submodel indices to the server. This is basically a \emph{private read} problem, which is equivalent to the private information retrieval (PIR) problem \cite{PIR_ORI}. In conventional PIR, a client wishes to retrieve a message out of a set of messages by communicating with the databases in the server without revealing the identity of the retrieved message to individual databases \cite{PIR,PIR_coded}. The second problem is, how can each client update/write-back these desired submodels still without disclosing the indices or the content of the updated submodels to the curious server. This is basically a \emph{private write} problem, which is tightly related to the oblivious random-access machine (ORAM) problem and the secure aggregation problem.

At present, there are a few different FSL approaches relying on different ideas. One class of approaches is based on ORAM. Assume that the storage in the server encompasses multiple data blocks with the same size. ORAM is introduced to hide the data access pattern from the server, namely, which blocks are read/written from/to the server \cite{ORAM}, by making sure that any two data access patterns are completely indistinguishable from the perspective of each individual database in the server. In general, the core idea behind ORAM is to shuffle and re-encrypt the storage data continuously once they are accessed. Most ORAM schemes are based on computational security \cite{DORAM,Path_ORAM}. Using the idea in ORAM as reference and the $X$-secure $T$-PIR scheme in \cite{XSTPIR,XSTPIR_MDS} as a building block, \cite{XSTFSL} puts forward an FSL approach to achieve information-theoretic security. Specifically, as the databases in the server are distributed \cite{DORAM}, the complete model is divided into submodels which are viewed as data blocks and encrypted by the server-side common randomness, and then stored across multiple distributed databases in the server. For each round of the FSL process, one client who is interested in accessing and updating a specific submodel participates in the training by sending two carefully-designed queries (a read query and a write query) to each database. Along this research line, a new technique called private read update write (PRUW) is proposed in order to further improve the total communication cost efficiency of FSL \cite{Sajani_FSL1}. In PRUW, a user downloads (reads), updates and uploads (writes) the increments back to the chosen data blocks while taking the privacy of the content downloaded, uploaded and their positions into account simultaneously. Through over-designing the PRUW with additional server-side common randomness in storage, the communication cost is decreased notably by combining all the updates into a single bit such that this bit can be decomposed and the corresponding updates can be placed as desired across the databases. Moreover, PRUW is extended by incorporating gradient sparsification where only a subset of the overall parameters in the full learning model is downloaded and updated \cite{Sajani_FSL_Trans, Sajani_FL_Trans}.

Another class of approaches is based on secure aggregation. As a critical building block of FL, secure aggregation aims to aggregate the locally trained model updates from a large number of clients at the server side in a secure manner, namely, no information about each client’s local training data is leaked to the others except that the aggregated result can be learned by the server \cite{SecAgg}. Most previous secure aggregation works concentrate on computational security, see e.g., \cite{SecAgg+, TurboAgg, FastSecAgg}. Using the secure aggregation idea and private set union (PSU) idea separately, an FSL approach is put forward in \cite{FSL} with computational security guarantees. Basically, in this approach, the server first calculates the union of the clients' desired submodels through a Bloom filter based PSU protocol. Then, through secure aggregation, the training model is updated by the clients within this submodel union at the sever side. The drawback of this approach is that the submodel union result is not accurate, and thus, the potential update efficiency of clients is not fully utilized. Recently, several secure aggregation protocols towards achieving information-theoretic security are proposed, see e.g., \cite{IT_SecAgg, IT_SecAgg_UGKey, IT_SecAgg_Region, LightSecAgg}. To overcome the shortcomings of the approach in \cite{FSL} and achieve an information-theoretic security, in our earlier work \cite{FSL-PSU}, we propose an FSL approach by unifying secure aggregation and PSU in the same framework. This idea is inspired by the fact that the fundamental features of secure aggregation and PSU are very similar. In \cite{FSL-PSU}, as in \cite{Prio} and PRUW, we consider the configuration of two databases at the server side, and then divide the clients into two groups. Thus, the second fundamental problem in FSL has close connections with PRUW and PSU; see detailed discussions in \cite{PIR_Extensions}.

In this work, we extend our previous approach in \cite{FSL-PSU} by considering more databases (more than two) at the server side. As the number of databases in the server is large now, several issues may arise, such as: a set of databases may be captured by an eavesdropper who can passively read database content and listen to database communications to capture database and client information \cite{EPIR}; a set of databases may fail \cite{Erasure_codes}. To solve the eavesdropping and database failure challenges in a distributed storage setting, \cite{Secure_RC} proposes secure regenerating codes, where the eavesdropper learns no useful information, and a replacement database can be constructed to replace the failed database by communicating with the remaining databases. Further, classical secret sharing refers to a setting where a secret is shared among multiple parties in such a way that any $t$ parties can recover the secret, but any fewer than $t$ parties learns nothing about the secret \cite{SS_IT}. In ramp secret sharing \cite{Ramp_SS}, a ramp zone is established such that, any $t$ parties can recover the secret, any $\tilde{t}$ parties learns nothing about the secret, and a set of parties whose cardinality is between $\tilde{t}$ and $t$ learns partial knowledge about the secret. This partial knowledge can be quantified by the mutual information and goes up as the cardinality of this set of parties in the ramp zone increases. 

For the distributed FSL problem in this paper, by combining the idea in ramp secret sharing \cite{Ramp_SS} together with the idea in secure regenerating codes \cite{Secure_RC}, we put forward a new customized coding scheme that we coin ramp secure regenerating code (RSRC) to develop an FSL scheme that is resilient to passive adversaries and database equipment failures. Hence, in our RSRC scheme, the useful information learned by the eavesdropper is quantifiable and controllable, and the performance of our RSRC-based distributed FSL scheme is adjustable. Further, a set of databases may be captured by an adversary who may actively overwrite the responses generated by its controlled databases \cite{BPIR}. In this case, the clients will receive erroneous information. Moreover, the databases may drop-out temporarily instead of collapsing permanently. As a consequence, the communication and computation at the server side can be delayed \cite{Speedup_ML}. Furthermore, the passive eavesdroppers and active adversaries may coexist simultaneously \cite{EBPIR}. Even worse, all four of these imperfections may happen simultaneously. In addition to these imperfections that may happen on the server side, we may have further problems on the clients side: Some clients may drop-out during the long-lasting FSL process or due to unstable client-database communication channels. Further, it is possible that some clients may not be able to transmit their generated answers to the server on time, which may cause the corresponding databases to conclude that clients have dropped-out. The late answers that come from these delayed clients may potentially disclose extra information about the local training data of these late-arriving clients to the databases \cite{SecAgg}. Our paper addresses all these issues mentioned above and deals with them altogether.

The performance of an FSL scheme is evaluated by three critical metrics: computation cost, communication cost, and storage cost. Towards achieving information-theoretic security, the schemes generally rely on operations in a finite field, which is the case in this work as well. As these operations are simple, the computation cost can be neglected. Moreover, since we are concentrating on distributed storage across the databases in this work, we consider only the server-side storage cost and neglect the client-side storage cost. Therefore, by storage cost, we only refer to the storage cost at the server side. In this paper, we put forward a new achievable scheme that is efficient in terms of communication cost and storage cost. We also prove that our proposed scheme is fully robust against permanent database failures, eavesdroppers, active adversaries, database drop-outs, client droup-outs, client late-arrivals, for one-round distributed FSL. This one-round distributed FSL scheme can be performed in an iterative manner until a predefined termination criterion is met.

\section{Problem Formulation} \label{Problem Formulation}
\begin{figure}[t]
\centering
\includegraphics[width=0.9\linewidth]{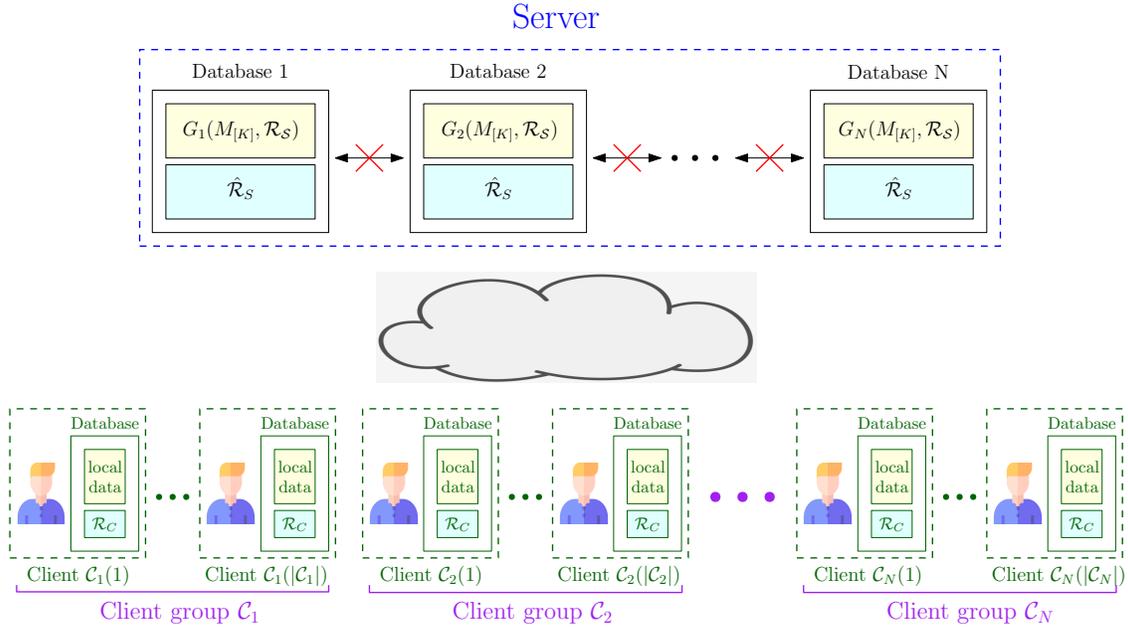}
\caption{Federated submodel learning (FSL) problem in distributed storage system.}
\label{System model}
\end{figure}

In this work, we consider a distributed FSL problem with one server that comprises $N$ independent databases and $C$ clients that are selected by the server to participate in one-round of the FSL process; see Fig.~\ref{System model}. By convention, each client at the user side establishes a direct secure and authenticated communication channel with each database at the server side.\footnote{Our distributed FSL scheme relies only
on the client-database communication for the sake of simplicity and stability.} In addition, the mutual communication among databases in the server is not required in this work. The full learning model\footnote{For any arbitrary positive integer $Z$, we use the notation $[Z] = \{1, 2, \dots , Z\}$ in this work for simplicity.} $M_{[K]} = \{M_1, M_2, \dots, M_K\}$ encompasses $K$ submodels, with each one consisting of $L$ i.i.d.~symbols that are uniformly selected from a finite field $\mathbb{F}_q$, and is stored across the databases. Thus, we have
\begin{align}
    H(M_k) &= L, \quad \forall k \label{Submodel Length} \\
    H(M_{[K]}) &= H(M_1) + H(M_2) + \dots + H(M_K)  = KL \label{Submodel IID}
\end{align}

For each $j \in [N]$, database $j$ takes as inputs the full model $M_{[K]}$ and the server-side common randomness $\mathcal{R}_S$, and stores the coded submodel information $G_j(M_{[K]},\mathcal{R}_S)$ by using its own encoder $G_j$. Some additional server-side common randomness $\hat{\mathcal{R}}_S$ in plain form is also stored across the databases to assist the execution of the FSL process. For each $i \in [C]$, client $i$ has its own data $\mathcal{D}_i$, which is used to train some submodels. Some necessary client-side common randomness $\mathcal{R}_C$ will be distributed to the clients before the FSL process starts. Following the client partition idea in \cite{FSL-PSU}, the large amount of $C$ clients are separated into $N$ groups according to their best communication channel bandwidth (or quality) with one specific database, i.e., if client $i$ has the optimum communication with database $j$ compared with the other databases, we assume that client $i$ belongs to the client group $\mathcal{C}_j$. As a consequence, we have $\mathcal{C}_{j_1} \cap \mathcal{C}_{j_2} = \emptyset$ for any $j_1, j_2 \in [N], j_1 \neq j_2$ and $\cup_{j \in [N]} \mathcal{C}_j = [C]$. Each client intends to update one or more submodels according to its local training data. Specifically, for $i \in [C]$, client $i$ wishes to update a set of submodels with its index set denoted by the random variable $\Gamma^{\langle i \rangle}$ (we use $\gamma^{\langle i \rangle}$ to denote the corresponding realization of $\Gamma^{\langle i \rangle}$). Moreover, for $i \in [C]$, we use the random variable $Y^{\langle i \rangle} = \{Y^{\langle i \rangle}_1,Y^{\langle i \rangle}_2,\dots,Y^{\langle i \rangle}_K\}$ to denote the corresponding incidence vector of $\Gamma^{\langle i \rangle}$ after being mapped to the alphabet as in \cite{PSI_journal,MP-PSI_journal}. 

Once a full round of the FSL process is complete, the submodels whose indices belong to the union $\Gamma = \Gamma^{\langle 1 \rangle} \cup \Gamma^{\langle 2 \rangle} \cup \cdots \cup \Gamma^{\langle C \rangle}$ are updated by the selected clients collaboratively while the remaining submodels remain the same. For $i \in [C]$, $k \in \Gamma$ and $l \in [L]$, the update $\Delta^{\langle i \rangle}_{k,l}$ is used to denote the corresponding increment generated in client $i$ for the submodel symbol $M_{k,l}$. If $k \notin \Gamma^{\langle i \rangle}$, the update $\Delta^{\langle i \rangle}_{k,l}$ is simply set as 0. Thus, for $k \in \Gamma$, the overall increment applied on the submodel symbol $M_{k,l}$ is $\sum_{i \in [C]} \Delta^{\langle i \rangle}_{k,l}$. The full increment sum vector over all submodels and over all symbols is defined as $\Delta_{\Gamma} = \{\sum_{i \in [C]} \Delta^{\langle i \rangle}_{k,l}, k \in \Gamma, l \in [L]\}$. Thus, the updated full learning model $M^{\prime}_{[K]}$ for any $l \in [L]$ is follows, 
\begin{align}
    M^{\prime}_{k,l} = 
    \begin{cases}
        M_{k,l} + \sum_{i \in [C]} \Delta^{\langle i \rangle}_{k,l}, & \text{if } k \in \Gamma\\
        M_{k,l}, & \text{otherwise}
    \end{cases}
\end{align}
Likewise, within the refreshed server-side common randomness $\mathcal{R}^{\prime}_S$ and $\hat{\mathcal{R}}^{\prime}_S$, the part coupled with the submodels in $M_\Gamma = \{M_k\}_{k \in \Gamma}$ should be updated. Thus, the storage of each database $j$ should be updated according to the up-to-date full model $M^{\prime}_{[K]}$ and refreshed server-side common randomness $\mathcal{R}^{\prime}_S, \hat{\mathcal{R}}^{\prime}_S$, while preserving its initial coding form. 

Let $\mathcal{M}_j$ denote all the information that can be attained by database $j$ including transmission information and storage information. Then, the FSL reliability in one-round of the FSL process is given by
\begin{align} \label{reliability}
    \text{[reliability]} \quad H(G_j(M^{\prime}_{[K]},\mathcal{R}^{\prime}_S),\hat{\mathcal{R}}^{\prime}_S|\mathcal{M}_j) = 0, \quad \forall j \in [N]
\end{align}

As introduced in \cite{FL}, the privacy constraint in FL typically requires that the aggregator learns nothing about clients’ individual inputs except for their sum. In FSL, since the full model is divided into multiple submodels, the privacy constraint needs to be tuned, namely, the aggregator learns nothing about clients’ local data except for their desired submodel union and submodel increment sum. Let $\mathcal{J} \subseteq [N]$ be an index set, then $\mathcal{M}_{\mathcal{J}} = \{\mathcal{M}_j\}_{j \in \mathcal{J}}$ is used to denote all the information involved in a set of databases whose indices belong to $\mathcal{J}$. Within this framework, we enforce that any set of databases with cardinality at most $J$ cannot infer any additional information about clients' local data $\mathcal{D}_{[C]} = \{\mathcal{D}_1,\mathcal{D}_2,\dots,\mathcal{D}_C\}$ beyond the union $\Gamma$ and the full increment sum vector $\Delta_{\Gamma}$, which is expressed by
\begin{align} \label{privacy}
    \text{[privacy]} \quad I(\mathcal{M}_{\mathcal{J}};\mathcal{D}_{[C]}|\Gamma,\Delta_{\Gamma}) = 0, \quad \forall \mathcal{J} \subseteq [N], ~ |\mathcal{J}| \leq J
\end{align}

Following the multi-user PIR/SPIR problem formulated in \cite{DoubleBlind_PIR,MultiBlind_SPIR}, it is also required that each participating client should not gain any knowledge about the other clients’ local data. Let $\mathcal{W}_i$ denote all the information that can be attained by client $i$ including transmission information and storage information, and let $\mathcal{D}_{\bar{i}}$ denote the set $\{\mathcal{D}_1,\dots,\mathcal{D}_{i-1},\mathcal{D}_{i+1},\dots,\mathcal{D}_C\}$. Then, we have the following inter-client privacy constraint,
\begin{align} \label{inter-client privacy}
    \text{[inter-client privacy]} \quad I(\mathcal{W}_{i};\mathcal{D}_{\bar{i}}) = 0, \quad \forall i \in [C]
\end{align}

Two different types of security threats to the FSL model are investigated in this paper. On the one hand, a passive eavesdropper can take control of any arbitrary $E$ databases at the server side. Let $\mathcal{E}$ be the set of indices corresponding to these $E$ databases where the cardinality of $\mathcal{E}$ is $E$. The eavesdropper can learn all the transmission information and storage information denoted by $\mathcal{M}_\mathcal{E}$. The eavesdropper is honest but curious in the sense that the goal of the eavesdropper is to obtain some additional information about the full learning model and clients' local data, but does not corrupt any transmissions. For simplicity, we assume that $E$ is smaller than or equal to $J$ in this work, the privacy constraint \eqref{privacy} implies that the eavesdropper cannot learn any knowledge about the clients' local data further than $\Gamma$ and $\Delta_{\Gamma}$. Hence, the information leakage to the eavesdropper can be measured only in the amount of up-to-date full learning model $M^{\prime}_{[K]}$. Not like the conventional configuration in which the eavesdropper should learn nothing about $M^{\prime}_{[K]}$ \cite{EPIR,XSTPIR}, we use the idea in \cite{ChaoTian_leakage,AleakyPIR} for reference and introduce a new parameter $\delta$, which is defined as the maximal fraction of latest full model information that can be learned by the eavesdropper. Thus, $\delta$ can be any rational number between $0$ and $1$.\footnote{If $\delta$ is equal to $1$, no eavesdropper security constraint is imposed in the problem formulation.} In the presence of a passive eavesdropper, we have the following security constraint,
\begin{align} \label{eavesdropper security}
     \text{[eavesdropper security]} \quad \frac{1}{KL} I(M^{\prime}_{[K]};\mathcal{M}_{\mathcal{E}}) \leq \delta, \quad \forall \mathcal{E} \subseteq [N], ~ |\mathcal{E}| = E
\end{align}

On the other hand, an active adversary can get command of any arbitrary $A$ databases at the server side to overwrite their transmissions to the clients. The adversary is Byzantine in the sense that the goal of the adversary is to intervene the normal running of the FSL process by generating arbitrarily erroneous information, without following the agreed upon protocol. The eavesdropper/adversary has an unlimited computational power and full knowledge of the FSL system. Neither the server nor the clients have any knowledge about the identities of the databases tapped in by the eavesdropper/adversary.

A basic one-round FSL achievable scheme under distributed coded storage is a one that satisfies the reliability constraint \eqref{reliability}, the privacy constraint \eqref{privacy}, the inter-client privacy constraint \eqref{inter-client privacy} and the eavesdropper security constraint \eqref{eavesdropper security}. A fully robust one-round FSL achievable scheme in distributed coded storage should satisfy these four basic constraints, especially the reliability constraint \eqref{reliability} at all times, even in the presence of active adversaries, database failures, database drop-outs, client drop-outs and client late-arrivals. Moreover, we also need to guarantee that this one-round distributed FSL scheme can be executed iteratively with no errors until a predefined termination criterion is satisfied. As discussed in the introduction, the performance of a fully robust FSL scheme is evaluated by two metrics: communication cost and storage cost, which are both measured in the number of $q$-ary bits. Our goal is to develop a fully robust FSL scheme for a given set of FSL system parameters such that the total communication cost and the total storage  are as small as possible.

\section{Main Result} 
The main contribution of this paper is a novel fully robust distributed FSL protocol. The performance of our protocol is evaluated in terms of the total communication cost and the total storage cost in each FSL round. The communication cost includes the cost incurred within the client-side and server-side common randomness generation. The main result of this paper is presented in the following theorem, which is proved in Section~\ref{Performance Evaluation}. 

\begin{theorem} \label{Main Theorem}
The total communication cost and the total storage cost of the proposed distributed FSL achievable scheme in one round are $\mathcal{O}(CK\!+\!C|\Gamma|L))$ and $\mathcal{O}(KL)$, respectively, where $C$ is the total number of participating clients, $K$ is the total number of submodels, and $|\Gamma|$ is the number of updated submodels in the given round.
\end{theorem}

\begin{remark}
    Our new distributed FSL protocol is constructed as a generalization of our previous two-database FSL scheme via PSU \cite{FSL-PSU}. Different than \cite{FSL-PSU}, here we have many databases at the server side. As in \cite{FSL-PSU}, we rely only on simple operations in a finite field at both client and server sides. Here, we achieve information-theoretic privacy for the clients as in \cite{FSL-PSU}, and additionally information-theoretic security against eavesdroppers. Unlike \cite{FSL-PSU}, here, the submodel information and server-side common randomness are stored across the databases in a coded form through our new RSRC technique. With RSRC, we achieve robustness against passive eavesdroppers, active adversaries and database failures, in addition to the existing resilience in \cite{FSL-PSU} against database drop-outs, client drop-outs and client late-arrivals.  
\end{remark}

\begin{remark}
    The communication cost in our new FSL protocol is order-wise the same as the one in \cite{FSL-PSU}. Hence, all the conclusions in \cite[Remark~6]{FSL-PSU} are applicable here. In particular, once a database fails, the order-wise communication cost incurred for the repair process is $\mathcal{O}(KL)$, which is not negligible compared with the communication cost in the normal FSL process. Note that when the RSRC mechanism is utilized, although the order-wise cost makes no difference, the communication cost of obtaining the desired submodel information at the beginning of the FSL-write phase, the communication cost of repairing the submodel information in the replacement database under a database failure, and the storage cost, can be decreased simultaneously, while permitting the eavesdropper to learn more information about the full learning model.
\end{remark}

\begin{remark}
    The storage cost in our new FSL protocol is $\mathcal{O}(KL)$, which is order-wise the same as storing the plain full learning model in the server. If we calculate the storage cost of the scheme in \cite{FSL-PSU}, it is also $\mathcal{O}(KL)$ including the server-side common randomness. Compared with previous FL or FSL approaches in \cite{SecAgg,FSL}, our storage cost $\mathcal{O}(KL)$ does not include the term $C^2$, which is incurred by the secret sharing scheme across the clients.
\end{remark}

\begin{remark}
    We analyze the required number of databases here. Note that each client needs to contact $D$ working databases out of $N$ databases in total to recover the desired submodels $M_\Gamma$, in the presence of an eavesdropper who controls $E$ databases, we must have $N > D > E$ in general. In order to preform the FSL process reliably, the total number of databases that drop-out or fail must be smaller than or equal to $N \!-\! D$. Furthermore, in the presence of an adversary who controls $A$ databases, we must have $N \geq \max(2A\!+\!D,(J\!+\!1)(2A\!+\!1))$ according to the analysis of active adversary robustness in Section~\ref{Full Robustness Verification}.
\end{remark}

\section{RSRC Technique} \label{RSRC}
In a distributed storage system, a set of messages are stored accross multiple databases either in a plain form or in a coded form. The secure regenerating code is developed such that in the presence of a passive eavesdropper or an active adversary, a client can recover the message (reconstruction process) and a replacement database can be built to replace the failed database (repair process) by communicating with some working databases in an efficient way \cite{Secure_RC}. To evaluate the performance of a secure regenerating code, three main metrics are considered: reconstruction communication cost that counts the total number of symbols transferred from the working databases to the client in the reconstruction process, repair communication cost that counts the total number of symbols transferred from the working databases to the replacement database in the repair process, storage cost that counts the total number of symbols needed for the customized storage across all the databases. Following our FSL system model in Section~\ref{Problem Formulation}, we assume that each client is permitted to contact $D$ working databases to download submodel recovery information (reconstruction process) and database repair information (repair process) if necessary. Note that the coded storage in database $j$ is $G_j(M_{[K]},\mathcal{R}_S)$ where $M_{[K]}$ and $\mathcal{R}_S$ denote all the message symbols and randomness symbols, respectively. The following three constraints must be guaranteed while applying a secure regenerating code to our FSL system. First, a client can recover $M_{[K]}$ by communicating with any $D$ working databases. This is referred to as the reconstruction constraint. Second, for all $j \in [N]$, a client can derive $G_j(M_{[K]},\mathcal{R}_S)$ by communicating with any $D$ remaining databases if database $j$ fails, and then forward this database repair information to the replacement database.\footnote{In a practical implementation, if this constraint is always satisfied, multiple clients can work together with each one routing part of $G_j(M_{[K]},\mathcal{R}_S)$ to the replacement database. Finally, the replacement database can still receive $G_j(M_{[K]},\mathcal{R}_S)$.} This is referred to as the repair constraint. Third, at most a fraction $\delta$ of $M_{[K]}$ can be learned by any arbitrary $E$ databases. This is referred to as the information leakage constraint.

In this paper, we propose a novel secure regenerating coding mechanism called RSRC, which is devised specifically for our FSL system. This secure regenerating coding scheme is inspired by the ramp secret sharing idea in \cite{Ramp_SS, Uniform_SS}.  Compared with previous secure regenerating code introduced in \cite{Secure_RC}, there are a number of innovative points in our coding scheme. First, the information leakage fraction $\delta$ is not limited to the choice of $0$. Second, in addition to the repair communication cost, reconstruction communication cost and storage cost are also considered for optimization. Third, in our coding scheme, we find that all of repair communication cost, reconstruction communication cost, and storage cost are certain simple functions of the parameter $\delta$. In our ultimate construction of our general FSL achievable scheme, we use our RSRC technique as an elementary building block.

\subsection{Construction and Performance of General RSRC} \label{General_RSRC}
Following the product-matrix code \cite{Product-Matrix_Codes}, we first define the encoding matrix $\Psi$. Encoding matrix $\Psi$ is basically an $N \times D$ Vandermonde matrix in the following form where all the elements $\{\psi_1,\psi_2,\psi_3,\dots,\psi_N\}$ included in this matrix are selected from a sufficiently large finite field $\mathbb{F}_q$ and are all distinct,
\begin{align} \label{General_Psi}
    \Psi = 
    \begin{bmatrix}
        1 & \psi_1 & \psi_1^2 & \cdots & \psi_1^{D-1} \\
        1 & \psi_2 & \psi_2^2 & \cdots & \psi_2^{D-1} \\
        1 & \psi_3 & \psi_3^2 & \cdots & \psi_3^{D-1} \\
        \vdots & \vdots & \vdots & \ddots & \vdots  \\  
        1 & \psi_N & \psi_N^2 & \cdots & \psi_N^{D-1} \\
    \end{bmatrix}_{N \times D}
\end{align}
Next, we define a message matrix $\Omega$ which is simply a $D \times D$ symmetric matrix. The $D(D+1)/2$ distinct symbols in $\Omega$ consist of two parts: message symbols $M_{[K]}$ and randomness symbols $\mathcal{R}_S$. Assume that the number of message symbols is $B$, then the number of randomness symbols is $D(D+1)/2 - B$. The code matrix $\zeta$ is obtained from the product of encoding matrix $\Psi$ and message matrix $\Omega$, i.e., $\zeta = \Psi \Omega$. For any $j \in [N]$, the $j$th row of $\zeta$ containing $D$ symbols denoted by $\zeta^T_j$ is stored in database $j$. As the size of $M_{[K]}$ is generally large, we apply this coding method multiple times in a duplicate way. The coded storage $G_j(M_{[K]},\mathcal{R}_S)$ is formed by concatenating all the generated $\zeta^T_j$. 

We use $C_1$ to denote the reconstruction communication cost, $C_2$ to denote the repair communication cost and $S$ to denote the storage cost in each database as the storage cost is uniform over all the available databases. Moreover, if we use $\ell_\lambda$ to measure the extent of the message information leakage in any $\lambda$ databases and $\mathcal{M}_\lambda$ to denote all the available information included in these $\lambda$ databases, we have $\ell_\lambda = \frac{I(M_{[K]};\mathcal{M}_\lambda)}{H(M_{[K]})}$. After normalizing $C_1, C_2, S$ by the total number of message symbols $B$, we derive the following theorem regarding the performance of our general RSRC scheme.

\begin{theorem} \label{RSRCtheorem}
    Given the values of $N$ and $D$ that satisfy $N > D \geq 2$, for any arbitrary integer $\lambda$ that satifies $0 < \lambda < D$ and any arbitrary rational number $\ell_\lambda$ that satisfies $0 \leq \ell_\lambda \leq 1$, an RSRC scheme with the following normalized performance can always be realized, 
    \begin{align} 
        \frac{C_1}{B} &\geq \begin{cases}
            \frac{D+\lambda+1}{D-\lambda+1}(1-\ell_\lambda), & \text{if } \ell_\lambda \leq \frac{2\lambda}{D+\lambda+1} \\
            1, & \text{otherwise}
        \end{cases} \label{C1cost} \\
        \frac{C_2}{B} &\geq  \begin{cases}
        \frac{2D}{(D-\lambda)(D-\lambda+1)}(1-\ell_\lambda), & \text{if } \ell_\lambda \leq \frac{2\lambda D - \lambda(\lambda-1)}{D(D+1)}\\
        \frac{2}{D+1}, & \text{otherwise}
        \end{cases} \label{C2cost}
        \\ 
        \frac{S}{B} &\geq \begin{cases}
        \frac{2D^2}{(D-\lambda)(D-\lambda+1)}(1-\ell_\lambda), & \text{if } \ell_\lambda \leq \frac{2\lambda D - \lambda(\lambda-1)}{D(D+1)}\\
        \frac{2D}{D+1}, & \text{otherwise}
        \end{cases} \label{Scost}
    \end{align}
\end{theorem}

\begin{remark}
    When $\ell_\lambda = 0$, the first inequality in Theorem~\ref{RSRCtheorem} reduces to the result in \cite{Secure_RC}.
\end{remark}

\begin{remark}
    According to the results in Theorem~\ref{RSRCtheorem}, when $0 \leq \ell_\lambda \leq \frac{2\lambda}{D+\lambda+1}$, the achieved normalized performance metrics of our RSRC scheme are all linear functions of the parameter $\ell_\lambda$. That means that, by allowing larger message information leakage, we can further reduce all of the costs simultaneously. When $\frac{2\lambda}{D+\lambda+1} \leq \ell_\lambda \leq \frac{2\lambda D - \lambda(\lambda-1)}{D(D+1)}$, the normalized $C_2$ and normalized $S$ can be further decreased if $\ell_\lambda$ is further increased, whereas the normalized $C_1$ reaches its minimum $1$. When $\frac{2\lambda D - \lambda(\lambda-1)}{D(D+1)} \leq \ell_\lambda \leq 1$, only the fraction $\frac{2\lambda D - \lambda(\lambda-1)}{D(D+1)}$ of message information is leaked, which is smaller than the given upper bound $\ell_\lambda$.
\end{remark}

\begin{Proof}
    We first determine the starting message matrix $\Omega$. As shown in Fig.~\ref{Message matrix}, we fill the upper left corner of size $(D\!-\!\lambda) \times (D\!-\!\lambda)$ with message symbols, whereas the remaining zone with randomness symbols. Thus, the total number of message symbols $B$ is $\frac{(D\!-\!\lambda)(D\!-\!\lambda\!+\!1)}{2}$ and the total number of randomness symbols is $\frac{D(D\!+\!1)}{2} \!-\! \frac{(D\!-\!\lambda)(D\!-\!\lambda\!+\!1)}{2} \!=\! \lambda D \!-\! \frac{\lambda(\lambda-1)}{2}$. We denote the current message matrix by $\Omega_1$. Following \cite[Thm.~11]{Secure_RC}, as a reduced version, any $\lambda$ databases gain no information about the message, i.e., $I(M_{[K]};\mathcal{M}_\lambda) = 0$ and $\ell_\lambda = 0$. From another perspective of information-theoretic security analysis, by just applying linear computation, $\lambda D$ coded symbols included in these databases can be transformed into $\lambda D$ equivalent symbols where $ \lambda D \!-\! \frac{\lambda(\lambda-1)}{2}$ symbols are associated with one distinct randomness symbol and $\frac{\lambda(\lambda\!-\!1)}{2}$ symbols are redundant.\footnote{This result can also be proved by simply using linear algebraic calculations where the conversion matrix is invertible. The number of redundant symbols comes from the symmetric property of the message matrix.} Afterwards, for the lower left corner with size $\lambda \times (D\!-\!\lambda)$, the existing randomness symbol is substituted one-by-one by new message symbols. Once a randomness symbol vanishes, a new symbol only involving message symbol appears. This is also the basic idea in the construction of ramp secret sharing on the basis of classical secret sharing \cite{Ramp_SS}. As a consequence, when the the lower left corner is filled with message symbols,\footnote{The upper right corner with size $(D\!-\!\lambda) \times \lambda$ is also filled with message symbols since the message matrix is always symmetric.} the total number of message symbols $B$ now becomes $\frac{(D\!-\!\lambda)(D\!+\!\lambda\!+\!1)}{2}$ and the total number of randomness symbols is $\frac{\lambda(\lambda+1)}{2}$. We denote this message matrix by $\Omega_2$. The green zone in Fig.~\ref{Message matrix} can be considered as a ramp zone in which the message information leakage increases gradually. At this point, $\lambda D$ coded symbols included in these databases are equivalent to $\lambda(D\!-\!\lambda)$ symbols only consisting of message symbols, $\frac{\lambda(\lambda+1)}{2}$ symbols associated with one distinct randomness symbol and $ \frac{\lambda(\lambda\!-\!1)}{2}$ redundant symbols. The message information leakage $\ell_\lambda$ in this scheme is $\frac{2\lambda(D\!-\!\lambda)}{D(D\!+\!1)\!-\!\lambda(\lambda+1)}$. Furthermore, if the whole message matrix is filled with message symbols without any randomness symbols, the total number of message symbols $B$ is thus $\frac{D(D\!+\!1)}{2}$. We denote this message matrix by $\Omega_3$. Out of $\lambda D$ coded symbols, $\frac{\lambda(\lambda\!-\!1)}{2}$ symbols are redundant. The message information leakage $\ell_\lambda$ in this scheme is $\frac{2\lambda D - \lambda(\lambda-1)}{D(D+1)}$. 
    \begin{figure}[t]
        \centering
        \includegraphics[width=0.45\linewidth]{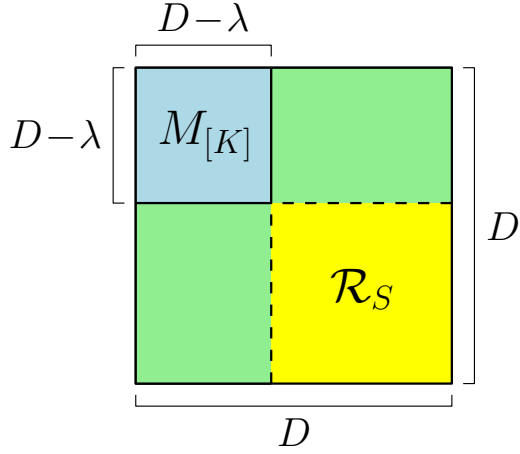}
        \caption{Structure of the $D \times D$ message matrix $\Omega$.}
        \label{Message matrix}
    \end{figure}   
    
    Regarding the message recovery in the reconstruction process, for the scheme with $\Omega_1$ or $\Omega_2$ or any message matrix in between, if all the coded symbols from these $D$ connected databases are put together to form a $D \times D$ matrix, we can only concentrate on the $D \times (D\!-\!\lambda)$ left submatrix as all the message symbols have already been included. Thus, as the encoding matrix $\Psi$ is a Vandermonde matrix, the client can download the full $D \times (D\!-\!\lambda)$ left submatrix except for the redundant upper right isosceles triangle (or lower right isosceles triangle) with side length $D \!-\! \lambda \!-\! 1$ to recover the message. Hence, the reconstruction communication cost $C_1$ is always $\frac{(D\!-\!\lambda)(D\!+\!\lambda\!+\!1)}{2}$. For the scheme with $\Omega_3$, the client is supposed to download the full $D \times D$ matrix except for the redundant upper right isosceles triangle (or lower right isosceles triangle) with side length $D\!-\!1$. Hence, the corresponding $C_1$ is $\frac{D(D\!+\!1)}{2}$. Regarding the database repair, for message matrix $\Omega$ in any form, the client only needs to download $1$ symbol from each database according to \cite[Thms.~6,~11]{Secure_RC}. Hence, the repair communication cost $C_2$ is always $D$. 
    
    For completeness, we also state the basic idea of the repair process. For all $j \in [N]$, let $\Psi^T_j$ denote the $j$th row of the encoding matrix $\Psi$. Then, the storage in database $j$ is $\zeta^T_j = \Psi^T_j \Omega$. Without loss of generality, let database $f$ fail. The corresponding storage in database $f$ is $\zeta^T_f = \Psi^T_f \Omega$, which is the exact form required in the replacement database. Note that a client can contact any $D$ working databases whose index set is denoted by $\mathcal{J}$. Then, for each $j \in \mathcal{J}$, database $j$ only needs to pass one symbol $\zeta^T_j \Psi_f$ to the client. After collecting $D$ symbols from the working databases, the client forwards the database $f$ repair information $\zeta^T_\mathcal{J} \Psi_f = \Psi^T_\mathcal{J} \Omega \Psi_f$ to the replacement database. Since the $D \times D$ submatrix $\Psi^T_\mathcal{J}$ of $\Psi$ is always invertible, the replacement database can rebuild $\Omega \Psi_f$, and then the required $\Psi^T_f \Omega$, because the message matrix $\Omega$ is symmetric. Note that the imported information to repair the failed database is equivalent to the original information stored in that database in terms of the message. As a consequence, even though a database repair operation is performed, the security analysis of information leakage constraint is not influenced. Regarding the storage cost, for any $D \times D$ message matrix $\Omega$, each database always needs to store $D^2$ symbols, i.e., $S = D^2$.

    Now, we employ the time-sharing idea to obtain the normalized performance result given in Theorem~\ref{RSRCtheorem}. For any rational number $\ell_\lambda$ that satisfies $0 \leq \ell_\lambda \leq \frac{2\lambda}{D+\lambda+1}$, we can utilize the scheme with $\Omega_1$ and the scheme with $\Omega_2$ in a time-sharing manner to achieve it. Let $\ell_1$ take the quotient form $\frac{p_1}{p_2}$ where $p_1$ and $p_2$ are both positive integers. We now use the first scheme $q_1$ times and the second scheme $q_2$ times such that the overall message leakage fraction $\ell_1$ is still $\frac{p_1}{p_2}$. Hence, we must have
    \begin{align}
        \frac{0q_1+\lambda(D-\lambda)q_2}{B} = \frac{0q_1+\lambda(D-\lambda)q_2}{\frac{(D-\lambda)(D-\lambda+1)}{2} \cdot q_1 + \frac{(D-\lambda)(D+\lambda+1)}{2} \cdot q_2} = \frac{p_1}{p_2}
    \end{align}
    If $q_1$ takes the value $2\lambda(D\!-\!\lambda)p_2 \!-\! (D\!-\!\lambda)(D\!+\!\lambda\!+\!1)p_1$, $q_2$ is equal to $(D\!-\!\lambda)(D\!-\!\lambda\!+\!1)p_1$. As a consequence, we can just use the first scheme $2\lambda(D\!-\!\lambda)p_2 \!-\! (D\!-\!\lambda)(D\!+\!\lambda\!+\!1)p_1$ times and the second scheme $(D\!-\!\lambda)(D\!-\!\lambda\!+\!1)p_1$ times. The total number of message symbols $B$ is
    \begin{align}
        B = \lambda(D-\lambda)q_2 \cdot \frac{p_2}{p_1} = \lambda(D-\lambda)^2(D-\lambda+1)p_2
    \end{align}
    The total reconstruction communication cost $C_1$ is
    \begin{align}
        C_1 = \frac{(D\!-\!\lambda)(D\!+\!\lambda\!+\!1)}{2} \cdot (q_1 + q_2) = \lambda(D\!-\!\lambda)^2(D\!+\!\lambda\!+\!1)(p_2-p_1)
    \end{align}
    After normalizing it by $B$, we have
    \begin{align}
        \frac{C_1}{B} = \frac{D+\lambda+1}{D-\lambda+1}(1-\ell_\lambda)
    \end{align}
    The total repair communication cost $C_2$ is
    \begin{align}
        C_2 = D \cdot (q_1 + q_2) = 2D\lambda(D\!-\!\lambda)(p_2-p_1)
    \end{align}
    After normalizing it by $B$, we have
    \begin{align}
        \frac{C_2}{B} = \frac{2D}{(D-\lambda)(D-\lambda+1)}(1-\ell_\lambda)
    \end{align}
    The storage cost $S$ is always $D$ multiple of $C_2$, which implies
    \begin{align}
        \frac{S}{B} = \frac{2D^2}{(D-\lambda)(D-\lambda+1)}(1-\ell_\lambda)
    \end{align}

    For any rational number $\ell_\lambda$ that satisfies $\frac{2\lambda}{D+\lambda+1} \leq \ell_\lambda \leq \frac{2\lambda D - \lambda(\lambda-1)}{D(D+1)}$, we can apply time-sharing idea once more by combining the scheme with $\Omega_2$ and the scheme with $\Omega_3$ now. Hence, we must have
    \begin{align}
        \frac{\lambda(D-\lambda)q_1+(\lambda D - \frac{\lambda(\lambda-1)}{2})q_2}{B} = \frac{\lambda(D-\lambda)q_1+(\lambda D - \frac{\lambda(\lambda-1)}{2})q_2}{\frac{(D-\lambda)(D+\lambda+1)}{2} \cdot q_1 + \frac{D(D+1)}{2} \cdot q_2} = \frac{p_1}{p_2}
    \end{align}
    If $q_1$ takes the value $(2\lambda D\!-\!\lambda^2\!+\!\lambda)p_2 \!-\! D(D\!+\!1)p_1$, $q_2$ is now equal to $(D\!-\!\lambda)(D\!+\!\lambda\!+\!1)p_1\!-\!2\lambda(D\!-\!\lambda)p_2$. As a consequence, we can just use the first scheme $(2\lambda D\!-\!\lambda^2\!+\!\lambda)p_2 \!-\! D(D\!+\!1)p_1$ times and the second scheme $(D\!-\!\lambda)(D\!+\!\lambda\!+\!1)p_1\!-\!2\lambda(D\!-\!\lambda)p_2$ times. The total number of message symbols $B$ is
    \begin{align}
        B = \frac{(D-\lambda)(D+\lambda+1)}{2} \cdot q_1 + \frac{D(D+1)}{2} \cdot q_2 = \frac{\lambda(\lambda+1)(D-\lambda)(D-\lambda+1)}{2}p_2
    \end{align}
    The total reconstruction communication cost $C_1$ is
    \begin{align}
        C_1 = \frac{(D-\lambda)(D+\lambda+1)}{2} \cdot q_1 + \frac{D(D+1)}{2} \cdot q_2 = \frac{\lambda(\lambda+1)(D-\lambda)(D-\lambda+1)}{2}p_2
    \end{align}
    After normalizing it by $B$, we have
    \begin{align}
        \frac{C_1}{B} = 1
    \end{align}
    The total repair communication cost $C_2$ is
    \begin{align}
        C_2 = D \cdot (q_1 + q_2) = \lambda(\lambda+1)D(p_2-p_1)
    \end{align}
    After normalizing it by $B$, we have
    \begin{align}
        \frac{C_2}{B} = \frac{2D}{(D-\lambda)(D-\lambda+1)}(1-\ell_\lambda)
    \end{align}
    The storage cost $S$ is still $D$ multiples of $C_2$, which implies
    \begin{align}
        \frac{S}{B} = \frac{2D^2}{(D-\lambda)(D-\lambda+1)}(1-\ell_\lambda)
    \end{align}

    For any rational number $\ell_\lambda$ that satisfies $\frac{2\lambda D - \lambda(\lambda-1)}{D(D+1)} \leq \ell_\lambda \leq 1$, we can use the scheme with $\Omega_3$ all the time. Hence, we must have
    \begin{align}
        \frac{C_1}{B} &= \frac{\frac{D(D+1)}{2}}{\frac{D(D+1)}{2}} = 1 \\
        \frac{C_2}{B} &= \frac{D}{\frac{D(D+1)}{2}} = \frac{2}{D+1} \\
        \frac{S}{B} &= \frac{D^2}{\frac{D(D+1)}{2}} = \frac{2D}{D+1}
    \end{align}
    concluding the proof.
\end{Proof}

\subsection{Examples to Illustrate the Basic Idea of RSRC}
Consider the special case of $N = 4,  D = 3$, all the symbols are operated in the finite field $\mathbb{F}_{13}$, and the encoding matrix $\Psi$ is 
\begin{align} \label{Psi}
    \Psi = 
    \begin{bmatrix}
        1 & 1 & 1  \\
        1 & 2 & 3 \\
        1 & 3 & 9 \\
        1 & 4 & 3
    \end{bmatrix}_{4 \times 3}
\end{align}
The message matrix $\Omega$ thus contains $9$ symbols and $6$ of them are distinct. If $\lambda = D = 3$, the study on $\ell_3$ is trivial. Due to the reconstruction constraint that a client can always recover the message $M_{[K]}$ by connecting to any $3$ working databases, these $3$ databases must include all the information about $M_{[K]}$, namely, $H(M_{[K]}|\mathcal{M}_3) = 0$ and $\ell_3 = 1$. Without loss of generality, we now assume that database $4$ fails.

\begin{example} \label{RSRC_ex1}
    We first consider the situation where $\lambda = 1$. If the message matrix $\Omega$ takes the following form where $M_1, M_2, M_3$ are i.i.d.~and uniformly selected message symbols from $\mathbb{F}_{13}$, $R_1, R_2, R_3$ are i.i.d.~and uniformly selected randomness symbols from $\mathbb{F}_{13}$,   
    \begin{align} \label{Omega1}
        \Omega =     
        \begin{bmatrix}
            M_1 & M_2 & R_1  \\
            M_2 & M_3 & R_2 \\
            R_1 & R_2 & R_3
        \end{bmatrix}
    \end{align}
    The message length $B$ is $3$ and the coded storage across the databases is as follows,
    \begin{align}
        &\mbox{DB} ~ 1: \quad M_1\!+\!M_2\!+\!R_1, \quad \ M_2\!+\!M_3\!+\!R_2, \quad \ R_1\!+\!R_2\!+\!R_3 \\
        &\mbox{DB} ~ 2: \quad  M_1\!+\!2M_2\!+\!3R_1, \ M_2\!+\!2M_3\!+\!3R_2, \ R_1\!+\!2R_2\!+\!3R_3 \\
        &\mbox{DB} ~ 3: \quad  M_1\!+\!3M_2\!+\!9R_1, \ M_2\!+\!3M_3\!+\!9R_2, \ R_1\!+\!3R_2\!+\!9R_3 \\
        &\mbox{DB} ~ 4: \quad  M_1\!+\!4M_2\!+\!3R_1, \ M_2\!+\!4M_3\!+\!3R_2, \ R_1\!+\!4R_2\!+\!3R_3
    \end{align}
    For any one database, as the value of the randomness symbol $R_3$ is unknown, the database cannot learn any knowledge about the randomness symbols $R_1, R_2$ from the third coded symbol. Furthermore, as none of $R_1, R_2$ are known, the database cannot learn any knowledge about the message symbols $M_1, M_2, M_3$ from the first two coded symbols. Therefore, each individual database learns nothing about the message set $M_{[K]}$, i.e., $I(M_{[K]};\mathcal{M}_1) = 0$ and $\ell_1 = 0$. Following the performance analysis of RSRC in the last subsection, a client can download $\{M_1\!+\!M_2\!+\!R_1,M_2\!+\!M_3\!+\!R_2\}$, $\{M_1\!+\!2M_2\!+\!3R_1,M_2\!+\!2M_3\!+\!3R_2\}$, $M_1\!+\!3M_2\!+\!9R_1$ from database $1$, $2$, $3$, respectively, to recover the message, and download $M_1\!+\!M_2\!+\!R_1\!+\!4(M_2\!+\!M_3\!+\!R_2)\!+\!3 (R_1\!+\!R_2\!+\!R_3)$, $ M_1\!+\!2M_2\!+\!3R_1\!+\!4(M_2\!+\!2M_3\!+\!3R_2)\!+\!3(R_1\!+\!2R_2\!+\!3R_3)$, $M_1\!+\!3M_2\!+\!9R_1\!+\!4(M_2\!+\!3M_3\!+\!9R_2) \!+\!3(R_1\!+\!3R_2\!+\!9R_3)$ from database $1$, $2$, $3$, respectively, to repair the failed database because of the following equality,
    \begin{align}
        &\begin{bmatrix}
            M_1\!+\!M_2\!+\!R_1\!+\!4(M_2\!+\!M_3\!+\!R_2)\!+\!3 (R_1\!+\!R_2\!+\!R_3) \\    M_1\!+\!2M_2\!+\!3R_1\!+\!4(M_2\!+\!2M_3\!+\!3R_2)\!+\!3(R_1\!+\!2R_2\!+\!3R_3) \\
            M_1\!+\!3M_2\!+\!9R_1\!+\!4(M_2\!+\!3M_3\!+\!9R_2) \!+\!3(R_1\!+\!3R_2\!+\!9R_3)
        \end{bmatrix} \notag \\
        &\qquad \quad = 
        \begin{bmatrix}
            1 & 1 & 1  \\
            1 & 2 & 3 \\
            1 & 3 & 9
        \end{bmatrix}
        \begin{bmatrix}
            M_1 & M_2 & R_1  \\
            M_2 & M_3 & R_2 \\
            R_1 & R_2 & R_3
        \end{bmatrix}
        \begin{bmatrix}
            1 \\
            4 \\
            3
        \end{bmatrix}
        = 
        \begin{bmatrix}
            1 & 1 & 1  \\
            1 & 2 & 3 \\
            1 & 3 & 9
        \end{bmatrix}
        \begin{bmatrix}
            M_1\!+\!4M_2\!+\!3R_1 \\
            M_2\!+\!4M_3\!+\!3R_2 \\
            R_1\!+\!4R_2\!+\!3R_3
        \end{bmatrix}
    \end{align}
    This scheme achieves $C_1=5$, $C_2=3$, $S=9$, which implies $\frac{C_1}{B} = \frac{5}{3}$, $\frac{C_2}{B} = 1$, $\frac{S}{B} = 3$.
    
    If one more message symbol is added to the message matrix $\Omega$ in place of the existing randomness symbol, the new $\Omega$ becomes 
    \begin{align} \label{Omega2}
        \Omega =     
        \begin{bmatrix}
            M_1 & M_2 & M_3  \\
            M_2 & M_4 & R_1 \\
            M_3 & R_1 & R_2
        \end{bmatrix}
    \end{align}
    The message length $B$ is $4$ and the coded storage across the databases is follows,
    \begin{align}
        &\mbox{DB} ~ 1: \quad M_1\!+\!M_2\!+\!M_3, \quad \ M_2\!+\!M_4\!+\!R_1, \quad \ M_3\!+\!R_1\!+\!R_2 \\
        &\mbox{DB} ~ 2: \quad M_1\!+\!2M_2\!+\!3M_3, \ M_2\!+\!2M_4\!+\!3R_1, \ M_3\!+\!2R_1\!+\!3R_2 \\
        &\mbox{DB} ~ 3: \quad M_1\!+\!3M_2\!+\!9M_3, \ M_2\!+\!3M_4\!+\!9R_1, \ M_3\!+\!3R_1\!+\!9R_2 \\
        &\mbox{DB} ~ 4: \quad M_1\!+\!4M_2\!+\!3M_3, \ M_2\!+\!4M_4\!+\!3R_1, \ M_3\!+\!4R_1\!+\!3R_2
    \end{align}
For any one database, due to the existence of the randomness symbols $R_1, R_2$, the only information concerning the message that can be learned by the database is the first coded symbol, which contains the ambiguity of $\frac{1}{4}H(M_{[K]})$. Therefore, we have $\ell_1 = \frac{1}{4}$ for each individual database. This scheme achieves the same performance, i.e., $C_1=5$, $C_2=3$, $S=9$, which implies $\frac{C_1}{B} = \frac{5}{4}$, $\frac{C_2}{B} = \frac{3}{4}$, $\frac{S}{B} = \frac{9}{4}$. Therefore, by allowing the leak of partial information about the messages to the database, the normalized values of $C_1$, $C_2$ and $S$ can be reduced. This partial information leakage idea can be considered as setting up a ramp zone where information leakage is not strictly prohibited.

If another message symbol is added to the current message matrix $\Omega$ in place of one of the two remaining randomness symbols, the new $\Omega$ becomes
\begin{align} \label{Omega3}
    \Omega =     
    \begin{bmatrix}
        M_1 & M_2 & M_3  \\
        M_2 & M_4 & M_5 \\
        M_3 & M_5 & R_1
    \end{bmatrix}
\end{align}
The message length $B$ is $5$ and the coded storage across the databases is as follows,
\begin{align}
    &\mbox{DB} ~ 1: \quad M_1\!+\!M_2\!+\!M_3, \quad \ M_2\!+\!M_4\!+\!M_5, \quad \ M_3\!+\!M_5\!+\!R_1 \\
    &\mbox{DB} ~ 2: \quad M_1\!+\!2M_2\!+\!3M_3, \ M_2\!+\!2M_4\!+\!3M_5, \ M_3\!+\!2M_5\!+\!3R_1 \\
    &\mbox{DB} ~ 3: \quad M_1\!+\!3M_2\!+\!9M_3, \ M_2\!+\!3M_4\!+\!9M_5, \ M_3\!+\!3M_5\!+\!9R_1 \\
    &\mbox{DB} ~ 4: \quad M_1\!+\!4M_2\!+\!3M_3, \ M_2\!+\!4M_4\!+\!3M_5, \ M_3\!+\!4M_5\!+\!3R_1 
\end{align}
Because of the existence of the randomness symbol $R_1$, each database can learn some information concerning the message from the first two coded symbols, which contains the ambiguity of $\frac{2}{5}H(M_{[K]})$. Therefore, we have $\ell_1 = \frac{2}{5}$ for each individual database. The performance of this scheme is still exactly the same as before, i.e., $C_1=5$, $C_2=3$, $S=9$, which implies $\frac{C_1}{B} = 1$, $\frac{C_2}{B} = \frac{3}{5}$, $\frac{S}{B} = \frac{9}{5}$.

For any rational number $\ell_1$ that satisfies $0 \leq \ell_1 \leq \frac{2}{5}$, we can utilize the scheme with $\Omega$ in \eqref{Omega1} $q_1$ times and the scheme with $\Omega$ in \eqref{Omega3} $q_2$ times in a time-sharing manner to achieve $\ell_1$, which can also be expressed in the form of $\frac{p_1}{p_2}$. Hence, we must have
\begin{align}
    \frac{0q_1+2q_2}{3q_1+5q_2} = \frac{p_1}{p_2}
\end{align}
If $q_1$ takes the value $2p_2-5p_1$, $q_2$ is now equal to $3p_1$. As a consequence, we can just use the first scheme $2p_2-5p_1$ times and the second scheme $3p_1$ times. After simple calculation, the overall message length $B$ is $6p_2$, the overall reconstruction communication cost $C_1$ is $10(p_2-p_1)$, the overall repair communication cost $C_2$ is $6(p_2-p_1)$, and the overall storage cost $S$ is $18(p_2-p_1)$. By normalizing these values by $B$, for $0 \leq \ell_1 \leq \frac{2}{5}$, we obtain
\begin{align} \label{ex1result1}
    \frac{C_1}{B} = \frac{5}{3}\cdot\frac{p_2-p_1}{p_2} = \frac{5}{3}(1 - \ell_1), \;\; \frac{C_2}{B} = \frac{p_2-p_1}{p_2} = 1 - \ell_1, \;\; \frac{S}{B} = 3\cdot\frac{p_2-p_1}{p_2} = 3(1 - \ell_1)
\end{align}
Note that by using these two schemes jointly to achieve $\ell_1 = \frac{1}{4}$, it has exactly the same normalized performance as the one obtained by using the scheme with $\Omega$ in \eqref{Omega2} directly.

If all the symbols in the message matrix $\Omega$ are message symbols without any randomness symbols, $\Omega$ is in the following form,
\begin{align} \label{Omega4}
    \Omega =     
    \begin{bmatrix}
        M_1 & M_2 & M_3  \\
        M_2 & M_4 & M_5 \\
        M_3 & M_5 & M_6
    \end{bmatrix}
\end{align}
Now, the message length $B$ becomes the maximal possible value $6$ and the coded storage across the databases is as follows,
\begin{align}
    &\mbox{DB} ~ 1: \quad M_1\!+\!M_2\!+\!M_3, \quad \ M_2\!+\!M_4\!+\!M_5, \quad \ M_3\!+\!M_5\!+\!M_6 \\
    &\mbox{DB} ~ 2: \quad M_1\!+\!2M_2\!+\!3M_3, \ M_2\!+\!2M_4\!+\!3M_5, \ M_3\!+\!2M_5\!+\!3M_6 \\
    &\mbox{DB} ~ 3: \quad M_1\!+\!3M_2\!+\!9M_3, \ M_2\!+\!3M_4\!+\!9M_5, \ M_3\!+\!3M_5\!+\!9M_6 \\
    &\mbox{DB} ~ 4: \quad M_1\!+\!4M_2\!+\!3M_3, \ M_2\!+\!4M_4\!+\!3M_5, \ M_3\!+\!4M_5\!+\!3M_6
\end{align}
Each individual database learns three coded symbols with each one only containing message symbols. Therefore, we have $\ell_1 = \frac{3}{6} = \frac{1}{2}$ for each database. For the message recovery, a client can download $\{M_1\!+\!M_2\!+\!M_3, M_2\!+\!M_4\!+\!M_5, M_3\!+\!M_5\!+\!M_6\}$, $\{M_1\!+\!2M_2\!+\!3M_3, M_2\!+\!2M_4\!+\!3M_5\}$, $M_1\!+\!3M_2\!+\!9M_3$ from database $1$, $2$, $3$, respectively. The corresponding performance of this scheme now becomes, $C_1=6$, $C_2=3$, $S=9$, which implies $\frac{C_1}{B} = 1$, $\frac{C_2}{B} = \frac{1}{2}$, $\frac{S}{B} = \frac{3}{2}$.

For any rational number $\ell_1$ that satisfies $\frac{2}{5} \leq \ell_1 \leq \frac{1}{2}$, we can utilize the time-sharing idea again by combining the scheme with $\Omega$ in \eqref{Omega3} and the scheme with $\Omega$ in \eqref{Omega4}. At this point, the first scheme is used $q_1$ times and the second scheme is used $q_2$ times such that the overall $\ell_1$ is equivalent to $\frac{p_1}{p_2}$, and we must have
\begin{align}
    \frac{2q_1+3q_2}{5q_1+6q_2} = \frac{p_1}{p_2}
\end{align}
If $q_1$ takes the value $3p_2-6p_1$, $q_2$ equals $5p_1-2p_2$. As a result, we can just apply the first scheme $3p_2-6p_1$ times and the second scheme $5p_1-2p_2$ times. After simple calculation, $B$ is $3p_2$, $C_1$ is $3p_2$, $C_2$ is $3(p_2-p_1)$ and $S$ is $9(p_2-p_1)$. By normalizing these values by $B$, for $\frac{2}{5} \leq \ell_1 \leq \frac{1}{2}$, we obtain
\begin{align} \label{ex1result2}
    \frac{C_1}{B} = \frac{3p_2}{3p_2} = 1, \;\; \frac{C_2}{B} = \frac{p_2-p_1}{p_2} = 1 - \ell_1, \;\; \frac{S}{B} = 3\cdot\frac{p_2-p_1}{p_2} = 3(1 - \ell_1)
\end{align}
By combining the results in \eqref{ex1result1}, \eqref{ex1result2} and the performance of the scheme using $\Omega$ in \eqref{Omega4} for $\frac{1}{2} \leq \ell_1 \leq 1$, we can get the full normalized performance in the case of $N=4$, $D=3$, $\lambda=1$, which matches the result in Theorem~\ref{RSRCtheorem}. 
\end{example}

\begin{example} \label{RSRC_ex2}
    We then consider the situation where $\lambda = 2$. The starting message matrix now takes the following form,
    \begin{align} \label{Omega5}
        \Omega =     
        \begin{bmatrix}
            M_1 & R_1 & R_2  \\
            R_1 & R_3 & R_4 \\
            R_2 & R_4 & R_5
        \end{bmatrix}
    \end{align}
    The message length $B$ is $1$ and the coded storage across the databases is as follows,
    \begin{align}
        &\mbox{DB} ~ 1: \quad M_1\!+\!R_1\!+\!R_2, \quad \ R_1\!+\!R_3\!+\!R_4, \quad \ R_2\!+\!R_4\!+\!R_5 \\
        &\mbox{DB} ~ 2: \quad M_1\!+\!2R_1\!+\!3R_2, \ R_1\!+\!2R_3\!+\!3R_4, \ R_2\!+\!2R_4\!+\!3R_5 \\
        &\mbox{DB} ~ 3: \quad M_1\!+\!3R_1\!+\!9R_2, \ R_1\!+\!3R_3\!+\!9R_4, \ R_2\!+\!3R_4\!+\!9R_5 \\
        &\mbox{DB} ~ 4: \quad M_1\!+\!4R_1\!+\!3R_2, \ R_1\!+\!4R_3\!+\!3R_4, \ R_2\!+\!4R_4\!+\!3R_5
    \end{align}
    For any two databases, according to the last two coded symbols from these two databases, neither of the randomness symbols $R_1$ and $R_2$ are decodable. Thus, for the first coded symbols from these two database, the message symbol $M_1$ is completely unknown to two databases. Therefore, we have $I(M_{[K]};\mathcal{M}_2) = 0$ and $\ell_2 = 0$. To recover the message, a client can simply download the first coded symbol from each database. The corresponding $C_1$ is $3$ while $C_2 = 3$ and $S = 9$ are not changed, that means $\frac{C_1}{B} = 3$, $\frac{C_2}{B} = 3$, $\frac{S}{B} = 9$. 

    As in Example~\ref{RSRC_ex1}, we import message symbols into the message matrix $\Omega$ gradually to replace the randomness symbols, and the new $\Omega$ first becomes
    \begin{align} \label{Omega6}
        \Omega =     
        \begin{bmatrix}
            M_1 & M_2 & R_1  \\
            M_2 & R_2 & R_3 \\
            R_1 & R_3 & R_4
        \end{bmatrix}
    \end{align}
    The message length $B$ is $2$ and the coded storage across the databases is follows,
    \begin{align}
        &\mbox{DB} ~ 1: \quad M_1\!+\!M_2\!+\!R_1, \quad \ M_2\!+\!R_2\!+\!R_3, \quad \ R_1\!+\!R_3\!+\!R_4 \\
        &\mbox{DB} ~ 2: \quad M_1\!+\!2M_2\!+\!3R_1, \ M_2\!+\!2R_2\!+\!3R_3, \ R_1\!+\!2R_3\!+\!3R_4 \\
        &\mbox{DB} ~ 3: \quad M_1\!+\!3M_2\!+\!9R_1, \ M_2\!+\!3R_2\!+\!9R_3, \ R_1\!+\!3R_3\!+\!9R_4 \\
        &\mbox{DB} ~ 4: \quad M_1\!+\!4M_2\!+\!3R_1, \ M_2\!+\!4R_2\!+\!3R_3, \ R_1\!+\!4R_3\!+\!3R_4
    \end{align}
    For any two databases, one useful value that only involves a linear combination of message symbols $M_1$ and $M_2$ can be attained, whereas the other information is useless in terms of the message. We use database $1$ and database $2$ as an example here, by subtracting the coded symbol in database $2$ from the triple coded symbol in database $1$ and the double coded symbol in database $1$ element-wisely, the six coded symbols from these two databases are equal to $2M_1\!+\!M_2$, $M_1\!-\!R_1$, $2M_2\!+\!R_2$, $M_2\!-\!R_3$, $R_1\!-\!R_4$ and redundant $2R_1\!+\!R_3$. Therefore, we have $\ell_2 = \frac{1}{2}$ for any two databases. Regarding the performance of this scheme, all metrics remain the same, i.e., $C_1 = 3$, $C_2 = 3$, $S = 9$, which implies $\frac{C_1}{B} = \frac{3}{2}$, $\frac{C_2}{B} = \frac{3}{2}$, $\frac{S}{B} = \frac{9}{2}$.

    When we move forward, the new $\Omega$ next becomes
    \begin{align} \label{Omega7}
        \Omega_3 =     
        \begin{bmatrix}
            M_1 & M_2 & M_3  \\
            M_2 & R_1 & R_2 \\
            M_3 & R_2 & R_3
        \end{bmatrix}
    \end{align}
    The message length $B$ is $3$ and the coded storage across the databases is follows,
    \begin{align}
        &\mbox{DB} ~ 1: \quad M_1\!+\!M_2\!+\!M_3, \quad \ M_2\!+\!R_1\!+\!R_2, \quad \ M_3\!+\!R_2\!+\!R_3 \\
        &\mbox{DB} ~ 2: \quad M_1\!+\!2M_2\!+\!3M_3, \ M_2\!+\!2R_1\!+\!3R_2, \ M_3\!+\!2R_2\!+\!3R_3 \\
        &\mbox{DB} ~ 3: \quad M_1\!+\!3M_2\!+\!9M_3, \ M_2\!+\!3R_1\!+\!9R_2, \ M_3\!+\!3R_2\!+\!9R_3 \\
        &\mbox{DB} ~ 4: \quad M_1\!+\!4M_2\!+\!3M_3, \ M_2\!+\!4R_1\!+\!3R_2, \ M_3\!+\!4R_2\!+\!3R_3 
    \end{align}
    Note that this construction is different from the one in \eqref{Omega1}, although the message length is the same. By downloading the first coded symbol from each database, the reconstruction communication cost $C_1$ in this scheme is now $3$ rather than $5$. For any two databases, two useful values  only involving a linear combination of message symbols can be attained, whereas the other information is still useless in terms of the message. If we look at database $1$ and database $2$, the six coded symbols from these two databases are equivalent to $2M_1\!+\!M_2$, $M_1\!-\!M_3$, $2M_2\!+\!R_1$, $M_2\!-\!R_2$, $M_3\!-\!R_3$ and redundant $2M_3\!+\!R_2$. Therefore, we have $\ell_2 = \frac{2}{3}$ for any two databases. The performance of this scheme is maintained, i.e., $C_1 = 3$, $C_2 = 3$, $S = 9$, which implies $\frac{C_1}{B} = 1$, $\frac{C_2}{B} = 1$, $\frac{S}{B} = 3$. 
    
    Following the time-sharing idea in Example~\ref{RSRC_ex1}, by unifying the scheme with $\Omega$ in \eqref{Omega5} and the scheme with $\Omega$ in \eqref{Omega7}, we obtain the following normalized performance for $0 \leq \ell_2 \leq \frac{2}{3}$,
    \begin{align} \label{ex2result1}
        \frac{C_1}{B} = 3(1 - \ell_2), \quad \frac{C_2}{B}  = 3(1 - \ell_2), \quad \frac{S}{B} = 9(1 - \ell_2)
    \end{align}
    When $\ell_2$ takes the value $\frac{1}{2}$, this is exactly the performance of the scheme that uses $\Omega$ in \eqref{Omega6}. If we keep importing message symbols and using the time-sharing idea, we can obtain the remaining normalized performance in this situation when $\frac{2}{3} \leq \ell_2 \leq \frac{5}{6}$, namely,
    \begin{align} \label{ex2result2}
        \frac{C_1}{B} = 1, \quad \frac{C_2}{B}  = 3(1 - \ell_2), \quad \frac{S}{B} = 9(1 - \ell_2)
    \end{align}
    The normalized reconstruction communication cost $\frac{C_1}{B}$ reaches the limit $1$ yet the other two values remain the same. By putting the results in \eqref{ex2result1}, \eqref{ex2result2} and the performance of the scheme using $\Omega$ in \eqref{Omega4} for $\frac{5}{6} \leq \ell_1 \leq 1$ together, we can derive the full performance for $N=4$, $D=3$, $\lambda=2$, which also matches the result in Theorem~\ref{RSRCtheorem}.
\end{example}

\section{Distributed FSL Motivating Example}
In this section, we consider a simple FSL setting as a toy example and provide a secure and robust achievable scheme in the presence of a passive eavesdropper and a database failure. The full learning model is divided into $K=4$ submodels, $M_{[4]}=\{M_1, M_2, M_3, M_4\}$ with each submodel consisting of $L=2$ symbols from the finite field $F_{13}$ is stored in a coded form across $N=4$ individual databases in the server. Any arbitrary $J = 2$ databases can collude with each other to learn the remote client data. In addition, $C=4$ random clients are selected by the server to update the submodels in this round of the FSL process. Each client should be able to obtain the required submodel information by communicating with any arbitrary $D = 3$ working databases. The desired submodel index set for each client is
\begin{align}
    &\mbox{Client} ~ 1 \in \mathcal{C}_1: \quad \Gamma^{\langle 1 \rangle} = \{1\} && \Rightarrow \quad Y^{\langle 1 \rangle} = [Y^{\langle 1 \rangle}_1 \:\: Y^{\langle 1 \rangle}_2 \:\: Y^{\langle 1 \rangle}_3 \:\: Y^{\langle 1 \rangle}_4]^T = [1 \:\: 0 \:\: 0 \:\: 0]^T \\
    &\mbox{Client} ~ 2 \in \mathcal{C}_1: \quad \Gamma^{\langle 2 \rangle} = \{1,3\} && \Rightarrow \quad Y^{\langle 2 \rangle} = [Y^{\langle 2 \rangle}_1 \:\: Y^{\langle 2 \rangle}_2 \:\: Y^{\langle 2 \rangle}_3 \:\: Y^{\langle 2 \rangle}_4]^T = [1 \:\: 0 \:\: 1 \:\: 0]^T \\
    &\mbox{Client} ~ 3 \in \mathcal{C}_2: \quad \Gamma^{\langle 3 \rangle} = \{1,4\} && \Rightarrow \quad Y^{\langle 3 \rangle} = [Y^{\langle 3 \rangle}_1 \:\: Y^{\langle 3 \rangle}_2 \:\: Y^{\langle 3 \rangle}_3 \:\: Y^{\langle 3 \rangle}_4]^T = [1 \:\: 0 \:\: 0 \:\: 1]^T \\
    &\mbox{Client} ~ 4 \in \mathcal{C}_3: \quad \Gamma^{\langle 4 \rangle} = \{1,3,4\} && \Rightarrow \quad Y^{\langle 4 \rangle} = [Y^{\langle 4 \rangle}_1 \:\: Y^{\langle 4 \rangle}_2 \:\: Y^{\langle 4 \rangle}_3 \:\: Y^{\langle 4 \rangle}_4]^T = [1 \:\: 0 \:\: 1 \:\: 1]^T  
\end{align}

\begin{example}
A passive eavesdropper can tap in on any arbitrary $E = 2$  databases to learn the storage data as well as all the communication data that comes in and goes out. Assuming that the submodel leakage parameter $\delta$ is $0.5$, the submodel information is then coded through the RSRC scheme with encoding matrix $\Psi$ in \eqref{Psi} and message matrix $\Omega$ in \eqref{Omega6} for better FSL performance. Thus, the storage across the databases including coded submodel information and extra uncoded server-side common randomness is initialized as in Table~\ref{E2_1}. At this point, database $4$ has a failure and cannot provide any reliable responses. In this example, the generation of client-side common randomness and server-side common randomness are skipped, but it will be introduced in detail in the general FSL achievable scheme.

 \begin{table}[ht]
\begin{center}
\begin{tabular}{|c|c|c|}
\hline
Database & \multicolumn{2}{c|}{Storage} \\
\hline
\multirow{4}{*}{DB 1} & $M_{1,1}\!+\!M_{1,2}\!+\!R_{1,1}, \ M_{1,2}\!+\!R_{1,2}\!+\!R_{1,3}, \ R_{1,1}\!+\!R_{1,3}\!+\!R_{1,4}$ & $\hat{R}_1, \hat{R}_{1,1}, \hat{R}_{1,2}$ \\ 
& $M_{2,1}\!+\!M_{2,2}\!+\!R_{2,1}, \ M_{2,2}\!+\!R_{2,2}\!+\!R_{2,3}, \ R_{2,1}\!+\!R_{2,3}\!+\!R_{2,4}$ & $\hat{R}_2, \hat{R}_{2,1}, \hat{R}_{2,2}$ \\ 
& $M_{3,1}\!+\!M_{3,2}\!+\!R_{3,1}, \ M_{3,2}\!+\!R_{3,2}\!+\!R_{3,3}, \ R_{3,1}\!+\!R_{3,3}\!+\!R_{3,4}$ & $\hat{R}_3, \hat{R}_{3,1}, \hat{R}_{3,2}$ \\ 
& $M_{4,1}\!+\!M_{4,2}\!+\!R_{4,1}, \ M_{4,2}\!+\!R_{4,2}\!+\!R_{4,3}, \ R_{4,1}\!+\!R_{4,3}\!+\!R_{4,4}$ & $\hat{R}_4, \hat{R}_{4,1}, \hat{R}_{4,2}$ \\ \hline
\multirow{4}{*}{DB 2} & $M_{1,1}\!+\!2M_{1,2}\!+\!3R_{1,1}, \ M_{1,2}\!+\!2R_{1,2}\!+\!3R_{1,3}, \ R_{1,1}\!+\!2R_{1,3}\!+\!3R_{1,4}$ & $\hat{R}_1, \hat{R}_{1,1}, \hat{R}_{1,2}$ \\ 
& $M_{2,1}\!+\!2M_{2,2}\!+\!3R_{2,1}, \ M_{2,2}\!+\!2R_{2,2}\!+\!3R_{2,3}, \ R_{2,1}\!+\!2R_{2,3}\!+\!3R_{2,4}$ & $\hat{R}_2, \hat{R}_{2,1}, \hat{R}_{2,2}$ \\ 
& $M_{3,1}\!+\!2M_{3,2}\!+\!3R_{3,1}, \ M_{3,2}\!+\!2R_{3,2}\!+\!3R_{3,3}, \ R_{3,1}\!+\!2R_{3,3}\!+\!3R_{3,4}$ & $\hat{R}_3, \hat{R}_{3,1}, \hat{R}_{3,2}$ \\ 
& $M_{4,1}\!+\!2M_{4,2}\!+\!3R_{4,1}, \ M_{4,2}\!+\!2R_{4,2}\!+\!3R_{4,3}, \ R_{4,1}\!+\!2R_{4,3}\!+\!3R_{4,4}$ & $\hat{R}_4, \hat{R}_{4,1}, \hat{R}_{4,2}$ \\ \hline
\multirow{4}{*}{DB 3} & $M_{1,1}\!+\!3M_{1,2}\!+\!9R_{1,1}, \ M_{1,2}\!+\!3R_{1,2}\!+\!9R_{1,3}, \ R_{1,1}\!+\!3R_{1,3}\!+\!9R_{1,4}$ & $\hat{R}_1, \hat{R}_{1,1}, \hat{R}_{1,2}$ \\ 
& $M_{2,1}\!+\!3M_{2,2}\!+\!9R_{2,1}, \ M_{2,2}\!+\!3R_{2,2}\!+\!9R_{2,3}, \ R_{2,1}\!+\!3R_{2,3}\!+\!9R_{2,4}$ & $\hat{R}_2, \hat{R}_{2,1}, \hat{R}_{2,2}$ \\ 
& $M_{3,1}\!+\!3M_{3,2}\!+\!9R_{3,1}, \ M_{3,2}\!+\!3R_{3,2}\!+\!9R_{3,3}, \ R_{3,1}\!+\!3R_{3,3}\!+\!9R_{3,4}$ & $\hat{R}_3, \hat{R}_{3,1}, \hat{R}_{3,2}$ \\ 
& $M_{4,1}\!+\!3M_{4,2}\!+\!9R_{4,1}, \ M_{4,2}\!+\!3R_{4,2}\!+\!9R_{4,3}, \ R_{4,1}\!+\!3R_{4,3}\!+\!9R_{4,4}$ & $\hat{R}_4, \hat{R}_{4,1}, \hat{R}_{4,2}$\\ \hline
\multirow{4}{*}{DB 4} & $M_{1,1}\!+\!4M_{1,2}\!+\!3R_{1,1}, \ M_{1,2}\!+\!4R_{1,2}\!+\!3R_{1,3}, \ R_{1,1}\!+\!4R_{1,3}\!+\!3R_{1,4}$ & $\hat{R}_1, \hat{R}_{1,1}, \hat{R}_{1,2}$ \\ 
& $M_{2,1}\!+\!4M_{2,2}\!+\!3R_{2,1}, \ M_{2,2}\!+\!4R_{2,2}\!+\!3R_{2,3}, \ R_{2,1}\!+\!4R_{2,3}\!+\!3R_{2,4}$ & $\hat{R}_2, \hat{R}_{2,1}, \hat{R}_{2,2}$ \\ 
& $M_{3,1}\!+\!4M_{3,2}\!+\!3R_{3,1}, \ M_{3,2}\!+\!4R_{3,2}\!+\!3R_{3,3}, \ R_{3,1}\!+\!4R_{3,3}\!+\!3R_{3,4}$ & $\hat{R}_3, \hat{R}_{3,1}, \hat{R}_{3,2}$ \\ 
& $M_{4,1}\!+\!4M_{4,2}\!+\!3R_{4,1}, \ M_{4,2}\!+\!4R_{4,2}\!+\!3R_{4,3}, \ R_{4,1}\!+\!4R_{4,3}\!+\!3R_{4,4}$ & $\hat{R}_4, \hat{R}_{4,1}, \hat{R}_{4,2}$ \\ \hline
\end{tabular}
\end{center}
\vspace{-1em}
\caption{Storage across the databases in the server when $D = 3$, $J = E = 2$ and $\delta = 0.5$.}
\label{E2_1}
\end{table}

\paragraph{FSL-PSU phase:}
In the first step of FSL-PSU phase, according to the client partition, the client answers are generated as follows\footnote{As in our previous work \cite{FSL-PSU}, the value in $\langle \rangle$ is used to denote the index of client and the value in $()$ is used to denote the index of database. In addition, the first subscript of the download $D$ or the answer $A$ is used to show it is within the FSL-PSU phase or FSL-write phase as the letter U stands for union and the letter W stands for write, whereas the second subscript is used to denote the step number within this phase.} where the symbols $c$ and $\{w^{\langle i \rangle}_k \!: i \in [4], k \in [4]\}$ are both client-side common randomness that satisfies $\sum_{i \in [4]} w^{\langle i \rangle}_k = 0$ for all $k \in [4]$ and is unknown to any $2$ databases at the server side,
\begin{align}
    &A_{U,1}^{\langle 1 \rangle,(1)} = \{c(Y^{\langle 1 \rangle}_1\!+\!w^{\langle 1 \rangle}_1), c(Y^{\langle 1 \rangle}_2\!+\!w^{\langle 1 \rangle}_2), c(Y^{\langle 1 \rangle}_3\!+\!w^{\langle 1 \rangle}_3), c(Y^{\langle 1 \rangle}_4\!+\!w^{\langle 1 \rangle}_4)\} \label{MP-PSU1 answer1} \\
    &A_{U,1}^{\langle 2 \rangle,(1)} = \{c(Y^{\langle 2 \rangle}_1\!+\!w^{\langle 2 \rangle}_1), c(Y^{\langle 2 \rangle}_2\!+\!w^{\langle 2 \rangle}_2), c(Y^{\langle 2 \rangle}_3\!+\!w^{\langle 2 \rangle}_3), c(Y^{\langle 2 \rangle}_4\!+\!w^{\langle 2 \rangle}_4)\} \label{MP-PSU1 answer2} \\
    &A_{U,1}^{\langle 3 \rangle,(2)} = \{c(Y^{\langle 3 \rangle}_1\!+\!w^{\langle 3 \rangle}_1), c(Y^{\langle 3 \rangle}_2\!+\!w^{\langle 3 \rangle}_2), c(Y^{\langle 3 \rangle}_3\!+\!w^{\langle 3 \rangle}_3), c(Y^{\langle 3 \rangle}_4\!+\!w^{\langle 3 \rangle}_4)\} \label{MP-PSU1 answer3} \\
    &A_{U,1}^{\langle 4 \rangle,(3)} = \{c(Y^{\langle 4 \rangle}_1\!+\!w^{\langle 4 \rangle}_1), c(Y^{\langle 4 \rangle}_2\!+\!w^{\langle 4 \rangle}_2), c(Y^{\langle 4 \rangle}_3\!+\!w^{\langle 4 \rangle}_3), c(Y^{\langle 4 \rangle}_4\!+\!w^{\langle 4 \rangle}_4)\} \label{MP-PSU1 answer4}
\end{align}
In the second step of FSL-PSU phase, after collecting the answers from its associated clients $1$ and $2$, database $1$ does the element-wise summation with the aid of its own extra uncoded server-side common randomness $\{\hat{R}_k \!: k \in [4]\}$ that is unknown to each individual client. This information is subsequently downloaded by a randomly selected client from the client group $\mathcal{C}_1$, say client $2$, 
\begin{align}
    D_{U,2}^{\langle 2 \rangle,(1)} = \{&c(Y^{\langle 1 \rangle}_1\!+\!Y^{\langle 2 \rangle}_1\!+\!w^{\langle 1 \rangle}_1\!+\!w^{\langle 2 \rangle}_1)\!+\!\hat{R}_1,c(Y^{\langle 1 \rangle}_2\!+\!Y^{\langle 2 \rangle}_2\!+\!w^{\langle 1 \rangle}_2\!+\!w^{\langle 2 \rangle}_2)\!+\!\hat{R}_2, \notag \\
    &c(Y^{\langle 1 \rangle}_3\!+\!Y^{\langle 2 \rangle}_3\!+\!w^{\langle 1 \rangle}_3\!+\!w^{\langle 2 \rangle}_3)\!+\!\hat{R}_3,c(Y^{\langle 1 \rangle}_4\!+\!Y^{\langle 2 \rangle}_4\!+\!w^{\langle 1 \rangle}_4\!+\!w^{\langle 2 \rangle}_4)\!+\!\hat{R}_4\} 
\end{align}
After further processing the received information $D_{U,2}^{\langle 2 \rangle,(1)}$ through additional client-side common randomness $\{w^{(j)}_k \!: j \in [3], k \in [4]\}$ that satisfies $\sum_{j \in [3]} w^{(j)}_k = 0$ for all $k \in [4]$ and is unknown to any $2$ databases, client $2$ forwards the following answer to all the functioning databases, i.e., database $1,2,3$,
\begin{align}
    A_{U,2}^{\langle 2 \rangle,([3])} = \{&c(Y^{\langle 1 \rangle}_1\!+\!Y^{\langle 2 \rangle}_1\!+\!w^{\langle 1 \rangle}_1\!+\!w^{\langle 2 \rangle}_1)\!+\!\hat{R}_1\!+\!w^{(1)}_1,c(Y^{\langle 1 \rangle}_2\!+\!Y^{\langle 2 \rangle}_2\!+\!w^{\langle 1 \rangle}_2\!+\!w^{\langle 2 \rangle}_2)\!+\!\hat{R}_2\!+\!w^{(1)}_2, \notag \\
    &c(Y^{\langle 1 \rangle}_3\!+\!Y^{\langle 2 \rangle}_3\!+\!w^{\langle 1 \rangle}_3\!+\!w^{\langle 2 \rangle}_3)\!+\!\hat{R}_3\!+\!w^{(1)}_3,c(Y^{\langle 1 \rangle}_4\!+\!Y^{\langle 2 \rangle}_4\!+\!w^{\langle 1 \rangle}_4\!+\!w^{\langle 2 \rangle}_4)\!+\!\hat{R}_4\!+\!w^{(1)}_4\}
\end{align}
Likewise, after downloading the following information $D_{U,2}^{\langle 3 \rangle,(2)}$ and $D_{U,2}^{\langle 4 \rangle,(3)}$ from database $2$ and database $3$, respectively, client $3$ and client $4$ forward the following generated answers $A_{U,2}^{\langle 3 \rangle,([3])}$ and $A_{U,2}^{\langle 4 \rangle,([3])}$ to all the functioning databases as well,  
\begin{align}
    D_{U,2}^{\langle 3 \rangle,(2)} = \{&c(Y^{\langle 3 \rangle}_1\!+\!w^{\langle 3 \rangle}_1)\!+\!\hat{R}_1, c(Y^{\langle 3 \rangle}_2\!+\!w^{\langle 3 \rangle}_2)\!+\!\hat{R}_2, \notag \\
    &c(Y^{\langle 3 \rangle}_3\!+\!w^{\langle 3 \rangle}_3)\!+\!\hat{R}_3, c(Y^{\langle 3 \rangle}_4\!+\!w^{\langle 3 \rangle}_4)\!+\!\hat{R}_4\}  \\
    D_{U,2}^{\langle 4 \rangle,(3)} = \{&c(Y^{\langle 4 \rangle}_1\!+\!w^{\langle 4 \rangle}_1)\!+\!\hat{R}_1, c(Y^{\langle 4 \rangle}_2\!+\!w^{\langle 4 \rangle}_2)\!+\!\hat{R}_2, \notag \\
    &c(Y^{\langle 4 \rangle}_3\!+\!w^{\langle 4 \rangle}_3)\!+\!\hat{R}_3, c(Y^{\langle 4 \rangle}_4\!+\!w^{\langle 4 \rangle}_4)\!+\!\hat{R}_4\} \\
    A_{U,2}^{\langle 3 \rangle,([3])} = \{&c(Y^{\langle 3 \rangle}_1\!+\!w^{\langle 3 \rangle}_1)\!+\!\hat{R}_1\!+\!w^{(2)}_1, c(Y^{\langle 3 \rangle}_2\!+\!w^{\langle 3 \rangle}_2)\!+\!\hat{R}_2\!+\!w^{(2)}_2, \notag \\
    &c(Y^{\langle 3 \rangle}_3\!+\!w^{\langle 3 \rangle}_3)\!+\!\hat{R}_3\!+\!w^{(2)}_3, c(Y^{\langle 3 \rangle}_4\!+\!w^{\langle 3 \rangle}_4)\!+\!\hat{R}_4\!+\!w^{(2)}_4\} \\
    A_{U,2}^{\langle 4 \rangle,([3])} = \{&c(Y^{\langle 4 \rangle}_1\!+\!w^{\langle 4 \rangle}_1)\!+\!\hat{R}_1\!+\!w^{(3)}_1, c(Y^{\langle 4 \rangle}_2\!+\!w^{\langle 4 \rangle}_2)\!+\!\hat{R}_2\!+\!w^{(3)}_2, \notag \\
    &c(Y^{\langle 4 \rangle}_3\!+\!w^{\langle 4 \rangle}_3)\!+\!\hat{R}_3\!+\!w^{(3)}_3, c(Y^{\langle 4 \rangle}_4\!+\!w^{\langle 4 \rangle}_4)\!+\!\hat{R}_4\!+\!w^{(3)}_4\}
\end{align}
For all $j \in [3]$, each an alive database $j$ is able to find the desired submodel union through element-wise summation after collecting the available answers $A_{U,2}^{\langle 2 \rangle,(j)}$, $A_{U,2}^{\langle 3 \rangle,(j)}$ and $A_{U,2}^{\langle 4 \rangle,(j)}$ from the routing clients $2, 3, 4$. For all $k \in [4]$, database $j$ can get the result $c(\sum_{i \in [4]} Y^{\langle i \rangle}_k) \!+\! 3\hat{R}_k$ since $\sum_{i \in [4]} w^{\langle i \rangle}_k = 0$ and $\sum_{j \in [3]} w^{(j)}_k = 0$, which implies the value of $c(\sum_{i \in [4]} Y^{\langle i \rangle}_k)$ since $\hat{R}_k$ is a known constant. It is straightforward to see that submodel $k$ is in the union if $c(\sum_{i \in [4]} Y^{\langle i \rangle}_k) \neq 0$, otherwise, $k$ is not in the union. Therefore, the union result $\Gamma = \{1,3,4\}$ is obtained by each alive database. Due to the limited information uploaded by the selected clients, it is easy to verify that the server can only learn this union and nothing beyond the union, even though all these databases can collude with each other.

\paragraph{FSL-write phase:}
First, by downloading information from $3$ functioning databases, each selected client needs to recover all the submodels in the submodel union $M_{\Gamma} = \{M_1, M_2, M_4\}$ to be updated in this FSL round. For all $i \in [3]$, we have
\begin{align}
    D_{W,1}^{\langle i \rangle,(1)} = \{&M_{1,1}\!+\!M_{1,2}\!+\!R_{1,1}, M_{3,1}\!+\!M_{3,2}\!+\!R_{3,1}, M_{4,1}\!+\!M_{4,2}\!+\!R_{4,1}\} \\
    D_{W,1}^{\langle i \rangle,(2)} = \{&M_{1,1}\!+\!2M_{1,2}\!+\!3R_{1,1}, M_{3,1}\!+\!2M_{3,2}\!+\!3R_{3,1}, M_{4,1}\!+\!2M_{4,2}\!+\!3R_{4,1}\} \\
    D_{W,1}^{\langle i \rangle,(3)} = \{&M_{1,1}\!+\!4M_{1,2}\!+\!3R_{1,1}, M_{3,1}\!+\!4M_{3,2}\!+\!3R_{3,1}, M_{4,1}\!+\!3M_{4,2}\!+\!9R_{4,1}\}  
\end{align}
Meanwhile, without loss of generality, client $4$ is utilized to route the required information of submodel $M_2$ to the replacement database as a substitution for the failed database $4$. Thus, 
\begin{align}
    D_{W,1}^{\langle 4 \rangle,(1)} \!=\! \{&D_{W,1}^{\langle i \rangle,(1)},     M_{2,1}\!+\!M_{2,2}\!+\!R_{2,1}\!+\!4(M_{2,2}\!+\!R_{2,2}\!+\!R_{2,3})\!+\! 3(R_{2,1}\!+\!R_{2,3}\!+\!R_{2,4})
    \} \\
    D_{W,1}^{\langle 4 \rangle,(2)} \!=\! \{&D_{W,1}^{\langle i \rangle,(2)},M_{2,1}\!+\!2M_{2,2}\!+\!3R_{2,1}\!+\! 4(M_{2,2}\!+\!2R_{2,2}\!+\!3R_{2,3})\!+\!3(R_{2,1}\!+\!2R_{2,3}\!+\!3R_{2,4})\} \\
    D_{W,1}^{\langle 4 \rangle,(3)} \!=\! \{&D_{W,1}^{\langle i \rangle,(3)}, M_{2,1}\!+\!3M_{2,2}\!+\!9R_{2,1}\!+\!4(M_{2,2}\!+\!3R_{2,2}\!+\!9R_{2,3})\!+\!3(R_{2,1}\!+\!3R_{2,3}\!+\!9R_{2,4})\}  
\end{align}
Due to the reconstruction constraint of the RSRC scheme, each client can reliably decode the desired submodels $M_\Gamma$ as well as the server-side common randomness symbols $R_{1,1}, R_{3,1}, R_{4,1}$. When the local training is done, the answers sent by the clients in the first step of FSL-write phase are as follows where the symbols $\{w^{\langle i \rangle}_{k,l} \!: i \in [4], k \in \Gamma, l \in [2]\}$ are client-side common randomness that satisfy $\sum_{i \in [4]} w^{\langle i \rangle}_{k,l} = 0$ for all $k \in \Gamma, l \in [2]$ and is unknown to any $2$ databases,
\begin{align}
    A^{\langle 1 \rangle,(1)}_{W,1} = \{&\Delta^{\langle 1 \rangle}_{1,1} \!+\! w^{\langle 1 \rangle}_{1,1}, \Delta^{\langle 1 \rangle}_{1,2} \!+\! w^{\langle 1 \rangle}_{1,2}, w^{\langle 1 \rangle}_{3,1}, w^{\langle 1 \rangle}_{3,2}, w^{\langle 1 \rangle}_{4,1}, w^{\langle 1 \rangle}_{4,2}\} \\
    A^{\langle 2 \rangle,(1)}_{W,1} = \{&\Delta^{\langle 2 \rangle}_{1,1} \!+\! w^{\langle 2 \rangle}_{1,1}, \Delta^{\langle 2 \rangle}_{1,2} \!+\! w^{\langle 2 \rangle}_{1,2}, \Delta^{\langle 2 \rangle}_{3,1} \!+\! w^{\langle 2 \rangle}_{3,1}, \Delta^{\langle 2 \rangle}_{3,2} \!+\! w^{\langle 2 \rangle}_{3,2}, w^{\langle 2 \rangle}_{4,1}, w^{\langle 2 \rangle}_{4,2}\} \\
    A^{\langle 3 \rangle,(2)}_{W,1} = \{&\Delta^{\langle 3 \rangle}_{1,1} \!+\! w^{\langle 3 \rangle}_{1,1}, \Delta^{\langle 3 \rangle}_{1,2} \!+\! w^{\langle 3 \rangle}_{1,2}, w^{\langle 3 \rangle}_{3,1}, w^{\langle 3 \rangle}_{3,2}, \Delta^{\langle 3 \rangle}_{4,1} \!+\! w^{\langle 3 \rangle}_{4,1}, \Delta^{\langle 3 \rangle}_{4,2} \!+\! w^{\langle 3 \rangle}_{4,2}\} \\
    A^{\langle 4 \rangle,(3)}_{W,1} = \{&\Delta^{\langle 4 \rangle}_{1,1} \!+\! w^{\langle 4 \rangle}_{1,1}, \Delta^{\langle 4 \rangle}_{1,2} \!+\! w^{\langle 4 \rangle}_{1,2}, \Delta^{\langle 4 \rangle}_{3,1} \!+\! w^{\langle 4 \rangle}_{3,1}, \Delta^{\langle 4 \rangle}_{3,2} \!+\! w^{\langle 4 \rangle}_{3,2}, \Delta^{\langle 4 \rangle}_{4,1} \!+\! w^{\langle 4 \rangle}_{4,1}, \Delta^{\langle 4 \rangle}_{4,2} \!+\! w^{\langle 4 \rangle}_{4,2}\}
\end{align}
Following the similar execution in the previous FSL-PSU phase, client $2$ downloads the following information from database $1$ in the second step of FSL-write phase where the symbols $\{\hat{R}_{k,l} \!: k \in \Gamma, l \in [2]\}$ are extra uncoded server-side common randomness that is unknown to each individual client,
\begin{align}
    D_{W,2}^{\langle 2 \rangle,(1)} \!=\! \{&\Delta^{\langle 1 \rangle}_{1,1} \!+\! \Delta^{\langle 2 \rangle}_{1,1} \!+\! w^{\langle 1 \rangle}_{1,1} \!+\! w^{\langle 2 \rangle}_{1,1} \!+\! \hat{R}_{1,1}, \Delta^{\langle 1 \rangle}_{1,2} \!+\! \Delta^{\langle 2 \rangle}_{1,2} \!+\! w^{\langle 1 \rangle}_{1,2} \!+\! w^{\langle 2 \rangle}_{1,2} \!+\! \hat{R}_{1,2}, \Delta^{\langle 2 \rangle}_{3,1} \!+\! w^{\langle 1 \rangle}_{3,1} \!+\! w^{\langle 2 \rangle}_{3,1} \!+\! \hat{R}_{3,1}, \notag \\ 
    &\Delta^{\langle 2 \rangle}_{3,2} \!+\! w^{\langle 1 \rangle}_{3,2} \!+\! w^{\langle 2 \rangle}_{3,2} \!+\! \hat{R}_{3,2}, w^{\langle 1 \rangle}_{4,1} \!+\! w^{\langle 2 \rangle}_{4,1} \!+\! \hat{R}_{4,1}, w^{\langle 1 \rangle}_{4,2} \!+\! w^{\langle 2 \rangle}_{4,2} \!+\! \hat{R}_{4,2} \}
\end{align}
Afterwards, client $2$ transmits the different coded answers to the different working databases after appending its own randomness $\{w^{(1)}_{k,l_0} \!: k \in \Gamma, l_0 \in [4]\}$ that is randomly selected only by client $2$ under a uniform distribution from $F_{13}$, and thus, completely unknown to all the databases in the server,
\begin{align}
    A_{W,2}^{\langle 2 \rangle,(1)} \!=\! \{&M_{1,1} \!+\! \Delta^{\langle 1 \rangle}_{1,1} \!+\! \Delta^{\langle 2 \rangle}_{1,1} \!+\! w^{\langle 1 \rangle}_{1,1} \!+\! w^{\langle 2 \rangle}_{1,1} \!+\! \hat{R}_{1,1} \!+\! M_{1,2} \!+\! \Delta^{\langle 1 \rangle}_{1,2} \!+\! \Delta^{\langle 2 \rangle}_{1,2} \!+\! w^{\langle 1 \rangle}_{1,2} \!+\! w^{\langle 2 \rangle}_{1,2} \!+\! \hat{R}_{1,2} \!+\! w^{(1)}_{1,1}, \notag \\
    & M_{1,2} \!+\! \Delta^{\langle 1 \rangle}_{1,2} \!+\! \Delta^{\langle 2 \rangle}_{1,2} \!+\! w^{\langle 1 \rangle}_{1,2} \!+\! w^{\langle 2 \rangle}_{1,2} \!+\! \hat{R}_{1,2} \!+\! w^{(1)}_{1,2} \!+\! w^{(1)}_{1,3}, w^{(1)}_{1,1} \!+\! w^{(1)}_{1,3} \!+\! w^{(1)}_{1,4}, \notag \\
    &M_{3,1} \!+\! \Delta^{\langle 2 \rangle}_{3,1} \!+\! w^{\langle 1 \rangle}_{3,1} \!+\! w^{\langle 2 \rangle}_{3,1} \!+\! \hat{R}_{3,1} \!+\! M_{3,2} \!+\! \Delta^{\langle 2 \rangle}_{3,2} \!+\! w^{\langle 1 \rangle}_{3,2} \!+\! w^{\langle 2 \rangle}_{3,2} \!+\! \hat{R}_{3,2} \!+\! w^{(1)}_{3,1}, \notag \\ 
    &M_{3,2} \!+\! \Delta^{\langle 2 \rangle}_{3,2} \!+\! w^{\langle 1 \rangle}_{3,2} \!+\! w^{\langle 2 \rangle}_{3,2} \!+\! \hat{R}_{3,2} \!+\! w^{(1)}_{3,2} \!+\! w^{(1)}_{3,3}, w^{(1)}_{3,1} \!+\! w^{(1)}_{3,3} \!+\! w^{(1)}_{3,4}, \notag \\
    &M_{4,1} \!+\! w^{\langle 1 \rangle}_{4,1} \!+\! w^{\langle 2 \rangle}_{4,1} \!+\! \hat{R}_{4,1} \!+\! M_{4,2} \!+\! w^{\langle 1 \rangle}_{4,2} \!+\! w^{\langle 2 \rangle}_{4,2} \!+\! \hat{R}_{4,2} \!+\! w^{(1)}_{4,1}, \notag \\
    & M_{4,2} \!+\! w^{\langle 1 \rangle}_{4,2} \!+\! w^{\langle 2 \rangle}_{4,2} \!+\! \hat{R}_{4,2} \!+\! w^{(1)}_{4,2} \!+\! w^{(1)}_{4,3}, w^{(1)}_{4,1} \!+\! w^{(1)}_{4,3} \!+\! w^{(1)}_{4,4} \} \\
    %%%%%%%%%%%%%%%%%%%%%%%%%%%%%%%%%%%%%%%%%%%%%%%%%%%%%%%%%%%%%%%
    A_{W,2}^{\langle 2 \rangle,(2)} \!=\! \{&M_{1,1} \!+\! \Delta^{\langle 1 \rangle}_{1,1} \!+\! \Delta^{\langle 2 \rangle}_{1,1} \!+\! w^{\langle 1 \rangle}_{1,1} \!+\! w^{\langle 2 \rangle}_{1,1} \!+\! \hat{R}_{1,1} \!+\! 2(M_{1,2} \!+\! \Delta^{\langle 1 \rangle}_{1,2} \!+\! \Delta^{\langle 2 \rangle}_{1,2} \!+\! w^{\langle 1 \rangle}_{1,2} \!+\! w^{\langle 2 \rangle}_{1,2} \!+\! \hat{R}_{1,2}) \!+\! 3w^{(1)}_{1,1}, \notag \\
    & M_{1,2} \!+\! \Delta^{\langle 1 \rangle}_{1,2} \!+\! \Delta^{\langle 2 \rangle}_{1,2} \!+\! w^{\langle 1 \rangle}_{1,2} \!+\! w^{\langle 2 \rangle}_{1,2} \!+\! \hat{R}_{1,2} \!+\! 2w^{(1)}_{1,2} \!+\! 3w^{(1)}_{1,3}, w^{(1)}_{1,1} \!+\! 2w^{(1)}_{1,3} \!+\! 3w^{(1)}_{1,4}, \notag \\
    &M_{3,1} \!+\! \Delta^{\langle 2 \rangle}_{3,1} \!+\! w^{\langle 1 \rangle}_{3,1} \!+\! w^{\langle 2 \rangle}_{3,1} \!+\! \hat{R}_{3,1} \!+\! 2(M_{3,2} \!+\! \Delta^{\langle 2 \rangle}_{3,2} \!+\! w^{\langle 1 \rangle}_{3,2} \!+\! w^{\langle 2 \rangle}_{3,2} \!+\! \hat{R}_{3,2}) \!+\! 3w^{(1)}_{3,1}, \notag \\ 
    &M_{3,2} \!+\! \Delta^{\langle 2 \rangle}_{3,2} \!+\! w^{\langle 1 \rangle}_{3,2} \!+\! w^{\langle 2 \rangle}_{3,2} \!+\! \hat{R}_{3,2} \!+\! 2w^{(1)}_{3,2} \!+\! 3w^{(1)}_{3,3}, w^{(1)}_{3,1} \!+\! 2w^{(1)}_{3,3} \!+\! 3w^{(1)}_{3,4}, \notag \\
    &M_{4,1} \!+\! w^{\langle 1 \rangle}_{4,1} \!+\! w^{\langle 2 \rangle}_{4,1} \!+\! \hat{R}_{4,1} \!+\! 2(M_{4,2} \!+\! w^{\langle 1 \rangle}_{4,2} \!+\! w^{\langle 2 \rangle}_{4,2} \!+\! \hat{R}_{4,2}) \!+\! 3w^{(1)}_{4,1}, \notag \\
    &M_{4,2} \!+\! w^{\langle 1 \rangle}_{4,2} \!+\! w^{\langle 2 \rangle}_{4,2} \!+\! \hat{R}_{4,2} \!+\! 2w^{(1)}_{4,2} \!+\! 3w^{(1)}_{4,3}, w^{(1)}_{4,1} \!+\! 2w^{(1)}_{4,3} \!+\! 3w^{(1)}_{4,4} \} \\
    %%%%%%%%%%%%%%%%%%%%%%%%%%%%%%%%%%%%%%%%%%%%%%%%%%%%%%%%%%%%%%%
    A_{W,2}^{\langle 2 \rangle,(3)} \!=\! \{&M_{1,1} \!+\! \Delta^{\langle 1 \rangle}_{1,1} \!+\! \Delta^{\langle 2 \rangle}_{1,1} \!+\! w^{\langle 1 \rangle}_{1,1} \!+\! w^{\langle 2 \rangle}_{1,1} \!+\! \hat{R}_{1,1} \!+\! 3(M_{1,2} \!+\! \Delta^{\langle 1 \rangle}_{1,2} \!+\! \Delta^{\langle 2 \rangle}_{1,2} \!+\! w^{\langle 1 \rangle}_{1,2} \!+\! w^{\langle 2 \rangle}_{1,2} \!+\! \hat{R}_{1,2}) \!+\! 9w^{(1)}_{1,1}, \notag \\
    & M_{1,2} \!+\! \Delta^{\langle 1 \rangle}_{1,2} \!+\! \Delta^{\langle 2 \rangle}_{1,2} \!+\! w^{\langle 1 \rangle}_{1,2} \!+\! w^{\langle 2 \rangle}_{1,2} \!+\! \hat{R}_{1,2} \!+\! 3w^{(1)}_{1,2} \!+\! 9w^{(1)}_{1,3}, w^{(1)}_{1,1} \!+\! 3w^{(1)}_{1,3} \!+\! 9w^{(1)}_{1,4}, \notag \\
    &M_{3,1} \!+\! \Delta^{\langle 2 \rangle}_{3,1} \!+\! w^{\langle 1 \rangle}_{3,1} \!+\! w^{\langle 2 \rangle}_{3,1} \!+\! \hat{R}_{3,1} \!+\! 3(M_{3,2} \!+\! \Delta^{\langle 2 \rangle}_{3,2} \!+\! w^{\langle 1 \rangle}_{3,2} \!+\! w^{\langle 2 \rangle}_{3,2} \!+\! \hat{R}_{3,2}) \!+\! 9w^{(1)}_{3,1}, \notag \\ 
    &M_{3,2} \!+\! \Delta^{\langle 2 \rangle}_{3,2} \!+\! w^{\langle 1 \rangle}_{3,2} \!+\! w^{\langle 2 \rangle}_{3,2} \!+\! \hat{R}_{3,2} \!+\! 3w^{(1)}_{3,2} \!+\! 9w^{(1)}_{3,3}, w^{(1)}_{3,1} \!+\! 3w^{(1)}_{3,3} \!+\! 9w^{(1)}_{3,4}, \notag \\
    &M_{4,1} \!+\! w^{\langle 1 \rangle}_{4,1} \!+\! w^{\langle 2 \rangle}_{4,1} \!+\! \hat{R}_{4,1} \!+\! 3(M_{4,2} \!+\! w^{\langle 1 \rangle}_{4,2} \!+\! w^{\langle 2 \rangle}_{4,2} \!+\! \hat{R}_{4,2}) \!+\! 9w^{(1)}_{4,1}, \notag \\
    &M_{4,2} \!+\! w^{\langle 1 \rangle}_{4,2} \!+\! w^{\langle 2 \rangle}_{4,2} \!+\! \hat{R}_{4,2} \!+\! 3w^{(1)}_{4,2} \!+\! 9w^{(1)}_{4,3}, w^{(1)}_{4,1} \!+\! 3w^{(1)}_{4,3} \!+\! 9w^{(1)}_{4,4} \} \\
    %%%%%%%%%%%%%%%%%%%%%%%%%%%%%%%%%%%%%%%%%%%%%%%%%%%%%%%%%%%%%%%
    A_{W,2}^{\langle 2 \rangle,(4)} \!=\! \{&M_{1,1} \!+\! \Delta^{\langle 1 \rangle}_{1,1} \!+\! \Delta^{\langle 2 \rangle}_{1,1} \!+\! w^{\langle 1 \rangle}_{1,1} \!+\! w^{\langle 2 \rangle}_{1,1} \!+\! \hat{R}_{1,1} \!+\! 4(M_{1,2} \!+\! \Delta^{\langle 1 \rangle}_{1,2} \!+\! \Delta^{\langle 2 \rangle}_{1,2} \!+\! w^{\langle 1 \rangle}_{1,2} \!+\! w^{\langle 2 \rangle}_{1,2} \!+\! \hat{R}_{1,2}) \!+\! 3w^{(1)}_{1,1}, \notag \\
    &M_{1,2} \!+\! \Delta^{\langle 1 \rangle}_{1,2} \!+\! \Delta^{\langle 2 \rangle}_{1,2} \!+\! w^{\langle 1 \rangle}_{1,2} \!+\! w^{\langle 2 \rangle}_{1,2} \!+\! \hat{R}_{1,2} \!+\! 4w^{(1)}_{1,2} \!+\! 3w^{(1)}_{1,3}, w^{(1)}_{1,1} \!+\! 4w^{(1)}_{1,3} \!+\! 3w^{(1)}_{1,4}, \notag \\
    &M_{3,1} \!+\! \Delta^{\langle 2 \rangle}_{3,1} \!+\! w^{\langle 1 \rangle}_{3,1} \!+\! w^{\langle 2 \rangle}_{3,1} \!+\! \hat{R}_{3,1} \!+\! 4(M_{3,2} \!+\! \Delta^{\langle 2 \rangle}_{3,2} \!+\! w^{\langle 1 \rangle}_{3,2} \!+\! w^{\langle 2 \rangle}_{3,2} \!+\! \hat{R}_{3,2}) \!+\! 3w^{(1)}_{3,1}, \notag \\ 
    &M_{3,2} \!+\! \Delta^{\langle 2 \rangle}_{3,2} \!+\! w^{\langle 1 \rangle}_{3,2} \!+\! w^{\langle 2 \rangle}_{3,2} \!+\! \hat{R}_{3,2} \!+\! 4w^{(1)}_{3,2} \!+\! 3w^{(1)}_{3,3}, w^{(1)}_{3,1} \!+\! 4w^{(1)}_{3,3} \!+\! 3w^{(1)}_{3,4}, \notag \\
    &M_{4,1} \!+\! w^{\langle 1 \rangle}_{4,1} \!+\! w^{\langle 2 \rangle}_{4,1} \!+\! \hat{R}_{4,1} \!+\! 4(M_{4,2} \!+\! w^{\langle 1 \rangle}_{4,2} \!+\! w^{\langle 2 \rangle}_{4,2} \!+\! \hat{R}_{4,2}) \!+\! 3w^{(1)}_{4,1}, \notag \\
    &M_{4,2} \!+\! w^{\langle 1 \rangle}_{4,2} \!+\! w^{\langle 2 \rangle}_{4,2} \!+\! \hat{R}_{4,2} \!+\! 4w^{(1)}_{4,2} \!+\! 3w^{(1)}_{4,3}, w^{(1)}_{4,1} \!+\! 4w^{(1)}_{4,3} \!+\! 3w^{(1)}_{4,4} \}
\end{align}
Likewise, the following information is downloaded by client $3$,
\begin{align}
    D_{W,2}^{\langle 3 \rangle,(2)} = \{&\Delta^{\langle 3 \rangle}_{1,1} \!+\! w^{\langle 3 \rangle}_{1,1} \!+\! \hat{R}_{1,1}, \Delta^{\langle 3 \rangle}_{1,2} \!+\! w^{\langle 3 \rangle}_{1,2} \!+\! \hat{R}_{1,2} , w^{\langle 3 \rangle}_{3,1} \!+\! \hat{R}_{3,1}, \notag\\
    &w^{\langle 3 \rangle}_{3,2} \!+\! \hat{R}_{3,2} , \Delta^{\langle 3 \rangle}_{4,1} \!+\! w^{\langle 3 \rangle}_{4,1} \!+\! \hat{R}_{4,1}, \Delta^{\langle 3 \rangle}_{4,2} \!+\! w^{\langle 3 \rangle}_{4,2} \!+\! \hat{R}_{4,2} \} 
\end{align}
Exactly like client $2$, client $3$ transmits the coded answers to the corresponding databases as follows where $\{w^{(2)}_{k,l_0} \!: k \in \Gamma, l_0 \in [4]\}$ is its own randomness,
\begin{align}
    A_{W,2}^{\langle 3 \rangle,(1)} \!=\! \{&\Delta^{\langle 3 \rangle}_{1,1} \!+\! w^{\langle 3 \rangle}_{1,1} \!+\! \hat{R}_{1,1} \!+\! \Delta^{\langle 3 \rangle}_{1,2} \!+\! w^{\langle 3 \rangle}_{1,2} \!+\! \hat{R}_{1,2} \!+\! w^{(2)}_{1,1}, \notag \\
    &\Delta^{\langle 3 \rangle}_{1,2} \!+\! w^{\langle 3 \rangle}_{1,2} \!+\! \hat{R}_{1,2} \!+\! w^{(2)}_{1,2} \!+\! w^{(2)}_{1,3}, w^{(2)}_{1,1} \!+\! w^{(2)}_{1,3} \!+\! w^{(2)}_{1,4}, \notag \\
    &w^{\langle 3 \rangle}_{3,1} \!+\! \hat{R}_{3,1} \!+\! w^{\langle 3 \rangle}_{3,2} \!+\! \hat{R}_{3,2} \!+\! w^{(2)}_{3,1}, \notag \\
    &w^{\langle 3 \rangle}_{3,2} \!+\! \hat{R}_{3,2} \!+\! w^{(2)}_{3,2} \!+\! w^{(2)}_{3,3}, w^{(2)}_{3,1} \!+\! w^{(2)}_{3,3} \!+\! w^{(2)}_{3,4}, \notag \\
    &\Delta^{\langle 3 \rangle}_{4,1} \!+\! w^{\langle 3 \rangle}_{4,1} \!+\! \hat{R}_{4,1} \!+\! \Delta^{\langle 3 \rangle}_{4,2} \!+\! w^{\langle 3 \rangle}_{4,2} \!+\! \hat{R}_{4,2} \!+\! w^{(2)}_{4,1}, \notag \\
    &\Delta^{\langle 3 \rangle}_{4,2} \!+\! w^{\langle 3 \rangle}_{4,2} \!+\! \hat{R}_{4,2} \!+\! w^{(2)}_{4,2} \!+\! w^{(2)}_{4,3}, w^{(2)}_{4,1} \!+\! w^{(2)}_{4,3} \!+\! w^{(2)}_{4,4} \} \\
    %%%%%%%%%%%%%%%%%%%%%%%%%%%%%%%%%%%%%%%%%%%%%%%%%%%%%%%%%%%%%%%
    A_{W,2}^{\langle 3 \rangle,(2)} \!=\! \{&\Delta^{\langle 3 \rangle}_{1,1} \!+\! w^{\langle 3 \rangle}_{1,1} \!+\! \hat{R}_{1,1} \!+\! 2(\Delta^{\langle 3 \rangle}_{1,2} \!+\! w^{\langle 3 \rangle}_{1,2} \!+\! \hat{R}_{1,2}) \!+\! 3w^{(2)}_{1,1}, \notag \\
    &\Delta^{\langle 3 \rangle}_{1,2} \!+\! w^{\langle 3 \rangle}_{1,2} \!+\! \hat{R}_{1,2} \!+\! 2w^{(2)}_{1,2} \!+\! 3w^{(2)}_{1,3}, w^{(2)}_{1,1} \!+\! 2w^{(2)}_{1,3} \!+\! 3w^{(2)}_{1,4}, \notag \\
    &w^{\langle 3 \rangle}_{3,1} \!+\! \hat{R}_{3,1} \!+\! 2(w^{\langle 3 \rangle}_{3,2} \!+\! \hat{R}_{3,2}) \!+\! 3w^{(2)}_{3,1}, \notag \\
    &w^{\langle 3 \rangle}_{3,2} \!+\! \hat{R}_{3,2} \!+\! 2w^{(2)}_{3,2} \!+\! 3w^{(2)}_{3,3}, w^{(2)}_{3,1} \!+\! 2w^{(2)}_{3,3} \!+\! 3w^{(2)}_{3,4}, \notag \\
    &\Delta^{\langle 3 \rangle}_{4,1} \!+\! w^{\langle 3 \rangle}_{4,1} \!+\! \hat{R}_{4,1} \!+\! 2(\Delta^{\langle 3 \rangle}_{4,2} \!+\! w^{\langle 3 \rangle}_{4,2} \!+\! \hat{R}_{4,2}) \!+\! 3w^{(2)}_{4,1}, \notag \\
    &\Delta^{\langle 3 \rangle}_{4,2} \!+\! w^{\langle 3 \rangle}_{4,2} \!+\! \hat{R}_{4,2} \!+\! 2w^{(2)}_{4,2} \!+\! 3w^{(2)}_{4,3}, w^{(2)}_{4,1} \!+\! 2w^{(2)}_{4,3} \!+\! 3w^{(2)}_{4,4} \} \\
    %%%%%%%%%%%%%%%%%%%%%%%%%%%%%%%%%%%%%%%%%%%%%%%%%%%%%%%%%%%%%%%
    A_{W,2}^{\langle 3 \rangle,(3)} \!=\! \{&\Delta^{\langle 3 \rangle}_{1,1} \!+\! w^{\langle 3 \rangle}_{1,1} \!+\! \hat{R}_{1,1} \!+\! 3(\Delta^{\langle 3 \rangle}_{1,2} \!+\! w^{\langle 3 \rangle}_{1,2} \!+\! \hat{R}_{1,2}) \!+\! 9w^{(2)}_{1,1}, \notag \\
    &\Delta^{\langle 3 \rangle}_{1,2} \!+\! w^{\langle 3 \rangle}_{1,2} \!+\! \hat{R}_{1,2} \!+\! 3w^{(2)}_{1,2} \!+\! 9w^{(2)}_{1,3}, w^{(2)}_{1,1} \!+\! 3w^{(2)}_{1,3} \!+\! 9w^{(2)}_{1,4}, \notag \\
    &w^{\langle 3 \rangle}_{3,1} \!+\! \hat{R}_{3,1} \!+\! 3(w^{\langle 3 \rangle}_{3,2} \!+\! \hat{R}_{3,2}) \!+\! 9w^{(2)}_{3,1}, \notag \\
    &w^{\langle 3 \rangle}_{3,2} \!+\! \hat{R}_{3,2} \!+\! 3w^{(2)}_{3,2} \!+\! 9w^{(2)}_{3,3}, w^{(2)}_{3,1} \!+\! 3w^{(2)}_{3,3} \!+\! 9w^{(2)}_{3,4}, \notag \\
    &\Delta^{\langle 3 \rangle}_{4,1} \!+\! w^{\langle 3 \rangle}_{4,1} \!+\! \hat{R}_{4,1} \!+\! 3(\Delta^{\langle 3 \rangle}_{4,2} \!+\! w^{\langle 3 \rangle}_{4,2} \!+\! \hat{R}_{4,2}) \!+\! 9w^{(2)}_{4,1}, \notag \\
    &\Delta^{\langle 3 \rangle}_{4,2} \!+\! w^{\langle 3 \rangle}_{4,2} \!+\! \hat{R}_{4,2} \!+\! 3w^{(2)}_{4,2} \!+\! 9w^{(2)}_{4,3}, w^{(2)}_{4,1} \!+\! 3w^{(2)}_{4,3} \!+\! 9w^{(2)}_{4,4} \} \\
    %%%%%%%%%%%%%%%%%%%%%%%%%%%%%%%%%%%%%%%%%%%%%%%%%%%%%%%%%%%%%%%
    A_{W,2}^{\langle 3 \rangle,(4)} \!=\! \{&\Delta^{\langle 3 \rangle}_{1,1} \!+\! w^{\langle 3 \rangle}_{1,1} \!+\! \hat{R}_{1,1} \!+\! 4(\Delta^{\langle 3 \rangle}_{1,2} \!+\! w^{\langle 3 \rangle}_{1,2} \!+\! \hat{R}_{1,2}) \!+\! 3w^{(2)}_{1,1}, \notag \\
    &\Delta^{\langle 3 \rangle}_{1,2} \!+\! w^{\langle 3 \rangle}_{1,2} \!+\! \hat{R}_{1,2} \!+\! 4w^{(2)}_{1,2} \!+\! 3w^{(2)}_{1,3}, w^{(2)}_{1,1} \!+\! 4w^{(2)}_{1,3} \!+\! 3w^{(2)}_{1,4}, \notag \\
    &w^{\langle 3 \rangle}_{3,1} \!+\! \hat{R}_{3,1} \!+\! 4(w^{\langle 3 \rangle}_{3,2} \!+\! \hat{R}_{3,2}) \!+\! 3w^{(2)}_{3,1}, \notag \\
    &w^{\langle 3 \rangle}_{3,2} \!+\! \hat{R}_{3,2} \!+\! 4w^{(2)}_{3,2} \!+\! 3w^{(2)}_{3,3}, w^{(2)}_{3,1} \!+\! 4w^{(2)}_{3,3} \!+\! 3w^{(2)}_{3,4}, \notag \\
    &\Delta^{\langle 3 \rangle}_{4,1} \!+\! w^{\langle 3 \rangle}_{4,1} \!+\! \hat{R}_{4,1} \!+\! 4(\Delta^{\langle 3 \rangle}_{4,2} \!+\! w^{\langle 3 \rangle}_{4,2} \!+\! \hat{R}_{4,2}) \!+\! 3w^{(2)}_{4,1}, \notag \\
    &\Delta^{\langle 3 \rangle}_{4,2} \!+\! w^{\langle 3 \rangle}_{4,2} \!+\! \hat{R}_{4,2} \!+\! 4w^{(2)}_{4,2} \!+\! 3w^{(2)}_{4,3}, w^{(2)}_{4,1} \!+\! 4w^{(2)}_{4,3} \!+\! 3w^{(2)}_{4,4} \} 
\end{align}
At the same time, the following information is downloaded by client $4$,
\begin{align}
    D_{W,2}^{\langle 4 \rangle,(3)} = \{&\Delta^{\langle 4 \rangle}_{1,1} \!+\! w^{\langle 4 \rangle}_{1,1} \!+\! \hat{R}_{1,1} , \Delta^{\langle 4 \rangle}_{1,2} \!+\! w^{\langle 4 \rangle}_{1,2} \!+\! \hat{R}_{1,2}, \Delta^{\langle 4 \rangle}_{3,1} \!+\! w^{\langle 4 \rangle}_{3,1} \!+\! \hat{R}_{3,1}, \notag\\
    &\Delta^{\langle 4 \rangle}_{3,2} \!+\! w^{\langle 4 \rangle}_{3,2} \!+\! \hat{R}_{3,2}, \Delta^{\langle 4 \rangle}_{4,1} \!+\! w^{\langle 4 \rangle}_{4,1} \!+\! \hat{R}_{4,1}, \Delta^{\langle 4 \rangle}_{4,2} \!+\! w^{\langle 4 \rangle}_{4,2} \!+\! \hat{R}_{4,2}
    \}
\end{align}
As client $4$ is also employed to route the information of submodel $2$ to the replacement database, it transmits the coded answers to the corresponding databases in the following way where $\{w^{(3)}_{k,l_0} \!: k \in \Gamma, l_0 \in [4]\}$ is its own randomness and $A_{W,2}^{\langle 4 \rangle,(4)}$ is particularly different,
\begin{align}
    A_{W,2}^{\langle 4 \rangle,(1)} \!=\! \{&\Delta^{\langle 4 \rangle}_{1,1} \!+\! w^{\langle 4 \rangle}_{1,1} \!+\! \hat{R}_{1,1} \!+\! \Delta^{\langle 4 \rangle}_{1,2} \!+\! w^{\langle 4 \rangle}_{1,2} \!+\! \hat{R}_{1,2} \!+\! w^{(3)}_{1,1}, \notag \\
    &\Delta^{\langle 4 \rangle}_{1,2} \!+\! w^{\langle 4 \rangle}_{1,2} \!+\! \hat{R}_{1,2} \!+\! w^{(3)}_{1,2} \!+\! w^{(3)}_{1,3}, w^{(3)}_{1,1} \!+\! w^{(3)}_{1,3} \!+\! w^{(3)}_{1,4} \notag \\
    &\Delta^{\langle 4 \rangle}_{3,1} \!+\! w^{\langle 4 \rangle}_{3,1} \!+\! \hat{R}_{3,1} \!+\! \Delta^{\langle 4 \rangle}_{3,2} \!+\! w^{\langle 4 \rangle}_{3,2} \!+\! \hat{R}_{3,1} \!+\! w^{(3)}_{3,1}, \notag\\
    &\Delta^{\langle 4 \rangle}_{3,2} \!+\! w^{\langle 4 \rangle}_{3,2} \!+\! \hat{R}_{3,2} \!+\! w^{(3)}_{3,2} \!+\! w^{(3)}_{3,3}, w^{(3)}_{3,1} \!+\! w^{(3)}_{3,3} \!+\! w^{(3)}_{3,4}, \notag\\
    &\Delta^{\langle 4 \rangle}_{4,1} \!+\! w^{\langle 4 \rangle}_{4,1} \!+\! \hat{R}_{4,1} \!+\! \Delta^{\langle 4 \rangle}_{4,2} \!+\! w^{\langle 4 \rangle}_{4,2} \!+\! \hat{R}_{4,2} \!+\! w^{(3)}_{4,1}, \notag \\
    &\Delta^{\langle 4 \rangle}_{4,2} \!+\! w^{\langle 4 \rangle}_{4,2} \!+\! \hat{R}_{4,2} \!+\! w^{(3)}_{4,2} \!+\! w^{(3)}_{4,3}, w^{(3)}_{4,1} \!+\! w^{(3)}_{4,3} \!+\! w^{(3)}_{4,4}\} \\
    %%%%%%%%%%%%%%%%%%%%%%%%%%%%%%%%%%%%%%%%%%%%%%%%%%%%%%%%%%%%%%%
    A_{W,2}^{\langle 4 \rangle,(2)} \!=\! \{&\Delta^{\langle 4 \rangle}_{1,1} \!+\! w^{\langle 4 \rangle}_{1,1} \!+\! \hat{R}_{1,1} \!+\! 2(\Delta^{\langle 4 \rangle}_{1,2} \!+\! w^{\langle 4 \rangle}_{1,2} \!+\! \hat{R}_{1,2}) \!+\! 3w^{(3)}_{1,1}, \notag \\
    &\Delta^{\langle 4 \rangle}_{1,2} \!+\! w^{\langle 4 \rangle}_{1,2} \!+\! \hat{R}_{1,2} \!+\! 2w^{(3)}_{1,2} \!+\! 3w^{(3)}_{1,3}, w^{(3)}_{1,1} \!+\! 2w^{(3)}_{1,3} \!+\! 3w^{(3)}_{1,4} \notag \\
    &\Delta^{\langle 4 \rangle}_{3,1} \!+\! w^{\langle 4 \rangle}_{3,1} \!+\! \hat{R}_{3,1} \!+\! 2(\Delta^{\langle 4 \rangle}_{3,2} \!+\! w^{\langle 4 \rangle}_{3,2} \!+\! \hat{R}_{3,1}) + 3w^{(3)}_{3,1}, \notag\\
    &\Delta^{\langle 4 \rangle}_{3,2} \!+\! w^{\langle 4 \rangle}_{3,2} \!+\! \hat{R}_{3,2} \!+\! 2w^{(3)}_{3,2} \!+\! 3w^{(3)}_{3,3}, w^{(3)}_{3,1} \!+\! 2w^{(3)}_{3,3} \!+\! 3w^{(3)}_{3,4}, \notag\\
    &\Delta^{\langle 4 \rangle}_{4,1} \!+\! w^{\langle 4 \rangle}_{4,1} \!+\! \hat{R}_{4,1} \!+\! 2(\Delta^{\langle 4 \rangle}_{4,2} \!+\! w^{\langle 4 \rangle}_{4,2} \!+\! \hat{R}_{4,2}) \!+\! 3w^{(3)}_{4,1}, \notag \\
    &\Delta^{\langle 4 \rangle}_{4,2} \!+\! w^{\langle 4 \rangle}_{4,2} \!+\! \hat{R}_{4,2} \!+\! 2w^{(3)}_{4,2} \!+\! 3w^{(3)}_{4,3}, w^{(3)}_{4,1} \!+\! 2w^{(3)}_{4,3} \!+\! 3w^{(3)}_{4,4}\} \\
    %%%%%%%%%%%%%%%%%%%%%%%%%%%%%%%%%%%%%%%%%%%%%%%%%%%%%%%%%%%%%%%
    A_{W,2}^{\langle 4 \rangle,(3)} \!=\! \{&\Delta^{\langle 4 \rangle}_{1,1} \!+\! w^{\langle 4 \rangle}_{1,1} \!+\! \hat{R}_{1,1} \!+\! 3(\Delta^{\langle 4 \rangle}_{1,2} \!+\! w^{\langle 4 \rangle}_{1,2} \!+\! \hat{R}_{1,2}) \!+\! 9w^{(3)}_{1,1}, \notag \\
    &\Delta^{\langle 4 \rangle}_{1,2} \!+\! w^{\langle 4 \rangle}_{1,2} \!+\! \hat{R}_{1,2} \!+\! 3w^{(3)}_{1,2} \!+\! 9w^{(3)}_{1,3}, w^{(3)}_{1,1} \!+\! 3w^{(3)}_{1,3} \!+\! 9w^{(3)}_{1,4} \notag \\
    &\Delta^{\langle 4 \rangle}_{3,1} \!+\! w^{\langle 4 \rangle}_{3,1} \!+\! \hat{R}_{3,1} \!+\! 3(\Delta^{\langle 4 \rangle}_{3,2} \!+\! w^{\langle 4 \rangle}_{3,2} \!+\! \hat{R}_{3,1}) + 9w^{(3)}_{3,1}, \notag\\
    &\Delta^{\langle 4 \rangle}_{3,2} \!+\! w^{\langle 4 \rangle}_{3,2} \!+\! \hat{R}_{3,2} \!+\! 3w^{(3)}_{3,2} \!+\! 9w^{(3)}_{3,3}, w^{(3)}_{3,1} \!+\! 3w^{(3)}_{3,3} \!+\! 9w^{(3)}_{3,4}, \notag\\
    &\Delta^{\langle 4 \rangle}_{4,1} \!+\! w^{\langle 4 \rangle}_{4,1} \!+\! \hat{R}_{4,1} \!+\! 3(\Delta^{\langle 4 \rangle}_{4,2} \!+\! w^{\langle 4 \rangle}_{4,2} \!+\! \hat{R}_{4,2}) \!+\! 9w^{(3)}_{4,1}, \notag \\
    &\Delta^{\langle 4 \rangle}_{4,2} \!+\! w^{\langle 4 \rangle}_{4,2} \!+\! \hat{R}_{4,2} \!+\! 3w^{(3)}_{4,2} \!+\! 9w^{(3)}_{4,3}, w^{(3)}_{4,1} \!+\! 3w^{(3)}_{4,3} \!+\! 9w^{(3)}_{4,4}\} \\
    %%%%%%%%%%%%%%%%%%%%%%%%%%%%%%%%%%%%%%%%%%%%%%%%%%%%%%%%%%%%%%%
    A_{W,2}^{\langle 4 \rangle,(4)} \!=\! \{&\Delta^{\langle 4 \rangle}_{1,1} \!+\! w^{\langle 4 \rangle}_{1,1} \!+\! \hat{R}_{1,1} \!+\! 4(\Delta^{\langle 4 \rangle}_{1,2} \!+\! w^{\langle 4 \rangle}_{1,2} \!+\! \hat{R}_{1,2}) \!+\! 3w^{(3)}_{1,1}, \notag \\
    &\Delta^{\langle 4 \rangle}_{1,2} \!+\! w^{\langle 4 \rangle}_{1,2} \!+\! \hat{R}_{1,2} \!+\! 4w^{(3)}_{1,2} \!+\! 3w^{(3)}_{1,3}, w^{(3)}_{1,1} \!+\! 4w^{(3)}_{1,3} \!+\! 3w^{(3)}_{1,4} \notag \\
    &\Delta^{\langle 4 \rangle}_{3,1} \!+\! w^{\langle 4 \rangle}_{3,1} \!+\! \hat{R}_{3,1} \!+\! 4(\Delta^{\langle 4 \rangle}_{3,2} \!+\! w^{\langle 4 \rangle}_{3,2} \!+\! \hat{R}_{3,1}) + 3w^{(3)}_{3,1}, \notag\\
    &\Delta^{\langle 4 \rangle}_{3,2} \!+\! w^{\langle 4 \rangle}_{3,2} \!+\! \hat{R}_{3,2} \!+\! 4w^{(3)}_{3,2} \!+\! 3w^{(3)}_{3,3}, w^{(3)}_{3,1} \!+\! 4w^{(3)}_{3,3} \!+\! 3w^{(3)}_{3,4}, \notag\\
    &\Delta^{\langle 4 \rangle}_{4,1} \!+\! w^{\langle 4 \rangle}_{4,1} \!+\! \hat{R}_{4,1} \!+\! 4(\Delta^{\langle 4 \rangle}_{4,2} \!+\! w^{\langle 4 \rangle}_{4,2} \!+\! \hat{R}_{4,2}) \!+\! 3w^{(3)}_{4,1}, \notag \\
    &\Delta^{\langle 4 \rangle}_{4,2} \!+\! w^{\langle 4 \rangle}_{4,2} \!+\! \hat{R}_{4,2} \!+\! 4w^{(3)}_{4,2} \!+\! 3w^{(3)}_{4,3}, w^{(3)}_{4,1} \!+\! 4w^{(3)}_{4,3} \!+\! 3w^{(3)}_{4,4}, \notag \\  &M_{2,1}\!+\!M_{2,2}\!+\!R_{2,1}\!+\!4(M_{2,2}\!+\!R_{2,2}\!+\!R_{2,3})\!+\! 3(R_{2,1}\!+\!R_{2,3}\!+\!R_{2,4}), \notag \\
    &M_{2,1}\!+\!2M_{2,2}\!+\!3R_{2,1}\!+\! 4(M_{2,2}\!+\!2R_{2,2}\!+\!3R_{2,3})\!+\!3(R_{2,1}\!+\!2R_{2,3}\!+\!3R_{2,4}), \notag \\ &M_{2,1}\!+\!3M_{2,2}\!+\!9R_{2,1}\!+\!4(M_{2,2}\!+\!3R_{2,2}\!+\!9R_{2,3})\!+\!3(R_{2,1}\!+\!3R_{2,3}\!+\!9R_{2,4})\} 
\end{align}

As it is easy to check that the privacy constraint \eqref{privacy} and the inter-client privacy constraint \eqref{inter-client privacy} are both inherited directly from our previous FSL-PSU scheme in \cite{FSL-PSU}, our emphasis here will be on the analysis of the reliability constraint \eqref{reliability}, the eavesdropper security constraint \eqref{eavesdropper security} and the database failure robustness.

Regarding the reliability constraint: After collecting the answers from the routing clients $2, 3, 4$ in the second step of the FSL-write phase, each database just does the element-wise summation. Without loss of generality, let us focus on the first coded submodel symbol $M_{1,1}\!+\!2M_{1,2}\!+\!3R_{1,1}$ in database $2$, then we have the following calculation where $\sum_{i \in [4]} w^{\langle i \rangle}_{1,1} = \sum_{i \in [4]} w^{\langle i \rangle}_{1,2} = 0$,
\begin{align}
    &M_{1,1} \!+\! \Delta^{\langle 1 \rangle}_{1,1} \!+\! \Delta^{\langle 2 \rangle}_{1,1} \!+\! w^{\langle 1 \rangle}_{1,1} \!+\! w^{\langle 2 \rangle}_{1,1} \!+\! \hat{R}_{1,1} \!+\! 2(M_{1,2} \!+\! \Delta^{\langle 1 \rangle}_{1,2} \!+\! \Delta^{\langle 2 \rangle}_{1,2} \!+\! w^{\langle 1 \rangle}_{1,2} \!+\! w^{\langle 2 \rangle}_{1,2} \!+\! \hat{R}_{1,2}) \!+\! 3w^{(1)}_{1,1} \notag \\
    &\quad \!+\! \Delta^{\langle 3 \rangle}_{1,1} \!+\! w^{\langle 3 \rangle}_{1,1} \!+\! \hat{R}_{1,1} \!+\! 2(\Delta^{\langle 3 \rangle}_{1,2} \!+\! w^{\langle 3 \rangle}_{1,2} \!+\! \hat{R}_{1,2}) \!+\! 3w^{(2)}_{1,1} \notag \\
    &\quad \!+\! \Delta^{\langle 4 \rangle}_{1,1} \!+\! w^{\langle 4 \rangle}_{1,1} \!+\! \hat{R}_{1,1} \!+\! 2(\Delta^{\langle 4 \rangle}_{1,2} \!+\! w^{\langle 4 \rangle}_{1,2} \!+\! \hat{R}_{1,2}) \!+\! 3w^{(3)}_{1,1}  \\
    &= M_{1,1} \!+\! \sum_{i \in [4]} \Delta^{\langle i \rangle}_{1,1} \!+\! 2(M_{1,2} \!+\! \sum_{i \in [4]} \Delta^{\langle i \rangle}_{1,2}) \!+\! 3(w^{(1)}_{1,1} \!+\! w^{(2)}_{1,1} \!+\! w^{(3)}_{1,1}) \!+\! 3\hat{R}_{1,1} \!+\! 6\hat{R}_{1,2}\\
    &= M^{\prime}_{1,1} \!+\! 2M^{\prime}_{1,2} \!+\! 3R^{\prime}_{1,1} \!+\! 3\hat{R}_{1,1} \!+\! 6\hat{R}_{1,2}
\end{align}
As $3\hat{R}_{1,1} \!+\! 6\hat{R}_{1,2}$ is a known constant to database $2$ and $w^{(1)}_{1,1} \!+\! w^{(2)}_{1,1} \!+\! w^{(3)}_{1,1}$ can be treated as $R^{\prime}_{1,1}$, this database is able to decode the value of $M^{\prime}_{1,1} \!+\! 2M^{\prime}_{1,2} \!+\! 3R^{\prime}_{1,1}$, which will be stored as a new coded submodel symbol for the next round of the FSL process. For the other needed symbols across the databases, the same calculation can be performed. In addition, in order to make the storage consistent across the databases and achieve perfect privacy in the next round, all the extra uncoded server-side common randomness also needs to be refreshed. When this round of the FSL process is complete, the updated storage in the server is shown in Table~\ref{E2_2}.

 \begin{table}[ht]
\begin{center}
\begin{tabular}{|c|c|c|}
\hline
Database & \multicolumn{2}{c|}{Storage} \\
\hline
\multirow{4}{*}{DB 1} & $M^{\prime}_{1,1}\!+\!M^{\prime}_{1,2}\!+\!R^{\prime}_{1,1}, \ M^{\prime}_{1,2}\!+\!R^{\prime}_{1,2}\!+\!R^{\prime}_{1,3}, \ R^{\prime}_{1,1}\!+\!R^{\prime}_{1,3}\!+\!R^{\prime}_{1,4}$ & $\hat{R}^{\prime}_1, \hat{R}^{\prime}_{1,1}, \hat{R}^{\prime}_{1,2}$ \\ 
& $M_{2,1}\!+\!M_{2,2}\!+\!R_{2,1}, \ M_{2,2}\!+\!R_{2,2}\!+\!R_{2,3}, \ R_{2,1}\!+\!R_{2,3}\!+\!R_{2,4}$ & $\hat{R}^{\prime}_2, \hat{R}^{\prime}_{2,1}, \hat{R}^{\prime}_{2,2}$ \\ 
& $M^{\prime}_{3,1}\!+\!M^{\prime}_{3,2}\!+\!R^{\prime}_{3,1}, \ M^{\prime}_{3,2}\!+\!R^{\prime}_{3,2}\!+\!R^{\prime}_{3,3}, \ R^{\prime}_{3,1}\!+\!R^{\prime}_{3,3}\!+\!R^{\prime}_{3,4}$ & $\hat{R}^{\prime}_3, \hat{R}^{\prime}_{3,1}, \hat{R}^{\prime}_{3,2}$ \\ 
& $M^{\prime}_{4,1}\!+\!M^{\prime}_{4,2}\!+\!R^{\prime}_{4,1}, \ M^{\prime}_{4,2}\!+\!R^{\prime}_{4,2}\!+\!R^{\prime}_{4,3}, \ R^{\prime}_{4,1}\!+\!R^{\prime}_{4,3}\!+\!R^{\prime}_{4,4}$ & $\hat{R}^{\prime}_4, \hat{R}^{\prime}_{4,1}, \hat{R}^{\prime}_{4,2}$ \\ \hline
\multirow{4}{*}{DB 2} & $M^{\prime}_{1,1}\!+\!2M^{\prime}_{1,2}\!+\!3R^{\prime}_{1,1}, \ M^{\prime}_{1,2}\!+\!2R^{\prime}_{1,2}\!+\!3R^{\prime}_{1,3}, \ R^{\prime}_{1,1}\!+\!2R^{\prime}_{1,3}\!+\!3R^{\prime}_{1,4}$ & $\hat{R}^{\prime}_1, \hat{R}^{\prime}_{1,1}, \hat{R}^{\prime}_{1,2}$ \\ 
& $M_{2,1}\!+\!2M_{2,2}\!+\!3R_{2,1}, \ M_{2,2}\!+\!2R_{2,2}\!+\!3R_{2,3}, \ R_{2,1}\!+\!2R_{2,3}\!+\!3R_{2,4}$ & $\hat{R}^{\prime}_2, \hat{R}^{\prime}_{2,1}, \hat{R}^{\prime}_{2,2}$ \\ 
& $M^{\prime}_{3,1}\!+\!2M^{\prime}_{3,2}\!+\!3R^{\prime}_{3,1}, \ M^{\prime}_{3,2}\!+\!2R^{\prime}_{3,2}\!+\!3R^{\prime}_{3,3}, \ R^{\prime}_{3,1}\!+\!2R^{\prime}_{3,3}\!+\!3R^{\prime}_{3,4}$ & $\hat{R}^{\prime}_3, \hat{R}^{\prime}_{3,1}, \hat{R}^{\prime}_{3,2}$ \\ 
& $M^{\prime}_{4,1}\!+\!2M^{\prime}_{4,2}\!+\!3R^{\prime}_{4,1}, \ M^{\prime}_{4,2}\!+\!2R^{\prime}_{4,2}\!+\!3R^{\prime}_{4,3}, \ R^{\prime}_{4,1}\!+\!2R^{\prime}_{4,3}\!+\!3R^{\prime}_{4,4}$ & $\hat{R}^{\prime}_4, \hat{R}^{\prime}_{4,1}, \hat{R}^{\prime}_{4,2}$ \\ \hline
\multirow{4}{*}{DB 3} & $M^{\prime}_{1,1}\!+\!3M^{\prime}_{1,2}\!+\!9R^{\prime}_{1,1}, \ M^{\prime}_{1,2}\!+\!3R^{\prime}_{1,2}\!+\!9R^{\prime}_{1,3}, \ R^{\prime}_{1,1}\!+\!3R^{\prime}_{1,3}\!+\!9R^{\prime}_{1,4}$ & $\hat{R}^{\prime}_1, \hat{R}^{\prime}_{1,1}, \hat{R}^{\prime}_{1,2}$ \\ 
& $M_{2,1}\!+\!3M_{2,2}\!+\!9R_{2,1}, \ M_{2,2}\!+\!3R_{2,2}\!+\!9R_{2,3}, \ R_{2,1}\!+\!3R_{2,3}\!+\!9R_{2,4}$ & $\hat{R}^{\prime}_2, \hat{R}^{\prime}_{2,1}, \hat{R}^{\prime}_{2,2}$ \\ 
& $M^{\prime}_{3,1}\!+\!3M^{\prime}_{3,2}\!+\!9R^{\prime}_{3,1}, \ M^{\prime}_{3,2}\!+\!3R^{\prime}_{3,2}\!+\!9R^{\prime}_{3,3}, \ R^{\prime}_{3,1}\!+\!3R^{\prime}_{3,3}\!+\!9R^{\prime}_{3,4}$ & $\hat{R}^{\prime}_3, \hat{R}^{\prime}_{3,1}, \hat{R}^{\prime}_{3,2}$ \\ 
& $M^{\prime}_{4,1}\!+\!3M^{\prime}_{4,2}\!+\!9R^{\prime}_{4,1}, \ M^{\prime}_{4,2}\!+\!3R^{\prime}_{4,2}\!+\!9R^{\prime}_{4,3}, \ R^{\prime}_{4,1}\!+\!3R^{\prime}_{4,3}\!+\!9R^{\prime}_{4,4}$ & $\hat{R}^{\prime}_4, \hat{R}^{\prime}_{4,1}, \hat{R}^{\prime}_{4,2}$\\ \hline
\multirow{4}{*}{DB 4} & $M^{\prime}_{1,1}\!+\!4M^{\prime}_{1,2}\!+\!3R^{\prime}_{1,1}, \ M^{\prime}_{1,2}\!+\!4R^{\prime}_{1,2}\!+\!3R^{\prime}_{1,3}, \ R^{\prime}_{1,1}\!+\!4R^{\prime}_{1,3}\!+\!3R^{\prime}_{1,4}$ & $\hat{R}^{\prime}_1, \hat{R}^{\prime}_{1,1}, \hat{R}^{\prime}_{1,2}$ \\ 
& $M_{2,1}\!+\!4M_{2,2}\!+\!3R_{2,1}, \ M_{2,2}\!+\!4R_{2,2}\!+\!3R_{2,3}, \ R_{2,1}\!+\!4R_{2,3}\!+\!3R_{2,4}$ & $\hat{R}^{\prime}_2, \hat{R}^{\prime}_{2,1}, \hat{R}^{\prime}_{2,2}$ \\ 
& $M^{\prime}_{3,1}\!+\!4M^{\prime}_{3,2}\!+\!3R^{\prime}_{3,1}, \ M^{\prime}_{3,2}\!+\!4R^{\prime}_{3,2}\!+\!3R^{\prime}_{3,3}, \ R^{\prime}_{3,1}\!+\!4R^{\prime}_{3,3}\!+\!3R^{\prime}_{3,4}$ & $\hat{R}^{\prime}_3, \hat{R}^{\prime}_{3,1}, \hat{R}^{\prime}_{3,2}$ \\ 
& $M^{\prime}_{4,1}\!+\!4M^{\prime}_{4,2}\!+\!3R^{\prime}_{4,1}, \ M^{\prime}_{4,2}\!+\!4R^{\prime}_{4,2}\!+\!3R^{\prime}_{4,3}, \ R^{\prime}_{4,1}\!+\!4R^{\prime}_{4,3}\!+\!3R^{\prime}_{4,4}$ & $\hat{R}^{\prime}_4, \hat{R}^{\prime}_{4,1}, \hat{R}^{\prime}_{4,2}$ \\ \hline
\end{tabular}
\end{center}
\vspace{-1em}
\caption{Updated storage across the databases in the server after one FSL training round when $D = 3$, $J = E = 2$ and $\delta = 0.5$.}
\label{E2_2}
\end{table}

Regarding the eavesdropper security constraint: Because the FSL-PSU phase has nothing to do with the updated full learning model $M^{\prime}_{[4]}$, we only need to consider the FSL-write phase. In terms of $M^{\prime}_{[4]}$, for all $j \in [4]$, it is easy to prove that all the storage data and communication data that can be obtained by each database $j$ is equivalent to the middle box in database $j$ in Table~\ref{E2_2}, i.e., $G_j(M^{\prime}_{\{1,3,4\}},\mathcal{R}^{\prime}_S)$ and $G_j(M_{\{2\}},\mathcal{R}_S)$ after cancelling the carefully-designed client-side common randomness. Therefore, the guarantee of eavesdropper security is directly inherited from the information leakage constraint of the RSRC scheme.

Regarding the database failure robustness: From the last $3$ symbols in $A_{W,2}^{\langle 4 \rangle,(4)}$, the replacement database can correctly decode the original storage for submodel $2$ in failed database $4$ due to the repair constraint of the RSRC scheme. For the updated submodels $1,3,4$, the desired storage can also be attained from the routing clients due to the satisfaction of the reliability constraint. The current replacement database is reflected in the database $4$ part in Table~\ref{E2_2}.
\end{example}

\section{General Distributed FSL Achievable Scheme}
Following the distributed FSL problem formulated in Section~\ref{Problem Formulation}, we provide our fully robust FSL achievable scheme for the general case in this section. Based on our previous work \cite{FSL-PSU}, the complete one-round FSL training is composed of four phases: client-side common randomness generation (FSL-CRG) phase that aims to distribute necessary client-side common randomness across all the selected clients, FSL-PSU phase that aims to privately determine the union of the submodel indices to be updated, FSL-write phase that aims to securely write the updated submodels in the union back to the databases, and server-side common randomness refresh (FSL-CRR) phase that aims to refresh the necessary server-side common randomness in preparation for the next round of the FSL process. In a practical implementation, the auxiliary FSL-CRG and FSL-CRR phases can be executed during the off-peak times because they are independent of the FSL-PSU and FSL-write phases. 

\subsection{FSL-CRG Phase} \label{FSL-CRG Phase}
All the databases in the server aim to collectively establish the desired client-side common randomness across the clients such that every client-side common randomness symbol is completely unknown to any set of $J$ colluding databases. The first type of client-side common randomness is a set of symbols $\{w_1,w_2,\dots,w_{\mathcal{L}}\}$ with a flexible set length $\mathcal{L}$. Within this set, each symbol is randomly and uniformly selected from $\mathbb{F}_q$ and the sum of these symbols is exactly $0$, i.e., $\sum_{i \in [\mathcal{L}]} w_i = 0$.  The second type of client-side common randomness is a random symbol $c$ that is uniform over the set $\mathbb{F}_q \backslash \{0\}$. 

For the first type, every $J\!+\!1$ databases can collaborate with each other to allocate the same set of client-side common randomness $\{w_1,w_2,\dots,w_{\mathcal{L}}\}$ across a small set of clients. To that end, each database in a database set of size $J\!+\!1$ first individually selects $\mathcal{L}\!-\!1$ random symbols from $\mathbb{F}_q$ under a uniform distribution, and then simply broadcasts them to $N$ distinct routing clients with indices $\theta_{[N]} = \{\theta_1, \theta_2, \dots, \theta_N\}$ randomly chosen from $N$ distinct client groups. After collecting all the random symbols from $J\!+\!1$ databases, these routing clients can just perform the element-wise summation over these $J\!+\!1$ random symbol sets of size $\mathcal{L}\!-\!1$ to obtain a new set of size $\mathcal{L}\!-\!1$. Subsequently, one more symbol is appended to the existing new set such that the set sum equals zero. At this point, this newly formed set can be used as $\{w_1,w_2,\dots,w_{\mathcal{L}}\}$ because the one-time pad encryption guarantees the privacy of this client-side common randomness set against any $J$ colluding databases. 

If the value of $\mathcal{L}$ is $N$, the symbols in this set can be used as $w^{(j)}_k$ or $w^{(j)}_{k,d_2}$ for the next two phases. However, if the value of $\mathcal{L}$ is $C$, for all $i \in [C\!-\!1] \backslash \theta_{[N]}$, each database in this database set also needs to send its $i$th random symbol to client $i$. For client $C$, if $C$ does not belong to $\theta_{[N]}$, these $J\!+\!1$ databases send all the generated random symbols to client $C$ like routing clients. Thus, client $C$ is able to calculate $w_C$. Now, the symbols in this set are ready to be used as $w^{\langle i \rangle}_k$ or $w^{\langle i \rangle}_{k,l}$ for the next two phases. As the required number of client-side common randomness sets is very large, the communication time in this phase can be further optimized by wisely constituting some database subset of size $J\!+\!1$ from a set of $N$ databases according to the actual situation. For example, if client $i$ has a high-bandwidth communication channel with a particular database and the bandwidth utilization ratio of this database with all the clients is currently low, client $i$ may select this database to participate in the client-side common randomness distribution. 

For the second type, we can also select a set of $J\!+\!1$ databases to participate. Each database first selects a random symbol from $\mathbb{F}_q \backslash \{0\}$ under a uniform distribution, and then simply broadcasts it to each selected client. Once the client receives all the random symbols from $J\!+\!1$ databases, it just calculates the product of these $J\!+\!1$ symbols within $\mathbb{F}_q$. This new product can be used as $c$ because the finite cyclic group $\mathbb{F}_q \backslash \{0\}$ under multiplication ensures the privacy of this client-side common randomness symbol against any $J$ colluding databases. This symbol $c$ can be used in the next FSL-PSU phase.

\subsection{FSL-PSU Phase} \label{FSL-PSU Phase}
The FSL-PSU phase in this work is similar to the private set union (FSL-PSU) phase in \cite[Sect.~5.2]{FSL-PSU} where nothing needs to be downloaded from the server to the clients at the beginning of FSL-PSU phase. After receiving required client-side common randomness in the last phase, each client uploads the index information of its desired submodels to the server in a private way. For any client $i$ that belongs to the client group $\mathcal{C}_j$, this client generates the following answer and sends it to database $j$ in the first step of FSL-PSU phase,
\begin{align} \label{AU1}
    A^{\langle i \rangle,(j)}_{U,1} = \{c(Y^{\langle i \rangle}_k \!+\! w^{\langle i \rangle}_k)\!: k \in [K]\}, \quad \forall i \in [C]
\end{align}
In the second step of FSL-PSU phase, for all $j \in [N]$, once database $j$ completes the collection of all the answers from its associated clients in $\mathcal{C}_j$, it produces a response via element-wise summation as follows where $\theta_j$ is the index of the randomly selected routing client in $\mathcal{C}_j$ and $\hat{R}_k$ is shared server-side common randomness,
\begin{align} \label{DU2}
    D^{\langle \theta_j \rangle,(j)}_{U,2} = \biggl\{c\sum_{i \in \mathcal{C}_j} (Y^{\langle i \rangle}_k \!+\! w^{\langle i \rangle}_k) \!+\! \hat{R}_k\!: k \in [K] \biggr\}, \quad \forall j \in [N] 
\end{align}
Then, this response is merely downloaded by its associated client $\theta_j$. After further processing this response, each client $\theta_j$ forwards the following answer to all the databases in the server,
\begin{align} \label{AU2}
    A^{\langle \theta_j \rangle,([N])}_{U,2} = \biggl\{c\sum_{i \in \mathcal{C}_j} (Y^{\langle i \rangle}_k \!+\! w^{\langle i \rangle}_k) \!+\! w^{(j)}_k \!+\! \hat{R}_k\!\!: k \in [K] \biggr\}, \quad \forall j \in [N] 
\end{align}

Each database $j$ receives the same $N$ answer sets. By summing these $N$ answer sets up element-wise, each database derives the value of the expression $c\sum_{i \in [C]} Y^{\langle i \rangle}_k$ for all $k \in [K]$ because server-side common randomness is known to the database and client-side common randomness is eliminated,
\begin{align}
    \sum_{j_0 \in [N]} A^{\langle \theta_{j_0} \rangle,(j)}_{U,2} &= \sum_{j_0 \in [N]} (c\sum_{i \in \mathcal{C}_{j_0}} (Y^{\langle i \rangle}_k \!+\! w^{\langle i \rangle}_k) \!+\! w^{(j_0)}_k \!+\! \hat{R}_k) \\
    &= c \sum_{j_0 \in [N]} \sum_{i \in \mathcal{C}_{j_0}} Y^{\langle i \rangle}_k \!+\! c \sum_{j_0 \in [N]} \sum_{i \in \mathcal{C}_{j_0}} w^{\langle i \rangle}_k \!+\! \sum_{j_0 \in [N]} w^{(j_0)}_k \!+\! \sum_{j_0 \in [N]} \hat{R}_k \\
    &= c \sum_{i \in [C]} Y^{\langle i \rangle}_k \!+\!  c \sum_{i \in [C]} w^{\langle i \rangle}_k \!+\! \sum_{j_0 \in [N]} w^{(j_0)}_k \!+\!\sum_{j_0 \in [N]} \hat{R}_k \\
    &= c \sum_{i \in [C]} Y^{\langle i \rangle}_k \!+\! N\hat{R}_k
\end{align}
For any arbitrary submodel index $k$, if at least one client wishes to update the submodel $k$, the value of the expression $c\sum_{i \in [C]} Y^{\langle i \rangle}_k$ cannot be zero. Otherwise, this value is equal to zero. Therefore, each individual database is able to determine the desired submodel union $\Gamma$ without any error by analyzing each submodel index one-by-one.

\subsection{FSL-Write Phase} \label{FSL-write Phase}
In the presence of an eavesdropper who can control any arbitrary $E$ databases and get at most the fraction $\delta$ of the up-to-date full learning model $M^\prime_{[K]}$, we can always find a RSRC-based FSL approach to satisfy the eavesdropper security constraint \eqref{eavesdropper security}. Here, the variables $\lambda$ and $\ell_\lambda$ in Section~\ref{General_RSRC} take the values $E$ and $\delta$, respectively. By permitting the dummy message symbols to fill the message matrix $\Omega$, if $0 \leq \delta \leq \frac{2E}{D+E+1}$, we can use the $\Omega_1$-based RSRC scheme and the $\Omega_2$-based RSRC scheme in a time-sharing manner to store the coded submodel information across the databases. If $\frac{2E}{D+E+1} \leq \delta \leq \frac{2DE - E(E-1)}{D(D+1)}$, we can use the $\Omega_2$-based RSRC scheme and the $\Omega_3$-based RSRC scheme in a time-sharing manner. Otherwise, if $\frac{2DE - E(E-1)}{D(D+1)} \leq \delta \leq 1$, we can just use the $\Omega_3$-based RSRC scheme.

Given the concrete form of $N \times D$ encoding matrix $\Psi$ in \eqref{General_Psi}, the concrete form of $D \times D$ message matrix $\Omega$ for the submodel $k$ is as follows without loss of generality,
\begin{align}
    \Omega = 
    \begin{bmatrix}
        X^k_{1,1} & X^k_{1,2} & X^k_{1,3} & \cdots & X^k_{1,D} \\
        X^k_{2,1} & X^k_{2,2} & X^k_{2,3} & \cdots & X^k_{2,D} \\
        X^k_{3,1} & X^k_{3,2} & X^k_{3,3} & \cdots & X^k_{3,D} \\
        \vdots & \vdots & \vdots & \ddots & \vdots  \\ 
        X^k_{D,1} & X^k_{D,2} & X^k_{D,3} & \cdots & X^k_{D,D}
    \end{bmatrix}
\end{align}
For the first $B$ message symbols in the submodel $k$, its corresponding storage in database $j$ consisting of $D$ symbols is in the following form,
\begin{align}
    \zeta^T_{j,k} = \biggl\{\sum_{d_1 \in [D]} \psi_j^{d_1-1} X^k_{d_1,d_2}\!: d_2 \in [D] \biggr\} , \quad \forall j \in [N],~\forall k \in [K]
\end{align}
where the symbol $X^k_{d_1,d_2}$ in $\Omega$ can be a message symbol in submodel $k$ or a randomness symbol depending on the realization of message matrix $\Omega$. 

When the FSL-PSU phase is finished, each database learns the desired submodel union $\Gamma$ privately. In this subsection, we focus on the update of the first $B$ symbols in each submodel whose index belongs to $\Gamma$. According to the statements in Section~\ref{General_RSRC}, all the remaining symbols can be fully updated through the repetition and time-sharing ideas. In the first step of the FSL-write phase,  each client needs to download $C_1$ coded symbols from $D$ working databases in order to recover the first $B$ message symbols in each submodel in $M_\Gamma$. For client $i$, let $\mathcal{N}_i$ denote the index set of $D$ databases it communicates with, then,
\begin{align} \label{DW1}
    D^{\langle i \rangle,(\mathcal{N}_i)}_{W,1} = \bigl\{ Z_{k,\alpha}\!: k \in \Gamma,~\alpha \in [C_1] \bigr\}, \quad \forall i \in [C]
\end{align}
where $Z_{k,\alpha}$ is picked from the set $\zeta^T_{j,k}$. Afterwards, this client generates the increments for its desired submodels whose index belongs to $\Gamma^{\langle i \rangle}$ when its local training is complete. Therefore, for any client $i$ in the client group $\mathcal{C}_j$, the answer transmitted from client $i$ to database $j$ is in the following form,
\begin{align} \label{AW1}
    A^{\langle i \rangle,(j)}_{W,1} = \bigl\{ \Delta^{\langle i \rangle}_{k,l} \!+\! w^{\langle i \rangle}_{k,l} \!: k \in \Gamma,~l \in [B]\bigr\}, \quad \forall i \in [C]
\end{align}
Like the second step of the FSL-PSU phase, each database $j$ now generates an answer to be downloaded by the randomly selected routing client $\theta_j$ from the client group $\mathcal{C}_j$ as follows,
\begin{align} \label{DW2}
    D^{\langle \theta_j \rangle,(j)}_{W,2} = \biggl\{ \sum_{i \in \mathcal{C}_j} (\Delta^{\langle i \rangle}_{k,l} \!+\! w^{\langle i \rangle}_{k,l}) \!+\! \hat{R}_{k,l} \!: k \in \Gamma,~l \in [B]\biggr\}, \quad \forall j \in [N]
\end{align}
Finally, each routing client $\theta_j$ needs to transfer different answers in different coded forms to all the available databases. Specifically, if $X^{k}_{d_1,d_2}$ is a submodel symbol, say $M_{k,l}$, then $\mathcal{X}^{k}_{d_1,d_2}$ is equal to $M_{k,l} \!+\! \sum_{i \in \mathcal{C}_1} (\Delta^{\langle i \rangle}_{k,l} \!+\! w^{\langle i \rangle}_{k,l}) \!+\! \hat{R}_{k,l}$ if $j = 1$ and is equal to $\sum_{i \in \mathcal{C}_j} (\Delta^{\langle i \rangle}_{k,l} \!+\! w^{\langle i \rangle}_{k,l}) \!+\! \hat{R}_{k,l}$ if $j = 2, \dots, N$. Otherwise, if $X^{k}_{d_1,d_2}$ is a randomness symbol, then $\mathcal{X}^{k}_{d_1,d_2}$ is a random value selected only by client $\theta_j$ under a uniform distribution from $\mathbb{F}_q$. Therefore, we have
\begin{align} \label{AW21}
    A^{\langle \theta_j \rangle,(j_0)}_{W,2} = \biggl\{\sum_{d_1 \in [D]} \psi_{j_0}^{d_1-1} \mathcal{X}^{k}_{d_1,d_2}\!: k \in \Gamma,~d_2 \in [D] \biggr\}, \quad \forall j \in [N],~\forall j_0 \in [N]
 \end{align}
Moreover, for some $d_2$, if all the symbols in the set $\{X^{k}_{d_1,d_2}\!: d_1 \in [D]\}$ are submodel symbols, a client-side common randomness needs to be appended, i.e., 
\begin{align} \label{AW22}
    A^{\langle \theta_j \rangle,(j_0)}_{W,2} = \biggl\{\sum_{d_1 \in [D]} \psi_{j_0}^{d_1-1} \mathcal{X}^{k}_{d_1,d_2}\!+\!w^{(j)}_{k,d_2}\!: k \in \Gamma,~d_2 \in [D] \biggr\}, \quad \forall j \in [N],~\forall j_0 \in [N]
\end{align}

\subsection{FSL-CRR Phase} \label{FSL-CRU phase}
In this phase, all the selected clients aim to jointly refresh the uncoded server-side common randomness $\hat{\mathcal{R}}_S$ coupled with  the submodels that are updated in this round of the FSL process. In other words, the server-side common randomness symbols $\{\hat{R}_k\!: k \in [K]\}$ and $\{\hat{R}_{k,l}\!: k \in \Gamma, l \in [L]\}$ need to be refreshed. As each server-side common randomness symbol should be unknown to any individual client, every pair of clients can collaborate with each other to complete this task. More specifically, both clients select a random symbol from $\mathbb{F}_q$ and then forward it to each available database. By simply adding these two received random symbols together, the databases can now share a new server-side common randomness symbol that can be used as refreshed $\hat{R}^\prime_k$ or $\hat{R}^\prime_{k,l}$ for the next round of the FSL process. Like the FSL-CRG phase, the required number of refreshed server-side common randomness symbols is also large. Hence, in a practical implementation, the communication time in this phase can also be optimized through partitioning the clients in a smart way.

\subsection{Basic Characteristics Verification} \label{Basic Characteristics Verification}
In this section, we verify the basic characteristics of a complete round of the FSL process including the four phases mentioned above.

\paragraph{Reliability:} According to the FSL-write phase in Section~\ref{FSL-write Phase}, for all $j \in [N]$, database $j$ selects a random client $\theta_j$ from its associated client group to forward the information in the second step of the FSL-write phase. Thus, each database $j$ receives the $N$ answer sets $\{A^{\langle \theta_1 \rangle,(j)}_{W,2}, A^{\langle \theta_2 \rangle,(j)}_{W,2}, \dots, A^{\langle \theta_N \rangle,(j)}_{W,2}\}$ from these $N$ randomly selected routing clients. By summing these $N$ answer sets up in an element-wise manner, for all $k \in \Gamma$ and all $d_2 \in [D]$, we have
\begin{align}
    \sum_{j_0 \in [N]} A^{\langle \theta_{j_0} \rangle,(j)}_{W,2} &= \sum_{j_0 \in [N]} \sum_{d_1 \in [D]} \psi_j^{d_1-1} \mathcal{X}^{k}_{d_1,d_2} = \sum_{d_1 \in [D]} \psi_j^{d_1-1} \sum_{j_0 \in [N]} \mathcal{X}^{k}_{d_1,d_2}
\end{align}
because the sum $\sum_{j \in [N]} w^{(j)}_{k,d_2}$ is equal to $0$. If $X^{k}_{d_1,d_2}$ is a submodel symbol, say $M_{k,l}$, 
\begin{align}
    \psi_j^{d_1-1} \sum_{j_0 \in [N]} \mathcal{X}^{k}_{d_1,d_2} &= \psi_j^{d_1-1} (M_{k,l} \!+\! \sum_{j_0 \in [N]} (\sum_{i \in \mathcal{C}_{j_0}} (\Delta^{\langle i \rangle}_{k,l} \!+\! w^{\langle i \rangle}_{k,l}) \!+\! \hat{R}_{k,l})) \\
    &= \psi_j^{d_1-1} (M_{k,l} \!+\! \sum_{i \in [C]} \Delta^{\langle i \rangle}_{k,l} \!+\! \sum_{i \in [C]} w^{\langle i \rangle}_{k,l} \!+\! N \hat{R}_{k,l}) \\
     &= \psi_j^{d_1-1} (M_{k,l} \!+\! \sum_{i \in [C]} \Delta^{\langle i \rangle}_{k,l} \!+\! N \hat{R}_{k,l}) \\
    &= \psi_j^{d_1-1} (M^\prime_{k,l} \!+\! N \hat{R}_{k,l}) 
\end{align}
At this point, each database $j$ is able to derive the value of $\psi_j^{d_1-1} M^\prime_{k,l}$ since the server-side common randomness $\hat{R}_{k,l}$ is known. Otherwise, if $X^{k}_{d_1,d_2}$ is a randomness symbol,
\begin{align}
    \psi_j^{d_1-1} \sum_{j_0 \in [N]} \mathcal{X}^{k}_{d_1,d_2} = \psi_j^{d_1-1} \mathcal{W}
\end{align}
where $\mathcal{W}$ is a random value that is completely unknown to the databases. For all $d_2 \in [D]$, by adding the available $\psi_j^{d_1-1} M^\prime_{k,l}$ and $\psi_j^{d_1-1} \mathcal{W}$ over $d_1 \in [D]$, the expected storage in database $j$ for the first $B$ symbols in the submodel $k$ is attained, which can be easily extended to all the submodel information in $M_{\Gamma}$ through repetition and time sharing. However, for all $k \in [K] \backslash \Gamma$, as these submodels are not updated at all, the corresponding storage in database $j$ is not changed. That means the required storage $G_j(M^{\prime}_{[K]},\mathcal{R}^{\prime}_S)$ is now achieved in all databases. Meanwhile, the additional plain server-side common randomness $\hat{\mathcal{R}}^{\prime}_S$ can be directly attained as we expect through the approach in Section~\ref{FSL-CRU phase}. Thus, the reliability constraint is satisfied.

\paragraph{Privacy:} For any set of databases with index set $\mathcal{J}$ that meets $|\mathcal{J}| \leq J$, the answer sets $\{A^{\langle \mathcal{C}_j \rangle,(j)}_{U,1}, A^{\langle \theta_{[N]} \rangle,(j)}_{U,2}, A^{\langle \mathcal{C}_j \rangle,(j)}_{W,1}, A^{\langle \theta_{[N]} \rangle,(j)}_{W,2}\!: j \in \mathcal{J}\}$ are received. With respect to the clients’ local data $\mathcal{D}_{[C]}$, it is easy to verify that these answer sets contain less information than the answer sets $\{A^{\langle \mathcal{C}_j \rangle,(j)}_{U,1},  A^{\langle \mathcal{C}_j \rangle,(j)}_{W,1}\!: j \in [N]\}$. Now, we can use the answer sets $\{A^{\langle \mathcal{C}_j \rangle,(j)}_{U,1}, A^{\langle \mathcal{C}_j \rangle,(j)}_{W,1}\!: j \in [N]\}$ as a base to analyze the privacy constraint. Note that each client-side common randomness symbol is completely unknown to this set of databases as it is generated by $J \!+\! 1$ clients collectively. In the answer sets $\{A^{\langle \mathcal{C}_j \rangle,(j)}_{U,1}: j \in [N]\}$, for all $k \in [K]$, the client-side common randomness symbol $w^{\langle i \rangle}_k$ can be used to protect the privacy of $Y^{\langle i \rangle}_k$, and the client-side common randomness symbol $c$ can be used to protect the privacy of $\sum_{i \in [C]} Y^{\langle i \rangle}_k$. As a result, these $|\mathcal{J}|$ databases can only learn the union $\Gamma$ and nothing beyond that.\footnote{The fact that both of $\sum_{i \in [C]} Y^{\langle i \rangle}_k$ and $\sum_{i \in [C]} \Delta^{\langle i \rangle}_{k,l}$ are always $0$ for $k \in [K] \backslash \Gamma$ is implied by the union $\Gamma$.} The concrete proof can be checked in the client's privacy proof in \cite[Subsect.~V.B]{MP-PSI_journal}. Then, we turn to the answer sets $\{A^{\langle \mathcal{C}_j \rangle,(j)}_{W,1}\!: j \in [N]\}$. As a reduced version, for all $k \in [K]$ and $l \in [L]$, the client-side common randomness symbol $w^{\langle i \rangle}_{k,l}$ can be used to protect the privacy of $\Delta^{\langle i \rangle}_{k,l}$. As a result, these $|\mathcal{J}|$ databases can only learn the the full increment sum $\Delta_{\Gamma}$ and nothing beyond that. Thus, any set of databases with cardinality less than or equal to $J$ cannot gain any additional information about $\mathcal{D}_{[C]}$ beyond $\Gamma$ and $\Delta_{\Gamma}$. Thus, the privacy constraint is satisfied.

\paragraph{Inter-Client Privacy:} Only one client from each client group is able to receive the information concerning the other clients' local data. For each client group $\mathcal{C}_j$, the routing client $\theta_j$ downloads $D^{\langle \theta_j \rangle,(j)}_{U,2}$ and $D^{\langle \theta_j \rangle,(j)}_{W,2}$ from database $j$. Due to the existence of the unknown server-side common randomness $\hat{R}_k$ and $\hat{R}_{k,l}$ in the downloads, client $\theta_j$ cannot learn any knowledge about the other clients' local data. Thus, the inter-client privacy constraint is satisfied.

\paragraph{Security Against the Eavesdropper:} To verify the security against the eavesdropper, we need to observe the storage information and the transmission information $\mathcal{M}_{\mathcal{E}}$ in any arbitrary $E$ databases whose index set is $\mathcal{E}$. Before an FSL training round begins, the storage information in each database $j$ is the coded submodel information $G_j(M_{[K]},\mathcal{R}_S)$ and additional server-side common randomness $\hat{\mathcal{R}}_S$, while the latter has nothing to do with the updated full learning model $M^{\prime}_{[K]}$. During this FSL training round, the transmission information that can be known by each database $j$ is $\{A^{\langle \mathcal{C}_j \rangle,(j)}_{U,1}, A^{\langle \theta_{[N]} \rangle,(j)}_{U,2}, A^{\langle \mathcal{C}_j \rangle,(j)}_{W,1}, A^{\langle \theta_{[N]}  \rangle,(j)}_{W,2}\}$. The first two answers only involve the information concerning clients' incidence vectors. The third answer is useless because of the coupled client-side common randomness symbol for each submodel increment symbol. Following the analysis of reliability constraint, the last answer is equivalent to $G_j(M^{\prime}_\Gamma,\mathcal{R}^{\prime}_S)$ due to the existence of carefully-designed client-side common randomness. Hence, for the submodels in $M_\Gamma$, we have the following relationship by using our specific RSRC-based approach in Section~\ref{FSL-write Phase},
\begin{align} 
    \frac{1}{|\Gamma|L} I(M^{\prime}_{\Gamma};\mathcal{M}_{\mathcal{E}}) \leq \delta, \quad \forall \mathcal{E} \subseteq [N], ~ |\mathcal{E}| = E
\end{align}
For the submodels in $M_{[K] \backslash \Gamma}$, the identity $M^\prime_{[K] \backslash \Gamma} = M_{[K] \backslash \Gamma}$ is true. Likewise, from the coded submodel information $G_j(M_{[K] \backslash \Gamma},\mathcal{R}_S)$, we have
\begin{align} 
    \frac{1}{(K\!-\!|\Gamma|)L} I(M^{\prime}_{[K] \backslash \Gamma};\mathcal{M}_{\mathcal{E}}) \leq \delta, \quad \forall \mathcal{E} \subseteq [N], ~ |\mathcal{E}| = E
\end{align}
Because $M^{\prime}_{[K]}$ and $M^{\prime}_{[K] \backslash \Gamma}$ are always  independent, we are able to derive the following outcome, which is sufficient to guarantee the eavesdropper security,
\begin{align}
    I(M^{\prime}_{[K]};\mathcal{M}_{\mathcal{E}}) =  I(M^{\prime}_{\Gamma};\mathcal{M}_{\mathcal{E}}) \!+\! I(M^{\prime}_{[K] \backslash \Gamma};\mathcal{M}_{\mathcal{E}}) \leq |\Gamma|L \cdot \delta \!+\! (K\!-\!|\Gamma|)L \cdot \delta = KL \cdot \delta
\end{align}
Thus, the security constraint against an eavesdropper is satisfied.

\subsection{Full Robustness Verification} \label{Full Robustness Verification}
In this subsection, we analyze the robustness of our proposed achievable scheme in the face of all kinds of non-ideal situations. If multiple such incidents happen simultaneously, we can simply incorporate the idea for each incident into the adjusted scheme one-by-one.

\paragraph{Client Drop-Outs Robustness:}
In this case, our expectation is that the training process proceeds normally only relying on the data from the active clients. Without loss of generality, for each client group $\mathcal{C}_j$, we assume that a subset of clients with indices $\Tilde{\mathcal{C}}_j$ drop-out. In the FSL-PSU phase, for all $j \in [N]$, the response $D^{\langle \theta_j \rangle,(j)}_{U,2}$ to be downloaded by client $\theta_j$ becomes 
\begin{align}
      D^{\langle \theta_j \rangle,(j)}_{U,2} = \biggl\{c\sum_{i \in \mathcal{C}_j \backslash \Tilde{\mathcal{C}}_j} (Y^{\langle i \rangle}_k \!+\! w^{\langle i \rangle}_k) \!+\! \hat{R}_k\!: k \in [K] \biggr\}, \quad \forall j \in [N] 
\end{align}
With the knowledge of the index set $\Tilde{\mathcal{C}}_j$ corresponding to the out-of-operation clients, each routing client $\theta_j$ can adjust the answer by additionally appending the value of $c\sum_{i \in \Tilde{\mathcal{C}}_j} w^{\langle i \rangle}_k$ corresponding to the missing client-side common randomness symbols for all $k \in [K]$. Thus, the answer to be forwarded to all the available databases becomes
\begin{align}
    A^{\langle \theta_j \rangle,([N])}_{U,2} = \biggl\{c\sum_{i \in \mathcal{C}_j \backslash \Tilde{\mathcal{C}}_j} Y^{\langle i \rangle}_k \!+\! c\sum_{i \in \mathcal{C}_j} w^{\langle i \rangle}_k \!+\! w^{(j)}_k \!+\! \hat{R}_k\!\!: k \in [K] \biggr\}, \quad \forall j \in [N] 
\end{align}
In this way, each database is still able to decode the union $\cup_{j \in [N]} \Gamma^{\langle \mathcal{C}_j \backslash \Tilde{\mathcal{C}}_j \rangle}$ for all the active clients from the element-wise summation results $\{c\sum_{i \in \cup_{j \in [N]} (\mathcal{C}_j \backslash \Tilde{\mathcal{C}}_j)} Y^{\langle i \rangle}_k \!: k \in [K]\}$.

In the FSL-write phase, for all $j \in [N]$, the response $D^{\langle \theta_j \rangle,(j)}_{W,2}$ to be downloaded becomes 
\begin{align} 
    D^{\langle \theta_j \rangle,(j)}_{W,2} = \biggl\{ \sum_{i \in \mathcal{C}_j \backslash \Tilde{\mathcal{C}}_j} (\Delta^{\langle i \rangle}_{k,l} \!+\! w^{\langle i \rangle}_{k,l}) \!+\! \hat{R}_{k,l} \!: k \in \Gamma,~l \in [B]\biggr\}, \quad \forall j \in [N]
\end{align}
Using the reliability constraint analysis in the last subsection as reference, if $X^{k}_{d_1,d_2}$ is a submodel symbol, say $M_{k,l}$, after eliminating the server-side common randomness, for each database $j$, we have
\begin{align}
    \psi_j^{d_1-1} \sum_{j_0 \in [N]} \mathcal{X}^{k}_{d_1,d_2} &= \psi_j^{d_1-1} (M_{k,l} \!+\! \sum_{j_0 \in [N]}   \sum_{i \in \mathcal{C}_{j_0} \backslash \Tilde{\mathcal{C}}_{j_0}} \!\!\!(\Delta^{\langle i \rangle}_{k,l} \!+\! w^{\langle i \rangle}_{k,l}) ) \\
    &= \psi_j^{d_1-1} (M_{k,l} \!+\! \sum_{i \in \cup_{j_0 \in [N]} (\mathcal{C}_{j_0} \backslash \Tilde{\mathcal{C}}_{j_0})} \!\!\!\!\!\!\!\! \Delta^{\langle i \rangle}_{k,l} \!+\! \sum_{i \in \cup_{j_0 \in [N]} (\mathcal{C}_{j_0} \backslash \Tilde{\mathcal{C}}_{j_0})} \!\!\!\!\!\!\!\! w^{\langle i \rangle}_{k,l} ) 
\end{align}
Otherwise, if $X^{k}_{d_1,d_2}$ is a randomness symbol, $\psi_j^{d_1-1} \sum_{j_0 \in [N]} \mathcal{X}^{k}_{d_1,d_2}$ is not changed. Hence, for all $k \in \Gamma$ and all $d_2 \in [D]$, each routing client $\theta_j$ first selects an empty value $\beta$, i.e., $\beta = 0$. Then, for each $d_1$ in the set $[D]$, as long as $X^{k}_{d_1,d_2}$ is some submodel symbol $M_{k,l}$, we add the value of $\psi_j^{d_1-1} \sum_{i \in \cup_{j_0 \in [N]} \Tilde{\mathcal{C}}_{j_0}} w^{\langle i \rangle}_{k,l}$ to the current $\beta$. After finishing the loop over $d_1$, each working database $j$ asks for the value of $\beta$ along with $A^{\langle \theta_j \rangle,(j_0)}_{W,2}$ from client $\theta_j$. Although we only consider the first $B$ symbols of each submodel here, this idea can be extended to all the updates in $M_\Gamma$ through repetition and time-sharing. In this way, each submodel in the union $\Gamma$ can be updated as desired through the training data from the active clients $\cup_{j \in [N]} \mathcal{C}_j \backslash \Tilde{\mathcal{C}}_j$.

\paragraph{Client Late-Arrivals Robustness:} If the answers generated by some clients in the first step of the FSL-PSU or FSL-write phases arrive at the server late, which is different from the wrong judgement made by the databases that these clients have dropped-out, the privacy constraint \eqref{privacy} is still satisfied, i.e., these late answers do not disclose any extra information about the local data possessed by these late-arriving clients. The fact that the reliability constraint is satisfied is inherited from the one in client drop-outs robustness. For each client group $\mathcal{C}_j$, we assume that a subset of clients with indices $\bar{\mathcal{C}}_j$ cause the answer late-arrivals. In the FSL-PSU phase, for all $j \in  [N]$, due to the wrong judgement, the broadcasting answer $A^{\langle \theta_j \rangle,([N])}_{U,2}$ takes the following form according to the analysis in client drop-outs robustness,
\begin{align} \label{late-arrival AU2}
    A^{\langle \theta_j \rangle,([N])}_{U,2} = \biggl\{c\sum_{i \in \mathcal{C}_j \backslash \bar{\mathcal{C}}_j} (Y^{\langle i \rangle}_k \!+\! w^{\langle i \rangle}_k) \!+\! w^{(j)}_k \!+\! \hat{R}_k\!\!: k \in [K] \biggr\}, \quad \forall j \in [N] 
\end{align}
It is easy to see that no information about the incidence vectors $Y^{\langle \bar{\mathcal{C}}_j \rangle}$ can be extracted by database $j$ from the late answers $ A^{\langle \bar{\mathcal{C}}_j \rangle,(j)}_{U,1}$ and the answers $A^{\langle \theta_{[N]} \rangle,(j)}_{U,2}$ in the form of  \eqref{late-arrival AU2} because of the extra client-side common randomness $w_k^{(j)}$. In the FSL-write phase, the privacy constraint can be guaranteed in the same way because of the randomness selected by the routing clients or the extra client-side common randomness $w^{(j)}_{k,d_2}$. This conclusion is still true for any set of databases with size smaller than or equal to $J$ as the randomness truly used here is completely unknown to any set of $J$ colluding databases.

\paragraph{Database Drop-Outs Robustness:} Under this situation, we rely on the remaining working databases to complete the training task normally. For instance, say database $f$ drops-out and cannot provide any helpful response to the clients in this FSL round. In the FSL-PSU phase, each working database $j \in [N] \backslash f$ can still receive the answers $A^{\langle \mathcal{C}_j \rangle,(j)}_{U,1}$ and $A^{\langle \theta_{[N] \backslash f} \rangle,(j)}_{U,2}$, but cannot receive any answer from the routing client $\theta_f$ in the client group $\mathcal{C}_f$. Thus, our adjusted aim is to derive the desired submodel union $\Gamma \backslash \Gamma^{\langle \mathcal{C}_f \rangle}$ instead. For all $k \in [K]$, in order to obtain the needed $c\sum_{i \in [C] \backslash \mathcal{C}_f} Y^{\langle i \rangle}_k$ from $A^{\langle \theta_{[N] \backslash f} \rangle,(j)}_{U,2}$, each working database $j$ can ask for the value of $c\sum_{i \in \mathcal{C}_f} (w^{\langle i \rangle}_k) \!+\! w^{(f)}_k$ from the routing client $\theta_j$ through one more communication step. After eliminating the known server-side common randomness $\hat{R}_k$, we have the following identity for all $k \in [K]$ as all the client-side common randomness can be cancelled,
\begin{align}
    \sum_{j_0 \in [N] \backslash f} (c\sum_{i \in \mathcal{C}_{j_0}} (Y^{\langle i \rangle}_k \!+\! w^{\langle i \rangle}_k) \!+\! w^{(j)}_k) \!+\! c\sum_{i \in \mathcal{C}_f} (w^{\langle i \rangle}_k) \!+\! w^{(f)}_k = c \!\! \sum_{i \in [C] \backslash \mathcal{C}_f} \!\! Y^{\langle i \rangle}_k
\end{align}
In the FSL-write phase, likewise, each working database $j \in [N] \backslash f$ can obtain the answers $A^{\langle \mathcal{C}_j \rangle,(j)}_{W,1}$ and $A^{\langle \theta_{[N] \backslash f} \rangle,(j)}_{W,2}$ without any answer from the routing client $\theta_f$ for the client group $\mathcal{C}_f$. Following the analysis of the reliability constraint in the last subsection, if $X^{k}_{d_1,d_2}$ is a submodel symbol, say $M_{k,l}$, we have the following result in each database $j$ when the known server-side common randomness is eliminated,
\begin{align}
    \psi_j^{d_1-1} \sum_{j_0 \in [N] \backslash f} \mathcal{X}^{k}_{d_1,d_2} &= \psi_j^{d_1-1} (M_{k,l} + \sum_{j_0 \in [N] \backslash f} \sum_{i \in \mathcal{C}_{j_0}} (\Delta^{\langle i \rangle}_{k,l} \!+\! w^{\langle i \rangle}_{k,l})) \\
    &= \psi_j^{d_1-1} (M_{k,l} + \sum_{i \in [C] \backslash \mathcal{C}_f} \!\!\Delta^{\langle i \rangle}_{k,l} + \sum_{i \in [C] \backslash \mathcal{C}_f} \!\! w^{\langle i \rangle}_{k,l})
\end{align}
Otherwise, if $X^{k}_{d_1,d_2}$ is a randomness symbol, and we still have
\begin{align}
    \psi_j^{d_1-1} \sum_{j_0 \in [N] \backslash f}  \mathcal{X}^{k}_{d_1,d_2} = \psi_j^{d_1-1} \mathcal{W}
\end{align}
Hence, for all $k \in \Gamma$ and all $d_2 \in [D]$, each routing client $\theta_j$ first selects an empty value $\beta = 0$. Then, for each $d_1$ in the set $[D]$, provided $X^{k}_{d_1,d_2}$ is some submodel symbol $M_{k,l}$, the value of $\psi_j^{d_1-1} \sum_{i \in \mathcal{C}_f} w^{\langle i \rangle}_{k,l}$ is added to the current $\beta$. When the loop is finished, each working database $j$ asks for the value of $\beta$ along with $A^{\langle \theta_j \rangle,(j_0)}_{W,2}$ from client $\theta_j$. Again, this idea works for all the submodel updates in the union $\Gamma \backslash \Gamma^{\langle \mathcal{C}_f \rangle}$ via repetition and time-sharing. That means even if database $f$ drops-out, the FSL process can proceed normally without collecting the updates from the clients in the client group $\mathcal{C}_f$. This remedy for single database drop-out can be extended to the case of multiple database drop-outs by replacing index $f$ with an index set.

\paragraph{Database Failure Robustness:} Once one of the available databases at the server side fails permanently rather than drops-out temporarily, our solution is to construct a replacement database such that the FSL protocol configured in the beginning is not affected by the database failure at all. If database $f$ fails, for the submodels in $M_\Gamma$, we can rely on $N \!-\! 1$ routing clients with indices $\theta_{[N] \backslash f}$ to transmit the submodel update information that is an encoding function $G_f$ of the latest submodels $M^\prime_\Gamma$ and their coupled refreshed server-side common randomness $\mathcal{R}^\prime_S$ to the replacement database. The concrete realization is inherited from the above-mentioned remedy for single database drop-out. Meanwhile, for the other submodels in $M_{[K] \backslash \Gamma}$, the clients can be used to forward the database $f$ repair information that is an encoding function $G_f$ of the previous submodels $M_{[K] \backslash \Gamma}$ and their coupled previous server-side common randomness $\mathcal{R}_S$ to the replacement database according to the database repair part in the proof of Theorem~\ref{RSRCtheorem}. Since the size of this repair information is generally large, a large number of clients can work in parallel, i.e., each client forwards a small part of the overall repair information to reduce the communication time. In addition, to make the plain server-side common randomness $\hat{\mathcal{R}}_S$ in the replacement database consistent with the other databases, all the plain server-side common randomness $\hat{\mathcal{R}}_S$ is refreshed through the approach in FSL-CRR phase whether it is coupled with submodels in $M_\Gamma$ or in $M_{[K] \backslash \Gamma}$. If multiple databases fail, the failed databases can be repaired one-by-one for each FSL round.  

\paragraph{Active Adversary Robustness:}
If there is an active adversary taking control of any arbitrary $A$ databases, the responses received by the clients from these $A$ corrupted databases will not be reliable any more. The core idea of our solution is to force the clients to download more responses than usual and then extract required information from different databases. In the original FSL-PSU phase, for all $j \in [N]$, each routing client $\theta_j$ needs to download $D^{\langle \theta_j \rangle,(j)}_{U,2}$ from database $j$. Now, each client $i$ in the client group $\mathcal{C}_j$ sends the answer in the form of \eqref{AU1} to $2A\!+\!1$ working databases in an efficient manner. Afterwards, these $2A\!+\!1$ working databases individually transmit the response in the form of \eqref{DU2} back to client $\theta_j$. Hence, client $\theta_j$ is able to decode the correct $D^{\langle \theta_j \rangle,(j)}_{U,2}$ according to the error correcting property of $[2A\!+\!1,1,2A\!+\!1]$ repetition code. In the original FSL-write phase, for all $i \in [C]$, each database $i$ needs to download $D^{\langle i \rangle,(\mathcal{N}_i)}_{W,1}$ from $D$ databases in order to recover the desired submodels $M_\Gamma$. If the maximal number of symbols downloaded from a database is $\eta$, then $D\!+\!2A$ working databases individually transmit $\eta$ symbols with the same positions to client $i$. Hence, client $i$ is able to decode $M_\Gamma$ without any error according to the error correcting property of $[2A\!+\!D,D,2A\!+\!1]$ Reed-Solomon code. Then, for all $j \in [N]$, each routing client $\theta_j$ needs to download $D^{\langle \theta_j \rangle,(j)}_{W,2}$ from database $j$. By utilizing $[2A\!+\!1,1,2A\!+\!1]$ repetition code again, the process is the same as the above-mentioned one for downloading $D^{\langle \theta_j \rangle,(j)}_{U,2}$. 

For the remaining auxiliary phases, in the FSL-CRR phase, there is no need for the clients to download any information from the databases. Therefore, the existence of active adversary has no influence on this phase. However, in the original FSL-CRG phase, every $J\!+\!1$ databases are collaborating with each other to distribute client-side common randomness symbols. As a malicious database can send the same symbol with different values to different clients in the broadcasting process, the procedure in the current FSL-CRG phase will be a bit more complicated. Now, each database will not broadcast its randomly selected symbol any more. Instead, the server-side partial common randomness is broadcast to the clients. To be more concrete, for the first type of client-side common randomness, by adjusting the FSL-CRR phase such that every pair of clients are forwarding the data to exactly $2A\!+\!1$ databases, a server-side partial common randomness symbol $\mathcal{R}_{S,1}$ can be owned by these $2A\!+\!1$ databases. Afterwards, these $2A\!+\!1$ databases broadcast this symbol to $N$ routing clients. Thus, each routing client can decode the desired $\mathcal{R}_{S,1}$ reliably through $[2A\!+\!1,1,2A\!+\!1]$ repetition code. Likewise, another $2A\!+\!1$ databases are employed to broadcast another server-side partial common randomness symbol $\mathcal{R}_{S,2}$ to all the routing clients. Following this way, each routing client can obtain the same set of server-side partial common randomness symbols $\{\mathcal{R}_{S,1},\mathcal{R}_{S,2},\dots,\mathcal{R}_{S,J\!+\!1}\}$ from $(J\!+\!1)(2A\!+\!1)$ different databases. By adding these $(J\!+\!1)$ symbols up, $N$ routing clients can share a randomness symbol that is unknown to any $J$ colluding databases. Therefore, the FSL-CRG phase can proceed as we desire even in the presence of an active adversary. By repeating this step $\mathcal{L}\!-\!1$ times, the required set $\{w_1,w_2,\dots,w_{\mathcal{L}\!-\!1}\}$ can be shared among these $N$ routing databases. By incorporating this idea into the original FSL-CRG phase, the necessary client-side common randomness can also be obtained by the other clients. The process of distributing the client-side common randomness symbol $c$ is very similar where all the selected clients are equally treated.

\subsection{Performance Evaluation} \label{Performance Evaluation}
We consider the performance of our proposed achievable scheme in this section. As we formulated in Section~\ref{Problem Formulation}, the evaluation consists of communication cost and storage cost.

\paragraph{Communication Cost:} First, we consider the basic  scheme without any adjustment for additional robustness. In the FSL-CRG phase, for each client-side common randomness symbol in the form of $w^{(j)}_k$ or $w^{(j)}_{k,d_2}$, the communication cost is $(J\!+\!1)N$. For each client-side common randomness symbol in the form of $w^{\langle i \rangle}_k$ or $w^{\langle i \rangle}_{k,l}$, the communication cost is at most $(J\!+\!1)(N\!+\!2)$. The communication cost for the client-side common randomness symbol $c$ is $(J\!+\!1)C$, which can be neglected as it is distributed only once. Therefore, the total communication cost in this phase is at most $(J\!+\!1)N(N\!-\!1)(K\!+\!D|\Gamma|L) \!+\! (J\!+\!1)(N\!+\!2)(C\!-\!1)(K\!+\!|\Gamma|L)$, which is approximately $\mathcal{O}(C(K\!+\!|\Gamma|L))$ since the values of $D,J,N$ are small compared with the values of $K,L$. In the FSL-PSU phase, the communication cost is $(C\!+\!N\!+\!N^2)K$, which is approximately $\mathcal{O}(CK)$. In the FSL-write phase, the communication cost for the recovery of the desired submodels $M_\Gamma$ across all the clients is $C\cdot \frac{C_1}{B}|\Gamma|L$ where $ \frac{C_1}{B}$ comes from \eqref{C1cost} in Theorem~\ref{RSRCtheorem}. For the remaining transmission, the communication cost is at most $(C \!+\! N) |\Gamma|L \!+\! N^2 D |\Gamma|L$. Therefore, the total communication cost in this phase is approximately $\mathcal{O}(C|\Gamma|L))$. In the FSL-CRR phase, for each server-side common randomness symbol whether it is in the form of $\hat{\mathcal{R}}_k$ or $\hat{R}_{k,l}$, the communication cost is always $2N$. Therefore, the total communication cost in this phase is $2N(K\!+\!|\Gamma|L)$, which is approximately $\mathcal{O}(K\!+\!|\Gamma|L)$. By summing the communication cost results from all four phases, the overall communication cost in our one-round distributed FSL achievable scheme is approximately $\mathcal{O}(C(K\!+\!|\Gamma|L))$.  

When the situations of client drop-outs, client late-arrivals or database drop-outs happen, it is easy to show that the overall communication cost will not be influenced significantly. Further, in the presence of an active adversary, as the value of $A$ is also small, the order of the overall communication cost will not change. It will still be $\mathcal{O}(C(K\!+\!|\Gamma|L))$. However, once the situation of database failure happens, for each failed database, the additional communication cost of transmitting the database repair information concerning $M_{[K] \backslash \Gamma}$ to the replacement database is $2 \frac{C_2}{B} (K\!-\!|\Gamma|)L$, where the coefficient $2$ comes from the fact that clients are used to route the information between databases and $\frac{C_2}{B}$ comes from \eqref{C2cost} in Theorem~\ref{RSRCtheorem}. In addition, the additional communication cost of refreshing server-side common randomness symbols $\{\hat{R}_{k,l}\!: k \in [K] \backslash \Gamma, l \in [L]\}$ is $2N(K\!-\!|\Gamma|)L$. Therefore, as the value of $|\Gamma|$ is generally much smaller than the value of $K$, the additional communication cost for the sake of database failure robustness is approximately $\mathcal{O}(KL)$. 

\paragraph{Storage Cost:} In order to ensure the security against eavesdropper \eqref{eavesdropper security} in the presence of a passive eavesdropper, without considering the server-side common randomness, the storage cost in each database is $ \frac{S}{B}KL$ where $ \frac{S}{B}$ comes from \eqref{Scost} in Theorem~\ref{RSRCtheorem}. In addition, the storage cost in each database for the plain server-side common randomness is $KL\!+\!L$. Therefore, the overall communication cost is $N(SKL\!+\!KL\!+\!L)$ which is approximately $\mathcal{O}(KL)$.

\section{Conclusion and Discussion}
In this paper, we proposed a new RSRC-based distributed FSL scheme that extends our previous two-database FSL scheme in \cite{FSL-PSU}. This new scheme has higher resilience than our previous scheme, while having the same order-wise communication cost and storage cost. More specifically, this new scheme is now fully robust against passive eavesdroppers, active adversaries, database failures, database drop-outs, client drop-outs and client late-arrivals. 

In this paper, we mainly considered the privacy and security from the perspective of the databases. In reality, it is also possible that the clients collude with each other or with the server \cite{IT_SecAgg, IT_ColludingUser}. Furthermore, the clients can also be malicious and return arbitrarily erroneous answers to the server. These are interesting research directions. MDS coding or Lagrange coding in \cite{LCC} across the clients can be utilized for this purpose. 

Regarding our RSRC technique that aims to reduce the reconstruction communication cost, repair communication cost, and storage cost simultaneously by allowing information leakage, we did not investigate a converse proof in this paper. It is quite likely that there exists a better coding scheme that outperforms our RSRC scheme in terms of part or all of the evaluation metrics. Another non-trivial point is how to group the databases and clients to improve the communication efficiency in the two auxiliary phases, i.e., FSL-CRG phase and FSL-CRR phase, in a practical implementation. These are basically optimization problems and would be interesting to explore in an actual distributed FSL configuration.

\bibliographystyle{unsrt}
\bibliography{Journal2023}

\begin{thebibliography}{10}

\bibitem{FL}
B.~McMahan and D.~Ramage.
\newblock Federated learning: Collaborative machine learning without
  centralized training data.
\newblock Available at
  https://ai.googleblog.com/2017/04/federated-learning-collaborative.html.

\bibitem{FL_Yangconcept}
Q.~Yang, Y.~Liu, T.~Chen, and Y.~Tong.
\newblock Federated machine learning: Concept and applications.
\newblock {\em ACM Trans. Intelligent Systems and Tech.}, 10(2):1--19, March
  2019.

\bibitem{SecAgg}
K.~Bonawitz, V.~Ivanov, et~al.
\newblock Practical secure aggregation for privacy preserving machine learning.
\newblock In {\em ACM SIGSAC Conf. Comput. Commun. Security}, page 1175–1191,
  2017.

\bibitem{FSL}
C.~Niu, F.~Wu, S.~Tang, L.~Hua, R.~Jia, C.~Lv, Z.~Wu, and G.~Chen.
\newblock Billion-scale federated learning on mobile clients: A submodel design
  with tunable privacy.
\newblock In {\em ACM MobiCom}, pages 1--14, 2020.

\bibitem{PIR_ORI}
B.~Chor, E.~Kushilevitz, O.~Goldreich, and M.~Sudan.
\newblock Private information retrieval.
\newblock {\em Journal of the ACM}, 45(6):965--981, November 1998.

\bibitem{PIR}
H.~Sun and S.~A. Jafar.
\newblock The capacity of private information retrieval.
\newblock {\em IEEE Trans. on Info. Theory}, 63(7):4075--4088, July 2017.

\bibitem{PIR_coded}
K.~Banawan and S.~Ulukus.
\newblock The capacity of private information retrieval from coded databases.
\newblock {\em IEEE Trans. on Info. Theory}, 64(3):1945--1956, March 2018.

\bibitem{ORAM}
O.~Goldreich and R.~Ostrovsky.
\newblock Software protection and simulation on oblivious {RAM}s.
\newblock {\em Journal of the ACM}, 43(3):431–473, May 1996.

\bibitem{DORAM}
S.~Lu and R.~Ostrovsky.
\newblock Distributed oblivious {RAM} for secure two-party computation.
\newblock In {\em Theory of Cryptography Conference}, page 377–396, March
  2013.

\bibitem{Path_ORAM}
E.~Stefanov, M.~Van Dijk, et~al.
\newblock Path {ORAM}: An extremely simple oblivious {RAM} protocol.
\newblock {\em Journal of the ACM}, 65(4):1–26, April 2018.

\bibitem{XSTPIR}
Z.~Jia, H.~Sun, and S.~A. Jafar.
\newblock Cross subspace alignment and the asymptotic capacity of ${X}$-secure
  ${T}$-private information retrieval.
\newblock {\em IEEE Trans. on Info. Theory}, 65(9):5783--5798, September 2019.

\bibitem{XSTPIR_MDS}
Z.~Jia and S.~A. Jafar.
\newblock {$X$}-secure {$T$}-private information retrieval from {MDS} coded
  storage with {B}yzantine and unresponsive servers.
\newblock {\em IEEE Trans. on Info. Theory}, 66(12):7427--7438, December 2020.

\bibitem{XSTFSL}
Z.~Jia and S.~A. Jafar.
\newblock {$X$}-secure {$T$}-private federated submodel learning with elastic
  dropout resilience.
\newblock {\em IEEE Trans. on Info. Theory}, 68(8):5418--5439, August 2022.

\bibitem{Sajani_FSL1}
S.~Vithana and S.~Ulukus.
\newblock Efficient private federated submodel learning.
\newblock In {\em IEEE ICC}, pages 3394--3399, May 2022.

\bibitem{Sajani_FSL_Trans}
S.~Vithana and S.~Ulukus.
\newblock Private read update write ({PRUW}) in federated submodel learning
  ({FSL}): Communication efficient schemes with and without sparsification.
\newblock Available at arXiv:2209.04421.

\bibitem{Sajani_FL_Trans}
S.~Vithana and S.~Ulukus.
\newblock Private read-update-write with controllable information leakage for
  storage-efficient federated learning with top $r$ sparsification.
\newblock Available at arXiv:2303.04123.

\bibitem{SecAgg+}
J.~Bell, K.~A. Bonawitz, A.~Gascón, T.~Lepoint, and M.~Raykova.
\newblock Secure single-server aggregation with (poly)logarithmic overhead.
\newblock In {\em Cryptology ePrint Archive}, 2020.

\bibitem{TurboAgg}
J.~So, B.~Guler, and A.~S. Avestimehr.
\newblock Turbo-aggregate: Breaking the quadratic aggregation barrier in secure
  federated learning.
\newblock In {\em Cryptology ePrint Archive}, 2020.

\bibitem{FastSecAgg}
S.~Kadhe and K.~Ramchandran N.~Rajaraman~abd O.~O.~Koyluoglu.
\newblock Fastsecagg: Scalable secure aggregation for privacy-preserving
  federated learning.
\newblock Available at arXiv:2009.11248.

\bibitem{IT_SecAgg}
Y.~Zhao and H.~Sun.
\newblock Information theoretic secure aggregation with user dropouts.
\newblock {\em IEEE Trans. on Info. Theory}, 68(11):7471--7484, November 2022.

\bibitem{IT_SecAgg_UGKey}
K.~Wan, H.~Sun, M.~Ji, and G.~Caire.
\newblock Information theoretic secure aggregation with uncoded groupwise keys.
\newblock Available at arXiv:2204.11364.

\bibitem{IT_SecAgg_Region}
Y.~Zhao and H.~Sun.
\newblock Secure summation: Capacity region, groupwise key, and feasibility.
\newblock Available at arXiv:2205.08458.

\bibitem{LightSecAgg}
J.~So, C.~He, et~al.
\newblock {LightSecAgg}: A lightweight and versatile design for secure
  aggregation in federated learning.
\newblock In {\em MLSys}, pages 694--720, 2022.

\bibitem{FSL-PSU}
Z.~Wang and S.~Ulukus.
\newblock Private federated submodel learning via private set union.
\newblock Available at arXiv:2301.07686.

\bibitem{Prio}
H.~Corrigan-Gibbs and D.~Boneh.
\newblock Prio: Private, robust, and scalable computation of aggregate
  statistics.
\newblock In {\em USENIX Conf. on Netw. Sys. Design and Impl.}, page 259–282,
  2017.

\bibitem{PIR_Extensions}
S.~Vithana, Z.~Wang, and S.~Ulukus.
\newblock Private information retrieval and its applications: An introduction,
  open problems, future directions.
\newblock Available at arXiv:2304.14397.

\bibitem{EPIR}
Q.~Wang, H.~Sun, and M.~Skoglund.
\newblock The capacity of private information retrieval with eavesdroppers.
\newblock {\em IEEE Trans. on Info. Theory}, 65(5):3198--3214, May 2019.

\bibitem{Erasure_codes}
K.~V. Rashmi, N.~B. Shah, D.~Gu, H.~Kuang, D.~Borthakur, and K.~Ramchandran.
\newblock A solution to the network challenges of data recovery in
  erasure-coded distributed storage systems: A study on the {F}acebook
  warehouse cluster.
\newblock In {\em USENIX Conf. on Hot Top. in Stor.e and File Sys.}, June 2013.

\bibitem{Secure_RC}
K.~V. Rashmi, N.~B. Shah, K.~Ramchandran, and P.~V. Kumar.
\newblock Information-theoretically secure erasure codes for distributed
  storage.
\newblock {\em IEEE Trans. on Info. Theory}, 64(3):1621--1646, March 2018.

\bibitem{SS_IT}
E.~D. Karnin, J.~W. Greene, and M.~E. Hellman.
\newblock On secret sharing systems.
\newblock {\em IEEE Trans. on Info. Theory}, 29(1):35--41, January 1983.

\bibitem{Ramp_SS}
H.~Yamamoto.
\newblock Secret sharing system using $(k, l, n)$ threshold scheme.
\newblock {\em Electronics and Communications in Japan, Part 1}, 69(9):46--54,
  1986.

\bibitem{BPIR}
K.~Banawan and S.~Ulukus.
\newblock The capacity of private information retrieval from {B}yzantine and
  colluding databases.
\newblock {\em IEEE Trans. on Info. Theory}, 65(2):1206--1219, February 2019.

\bibitem{Speedup_ML}
K.~Lee, M.~Lam, R.~Pedarsani, D.~Papailiopoulos, and K.~Ramchandran.
\newblock Speeding up distributed machine learning using codes.
\newblock {\em IEEE Trans. on Info. Theory}, 64(3):1514--1529, March 2018.

\bibitem{EBPIR}
Q.~Wang and M.~Skoglund.
\newblock On {PIR} and symmetric {PIR} from colluding databases with
  adversaries and eavesdroppers.
\newblock {\em IEEE Trans. on Info. Theory}, 65(5):3183--3197, May 2019.

\bibitem{PSI_journal}
Z.~Wang, K.~Banawan, and S.~Ulukus.
\newblock Private set intersection: A multi-message symmetric private
  information retrieval perspective.
\newblock {\em IEEE Trans. on Info. Theory}, 68(3):2001--2019, March 2022.

\bibitem{MP-PSI_journal}
Z.~Wang, K.~Banawan, and S.~Ulukus.
\newblock Multi-party private set intersection: An information-theoretic
  approach.
\newblock {\em IEEE Jour. on Selected Areas in Info. Theory}, 2(1):366--379,
  March 2021.

\bibitem{DoubleBlind_PIR}
Y.~Lu, Z.~Jia, and S.~A. Jafar.
\newblock Double blind {$T$}-private information retrieval.
\newblock {\em IEEE Jour. on Selected Areas in Info. Theory}, 2(1):428--440,
  March 2021.

\bibitem{MultiBlind_SPIR}
J.~Zhu, Q.~Yan, and X.~Tang.
\newblock Multi-user blind symmetric private information retrieval from coded
  servers.
\newblock {\em IEEE Jour. on Sel, Areas in Commun.}, 40(3):815--831, March
  2022.

\bibitem{ChaoTian_leakage}
T.~Guo, R.~Zhou, and C.~Tian.
\newblock On the information leakage in private information retrieval systems.
\newblock {\em IEEE Trans. on Info. Forensics and Security}, 15:2999--3012,
  2020.

\bibitem{AleakyPIR}
I.~Samy, M.~Attia, R.~Tandon, and L.~Lazos.
\newblock Asymmetric leaky private information retrieval.
\newblock {\em IEEE Trans. on Info. Theory}, 67(8):5352--5369, August 2021.

\bibitem{Uniform_SS}
M.~Yoshida, T.~Fujiwara, and M.~P.~C. Fossorier.
\newblock Optimal uniform secret sharing.
\newblock {\em IEEE Trans. on Info. Theory}, 65(1):436--443, January 2019.

\bibitem{Product-Matrix_Codes}
K.~V. Rashmi, N.~B. Shah, and P.~V. Kumar.
\newblock Optimal exact-regenerating codes for distributed storage at the {MSR}
  and {MBR} points via a product-matrix construction.
\newblock {\em IEEE Trans. on Info. Theory}, 57(8):5227--5239, August 2011.

\bibitem{IT_ColludingUser}
Z.~Li, Y.~Zhao, and H.~Sun.
\newblock Weakly secure summation with colluding users.
\newblock Available at arXiv:2304.09771.

\bibitem{LCC}
Q.~Yu, S.~Li, N.~Raviv, S.~M.~M. Kalan, M.~Soltanolkotabi, and S.~A.
  Avestimehr.
\newblock Lagrange coded computing: Optimal design for resiliency, security,
  and privacy.
\newblock In {\em AISTATS}, pages 1215--1225, April 2019.

\end{thebibliography}

\end{document}